\documentclass[twocolumn, nofootinbib, showpacs, superscriptaddress,floatfix]{revtex4} 
\usepackage{amsfonts}
\usepackage{amsmath}
\usepackage{ulem}
\usepackage{dsfont}
\usepackage[pdftex]{graphicx}
\usepackage{epstopdf}
\usepackage{textcomp} 
\usepackage{wasysym} 
\usepackage{graphicx}
\usepackage{bm}
\usepackage{cancel}
\usepackage{float}
\usepackage{placeins}
\usepackage[varg]{txfonts} 
\usepackage{xcolor}
\usepackage{comment}
\usepackage{multirow}
\usepackage{hyperref}
\hypersetup{
    colorlinks=true,
    linkcolor=red,
    citecolor=blue,
} 

\specialcomment{supplement}{\begingroup\color{cyan}\footnotesize}{\endgroup}
\specialcomment{formula}{\begingroup\footnotesize}{\endgroup}
\excludecomment{details}

\newcommand{\highlight}[1]{#1}

\newcommand{\sigx}{{\sigma_{\! \! x}}}
\newcommand{\sigu}{{\sigma_{\! \! u}}}
\newcommand{\sigq}{{\sigma_{\! \! q}}}

\newcommand{\Mpc}{{\,\mathrm{Mpc}/h}}

\newcommand{\del}[0]{\partial }
\newcommand{\sH}[0]{{\mathcal{H}}}
\newcommand{\hvz}{\hat{\v{z}}}
\newcommand{\vol}[2]{\hspace{-0.8mm}\mbox{$\text{d}^{\hspace{-0.0mm}#1}$}\hspace{-0.2mm}#2\hspace{0.8mm}\ }
\newcommand{\varvol}[2]{\hspace{-0.0mm}\mbox{$\text{d}^{\hspace{-0.0mm}#1}$}\hspace{-0.2mm}#2\hspace{0.8mm}\!}
\renewcommand{\d}{\text{d}}
\renewcommand{\v}[1]{\bm{#1} }
\newcommand{\vx}[0]{\bm{x} }

\newcommand{\vz}[0]{\bm{z} }
\newcommand{\vp}[0]{\bm{p} }
\newcommand{\vk}[0]{\bm{k} }
\newcommand{\vq}[0]{\bm{q} }

\newcommand{\lgM}{\mathrm{lgM} }
\begin{document}
\title{Edgeworth streaming model for redshift space distortions}
\author{Cora Uhlemann} 
\email{cora.uhlemann@physik.lmu.de}
\affiliation{Arnold Sommerfeld Center for Theoretical Physics, Ludwig-Maximilians-Universit\"at, Theresienstrasse 37, 80333 Munich, Germany} 
\affiliation{Excellence Cluster Universe, Boltzmannstrasse 2, 85748 Garching, Germany} 
\author{Michael Kopp} 
\email{kopp.michael@ucy.ac.cy}
\affiliation{Arnold Sommerfeld Center for Theoretical Physics, Ludwig-Maximilians-Universit\"at, Theresienstrasse 37, 80333 Munich, Germany} 
\affiliation{Excellence Cluster Universe, Boltzmannstrasse 2, 85748 Garching, Germany} 
\affiliation{University Observatory, Ludwig-Maximilians University Munich, Scheinerstrasse 1, 81679 Munich, Germany} 
\author{Thomas Haugg} 
\email{thomas.haugg@physik.lmu.de}
\affiliation{Arnold Sommerfeld Center for Theoretical Physics, Ludwig-Maximilians-Universit\"at, Theresienstrasse 37, 80333 Munich, Germany} 

\begin{abstract}
We derive the Edgeworth streaming model (ESM) for the redshift space correlation function starting from an arbitrary distribution function for biased tracers of dark matter by considering its two-point statistics and show that it reduces to the Gaussian streaming model (GSM) when neglecting non-Gaussianities. We test the accuracy of the GSM and ESM independent of perturbation theory using the Horizon Run 2 $N$-body halo catalog.
While the monopole of the redshift space halo correlation function is well described by the GSM, higher multipoles improve upon including the leading order non-Gaussian correction in the ESM: the GSM quadrupole breaks down on scales below 30 Mpc$/h$ whereas the ESM stays accurate to 2\% within statistical errors down to 10 Mpc$/h$.
To predict the scale dependent functions entering the streaming model we employ Convolution Lagrangian perturbation theory (CLPT) based on the dust model and local Lagrangian bias.
Since dark matter halos carry an intrinsic length scale given by their Lagrangian radius, we extend CLPT to the coarse-grained dust model and consider two different smoothing approaches operating in Eulerian and Lagrangian space, respectively. The coarse-graining in Eulerian space features modified fluid dynamics different from dust while the coarse-graining in Lagrangian space is performed in the initial conditions with subsequent single streaming dust dynamics, implemented by smoothing the initial power spectrum in the spirit of the truncated Zel'dovich approximation.  Finally, we compare the predictions of the different coarse-grained models for the streaming model ingredients to $N$-body measurements and comment on the proper choice of both the tracer distribution function and the smoothing scale. Since the perturbative methods we considered are not yet accurate enough on small scales, the GSM is sufficient when applied to perturbation theory. 
\end{abstract}
\maketitle

\section{Introduction}
Redshift space distortions observed in galaxy surveys provide a unique insight into the build-up of cosmological structure by gravitational clustering of dark matter and its tracers such as halos and galaxies. Indeed, the redshift space two point correlation function carries valuable information on both, the real-space clustering and the peculiar velocity field since the observed redshift depends not only on distance but also on deviations from the overall Hubble flow. Peculiar velocities are generated by and hence correlated with the clustering of matter.


There are two main effects in redshift space, \highlight{a term introduced in \cite{ST77}}, that affect the correlation function on large and small scales, respectively. On large scales the peculiar velocity associated with the coherent infall into overdense regions squashes structures and enhances the correlation function along the line of sight which is captured by linear theory and known as the Kaiser effect \cite{K87}. On small scales, the elongation of nonlinear structures along the line of sight, the so-called `Fingers of God' effect \highlight{coined in \cite{TF78} and first described in \cite{J72}}, leads to a suppression of the correlation function. Based on this observation one of the first streaming models was developed in \cite{P80} by assuming an exponential relative or pairwise velocity distribution with a scale-independent dispersion. 
\highlight{Dispersion models \cite{PVGH94,PD94} aimed to phenomenologically combine effects of linear clustering and small-scale velocity dispersion which act both multiplicative onto the redshift space power spectrum when their correlation is neglected. However, they have been shown to be unable to properly account for redshift space distortions over a vast range of scales by means of $N$-body simulations \cite{HC98,TN11}.}

To reunite the two disparate results for large and small scales, given by the linear theory \cite{K87} and the streaming model for nonlinear scales \cite{P80}, the so-called Gaussian streaming model (GSM) was introduced in \cite{F94}. To obtain the GSM, the matter correlation function in redshift space was derived by considering the joint probability distribution of density and velocity. Assuming that the density is a Gaussian random field and the velocity is related to density as in linear perturbation theory one obtains a simple expression for the redshift space correlation function. It is given by a convolution of the real space correlation function and an approximately Gaussian pairwise velocity distribution whose mean and variance are given by the scale-dependent mean and variance of the pairwise velocity. The GSM obtained via this approach can be understood as generalization of the streaming model originally introduced in \cite{P80} to a scale-dependent rather than constant velocity dispersion which correctly reproduces the linear theory result \cite{K87}. \highlight{The GSM, derived for the special case of Gaussian fluctuations in \cite{F94}, has been generalized to fully non-Gaussian fields in \cite{S04}. Furthermore, therein a connection between the redshift-space clustering and the pairwise velocity moments has been established.}

Furthermore, it has been shown recently in \cite{BCG15} that the assumption that the pairwise velocity distribution is locally Gaussian, with its mean and variance themselves Gaussian distributed allows to accurately recover the non-Gaussian pairwise velocity distribution measured in simulations. This approach is different from the one presented here, where we assume that the mean and variance are not random variables but functions of separation that are either determined from data or inferred from theory.

We start from a phase space distribution function for dark matter or its tracers, similar as done in \cite{VSMOB12,VSOD13}. Indeed, our formulation relates the distribution function approach studied in Fourier space in \cite{VSMOB12,VSOD13} and the Gaussian streaming model for redshift space distortions operating in configuration space. We decide to work in configuration rather than Fourier space. A practical reason is that our formulation of perturbation theory will naturally produce expressions in real space. Another argument is the fact that small spatial scales in the correlation function can be strongly affected by late-time baryonic physics, while large scales, most importantly the baryon acoustic oscillations (BAO) peak, are not affected, see \cite{AWSH13}. Therefore although late-time baryonic physics is confined to small $r$, it appears spread out in Fourier space. \highlight{A generalized dispersion model, taking nonlinear couplings between density and velocity fields into account, has been proposed in \cite{TNB13} to provide consistent predictions for power spectra and correlation functions at the same time.}



To predict halo correlation functions in redshift space the GSM has been combined with perturbation theory to extract the streaming model ingredients, namely the real space correlation and the mean pairwise velocity and its dispersion, in \cite{RW11,WRW14,W14}. A test of different analytic and phenomenological streaming models combined with perturbation theory, performed in \cite{WR14}, showed that they reasonably fit the simulations on intermediate scales $40\Mpc\lesssim s \lesssim80\Mpc$ while all models fail at small scales with Lagrangian schemes having the best performance around the scale of BAO.

It is well known that no perturbative framework is able to accurately describe the fully nonlinear regime of structure formation. Fortunately, dark matter halos and their progenitors, which we denote by proto-halos, can themselves be treated as large cold dark matter (CDM) particles and therefore described by a pressureless dust fluid. The motion of these proto-halos is mostly determined by the large scale gravitational field and therefore much better describable with perturbation theory. The pressureless CDM fluid is described by a coupled system of differential equations consisting of continuity, Euler and Poisson equations. These equations can be solved perturbatively -- either in the Eulerian frame (SPT)\cite{B01} where everything is expanded in terms of density and velocity or in the Lagrangian frame (LPT) \cite{B92} where fluid-trajectories or displacement fields are considered. 
It is clear that the fluid description should be applied only on scales larger than the particle size, in case of proto-halo ``particles'' this is the Lagrangian size of the halos. Therefore it is natural to implement the Lagrangian halo size as a physically meaningful coarse-graining scale into the fluid description for (proto-)halos \cite{BM96}. This approach is to be seen in contrast to the so-called effective field theory of LSS \cite{PSZ13} for dark matter where the dependence of dark matter properties on the smoothing scale is unphysical and removed through renormalization. In order to model the trajectories of proto-halos we study in this paper a coarse-grained dust model in terms of the displacement field within Lagrangian perturbation theory (LPT). 

A big advantage of Lagrangian schemes \cite{B92} is the clearer physical picture they offer for the study of halo correlation functions, which are a key ingredient of the halo model \cite{CS02}  that is widely used in the analysis of galaxy, cluster and lensing surveys. In order to understand halo correlations one needs to understand the bias between the halo field and the underlying dark matter field. But halo bias is best understood using the spherical collapse model and excursion set theory \cite{Bondetal,Laceyetal}, both of which operate in the initial conditions and therefore in Lagrangian space, where they locally identify proto-halos within the initial density field and assign mass and collapse time to them. Therefore once the clustered or biased field of proto-halos is known it can be propagated to Eulerian space using a Lagrangian method.
Another advantage of Lagrangian methods concerns the convergence properties and the accuracy of the correlation function on the scales of interest, like the BAO scale or the mildly nonlinear scales. It is known that LPT performs much better on those scales, see the first Figure of \cite{Ta14}; a higher precision is achieved with a smaller order in perturbation theory. The better convergence properties of the LPT displacement field compared to standard perturbation theory (SPT) in Eulerian space are mainly due to fact that the relation between the density contrast and the displacement field is nonlinear and can be handled nonperturbatively. 

In first order LPT it is possible to analytically compute the density correlation function from the first order displacement field in a nonperturbative fashion which is called Zel'dovich approximation (ZA), see \cite{Z70}. In the ZA particles are displaced  along straight trajectories, parametrized by the linear growth function, in a direction determined by their initial velocity. 
Despite its simplicity, the ZA is capable of accurately describing gravitational dynamics over a surprisingly wide range of scales \cite{C93, Ta14}. In \cite{C93} the so-called truncated Zel'dovich approximation (TZA) was proposed as phenomenological method to improve the agreement between Zel'dovich and proper $N$-body simulations by artificially smoothing the initial power spectrum at the nonlinear scale of the final time of the simulation. The effect of the smoothing is to decrease the velocity in high density regions thereby reducing the amount of shell-crossing events and subsequent erasure of overdensities. Therefore, counterintuitively, smoothing the initial power spectrum, which reduces the initial power on small scales, actually can increase the final power on those scales. Focusing on statistical properties of the nonlinearly evolved density field like the power spectrum, the TZA amounts to smoothing the linear initial power spectrum without affecting the dynamics itself. A detailed study and comparison between different filters in \cite{M94} revealed that a Gaussian filtering scheme leads to best agreement with $N$-body data and considerable improvement over sharp k-truncation as originally suggested in \cite{C93} and top-hat filtering as studied in \cite{P11}. 

It is known that the Post-Zel'dovich approximation (PZA), where the displacement fields are calculated from second order LPT, improves over the ZA. Accordingly, the truncated Post-Zel'dovich approximation (TPZA) with a smoothed initial power spectrum performs even better than TZA, compare \cite{BMW94,WGB95}. We apply the framework of Convolution Lagrangian perturbation theory (CLPT) developed in \cite{CRW13} which recovers the ZA at lowest order while providing an approximation to PZA at higher order. CLPT can be understood as a partial resummation of the formalism presented in \cite{M08} providing a nonperturbative resummation of LPT that incorporates nonlinear halo bias. 
We will compare two different smoothing approaches within CLPT, namely a coarse-graining in Eulerian space (cgCLPT) with a coarse-graining in Lagrangian space implemented by smoothing the initial power spectrum in the spirit of the truncated Zel'dovich approximation (TCLPT). Those two procedures are distinct since a coarse-graining in Eulerian space also modifies the underlying dynamics becoming manifest beyond linear order in Lagrangian space, see \cite{UK14}, while our coarse-graining in Lagrangian space only affects the initial conditions.

\paragraph*{Structure}

This paper is organized as follows: In Sec.\,\ref{sec:GSM} we derive the Edgeworth streaming model (ESM) for the redshift space correlation function starting from an arbitrary distribution function for biased tracers of dark matter by considering its two-point statistics and show that it reduces to the Gaussian streaming model (GSM) when neglecting non-Gaussianities in the pairwise velocity distribution. We then demonstrate the accuracy of the GSM and ESM on the basis of $N$-body simulations employing the Horizon Run 2 halo catalog. In Sec.\,\ref{sec:GSM-dust} we built up on existing work and describe how the ingredients of the streaming models can be inferred from the dust model and propose two different coarse-grained generalizations of the fluid description. In Sec.\,\ref{sec:GSM-CLPT} we  compute the real-space halo correlation function and the halo velocity statistics for the dust model employing Convolution Lagrangian Perturbation Theory (CLPT) with two different coarse-graining schemes, an Eulerian (cgCLPT) and a Lagrangian (TCLPT) one. We conclude and describe possible further interesting lines of study in Sec.\,\ref{sec:outlook}. A list of  abbreviations commonly used within this work can be found in App.\,\ref{AppAbb}.

\section{Edgeworth Streaming Model}
\label{sec:GSM}
In order to infer predictions for the halo correlation function in redshift space we use the Gaussian streaming model, originally derived in \cite{F94} and studied in \cite{RW11} for the dust model. Starting from an arbitrary distribution of proto-halos we present a self-contained derivation of the Gaussian streaming model (GSM) from general assumptions which allows to include non-Gaussian corrections leading to the Edgeworth streaming model (ESM). We test the accuracy of the GSM and ESM using $N$-body simulation data from the Horizon Run 2 (HR2) \cite{KPetal09, KPetal11} independent of perturbation theory. We then describe in Sec.\,\ref{sec:GSM-dust} how the ingredients of the streaming models can be inferred from the dust model and it's coarse-grained generalization and present the CLPT computation and results in Sec.\,\ref{sec:GSM-CLPT}.

\subsection{Derivation of the ESM}
Let the phase space distribution function of dark matter tracers $X$ (like galaxies, clusters or halos) be given by $f_X(\v{r},\v{u},t)$. In this section we do not make any assumptions about its dynamics or statistical properties apart from that it is spatially statistically homogeneous
\begin{equation}
\langle f_X(\v{r}_1,\v{u}_1,t) f_X(\v{r}_2,\v{u}_2,t) \rangle = \langle f_X f_X \rangle (\v{r}= \v{r}_2 - \v{r}_1 ,  \v{u}_1,\v{u}_2,t)\,.
\end{equation}
In addition we assume that the tracer density field
\begin{align}
1+\delta_X(\v{r},t) = \int \vol{3}{u} f_X(\v{r},\v{u},t)
\end{align}
\highlight{and higher moments} $ \int \vol{3}{u} \!\! f_X(\v{r},\v{u},t) u_{i_1}...u_{i_n}$ are statistically  homogeneous and isotropic.

The observed position of a tracer $\v{s}_{\rm obs}$ -- its angle on the sky $\hat{\v{n}}_{\rm obs}$ and its observed redshift $z_{\rm obs}$ -- corresponds to a point on the observer's past light cone. As a first step towards calculating tracer correlations on the past light cone, we will make two common simplifying assumptions. First, since we are interested in equal-time correlation functions, we will approximate the light cone in the neighbourhood of $t$ by the $t$=const slice. Secondly, we use the distant observer approximation, where the line of sight is assumed to be a fixed direction $\hvz$ which is without loss of generality chosen as the direction of the $z-$axis, to relate the observed redshift-space position $\v{s}$ of a dark matter tracer to its real-space position $\v{r}$ . Those approximations are despite their simplicity sufficient even for modern wide-area surveys within the level of current error bars, see e.g. Fig.\,10 in \cite{SPR12}. For a general definition of redshift space and a discussion of wide-angle effects in linear perturbation theory we refer to \cite{M00}. In the distant observer approximation the observed comoving distance in redshift space $\v{s}$ is affected by the peculiar velocity $\v{v}\cdot \hvz =v_z$ of the tracer along the line of sight via
\begin{subequations}
\label{RSspace}
\begin{equation}
\label{defs}
\v{s} = \v{r}+ \sH^{-1} (\v{v}\cdot \hvz)\ \hvz \,,
\end{equation}
where $\mathcal{H} = a H = \dot{a}$ and $\v{u}=a\v{v}$. The observed position of the tracer perpendicular to the line of sight $\v{s}_\perp$ remains unaffected if we neglect gravitational lensing. In contrast, its coordinate $s_{||}$ parallel to the line of sight $\hvz$  depends on the peculiar velocity $v_z$ 
\begin{equation}
\v{s}_\perp = \v{r}_\perp \quad \,, \quad s_{||}=\v{s}\cdot \hvz =r_{||} + \sH^{-1} v_z\,.
\end{equation}
\end{subequations}
Since objects cannot disappear going from real space to redshift space (assuming that all objects remain observable) we have the following relation between the densities in real and redshift space
\begin{align}
(1+\delta_X(\v{s},t))\, \varvol{3}{s} &= (1+\delta_X(\v{r},t))\, \varvol{3}{r} \,.\label{realtoredshift}
\end{align}
Although the correction to the real space position in redshift space is very small $\sH^{-1} v_z \ll r_{||}$, the clustering is affected considerably since the change of volume measure between real and redshift space, given by the Jacobian between $\varvol{3}{s}$ and $\varvol{3}{r}$, involves the gradient of $v_z$ in linear perturbation theory \cite{K87}.
In the distant observer approximation, the tracer density fluctuation in redshift space \eqref{realtoredshift} can be equivalently written as
\begin{align} \label{TracerflucGeneralf}
1+\delta_X(\v{s},t) = \int \vol{3}{r} \!\!\!\int \vol{3}{u} f_X(\v{r},\v{u},t) \delta_{\rm D} \left(\v{s} - \v{r} - \frac{\v{u}\cdot \hvz}{a^2 H} \hvz\right)\,,
 \end{align} 
which holds even for the case where the tracer velocity $\v{v}$ is not a single valued function of $\v{r}$ but instead has multiple streams or a continuous distribution, see also \cite{SMc11}. Later, in Sec.\,\ref{sec:GSM-dust}, we will consider the special case of single streaming tracers described by the dust model for which this relation simplifies to \eqref{deltasr}.
 
We are interested in the redshift space two-point correlation function
\begin{equation}
\label{2ptcorrfctgeneral}
1+\xi_X(\v{s},t) =  \Big\langle (1+\delta_X(\v{s}_1)) (1+\delta_X(\v{s}_2)) \Big\rangle \,,
\end{equation}
where $\v{s}=\v{s}_2-\v{s}_1$. By inserting \eqref{TracerflucGeneralf} in \eqref{2ptcorrfctgeneral} and re-expressing the delta functions in Fourier space and integrating over $\v{R}=\v{r}_1+\v{r}_2$ and one momentum variable the correlation function can be brought into the following form
 \begin{subequations} \label{xizspacegeneral}
\begin{align}
\label{xizspaceexprgeneral}
1+\xi_X(\v{s},t) &=\int \vol{3}{r}\!\! \int \frac{\vol{3}{k}}{(2\pi)^3} e^{i \v{k}\cdot (\v{r} - \v{s}) } Z\Big(\v{r}, \v{J} = (\v{k}\cdot\hvz)\ \hvz , t\Big)\,,\\
\notag Z(\v{r}, \v{J}, t) &= \int \vol{3}{u_1}\!\!\!\!\!\int \vol{3}{u_2} \langle f_X f_X \rangle (\v{r},  \v{u}_1,\v{u}_2,t)\\
&\qquad\qquad\qquad \times \exp\left[i \frac{(\v{u}_2 - \v{u}_1) \cdot \v{J}}{a^2 H}\right]\label{defZgeneral}\,,
\end{align}
\end{subequations}
where $Z$ is the pairwise generating function.
Next we Taylor expand $W(\v{J}):= \ln Z$ around $\v{J}=0$
\begin{equation} \label{WexpinJ}
W(\v{J}) = \sum_{n=0}^{\infty} \frac{1}{n!} \v{\kappa}_n(i \v{J})^n \ , \ \v{\kappa}_n:= \frac{\partial^n W}{(\partial i \v{J})^n}\Bigg|_{J=0}\,.
\end{equation}
Keeping only the terms up to third order $n=3$ we obtain
\begin{subequations}
\label{Wexpansion}
\begin{align}
W(\v{J}) \simeq &\ln (1+ \xi_X(r,t))+ i \v{v}_{12} \cdot \v{J} - \frac{1}{2}\v{J}^{T} \v{\sigma}^2_{12} \v{J} \label{W2}\\
&  - \frac{i}{6} \Lambda_{12}^{ijk} J^{i}J^{j}J^k \,, \label{W3}
\end{align}
\end{subequations}
with the cumulants $\v{\kappa}_n$ as expansion coefficients 
\begin{subequations}
\label{GSMparam}
\begin{align}
1+ \xi_X(r,t) &:= \exp \kappa_0 = Z\, |_{J=0} \,,\label{xirel}\\
\v{v}_{12}(\v{r},t) & := \v{\kappa}_1  = \frac{\frac{\partial Z}{(\partial i \v{J})}\big|_{J=0}}{(1+ \xi_X(r,t))} \,,\label{v12rel}\\
\v{\sigma}^2_{12}(\v{r},t) & :=  \v{\kappa}_2  = \frac{\frac{\partial^2 Z}{(i\partial \v{J})^2}\big|_{J=0}}{(1+ \xi_X(r,t))} -\frac{\frac{\partial Z}{(i\partial \v{J})} \frac{\partial Z}{(i\partial \v{J})}\big|_{J=0}}{(1+ \xi_X(r,t))^2}  \notag\\
&=  \tilde{\v{\sigma}}^2_{12}(\v{r},t) -\v{v}_{12}(\v{r},t) \v{v}_{12}(\v{r},t)\label{sigma12rel}\,,\\
\Lambda^{ijk}_{12}(\v{r},t) &:= \kappa_3^{ijk} = \tilde{\Lambda}_{12}^{ijk} - (\sigma^2_{12})^{(ij} v^{k)}_{12}-  v^{i}_{12}v^{j}_{12}v^{k}_{12}\,, \label{Lambda12rel}
\end{align}
\end{subequations}
where $A^{(ij}B^{k)}:=A^{ij}B^k+A^{jk}B^i+A^{ki}B^j$. Since we have to evaluate all expressions at $\v{J} = (\v{k}\cdot\hvz)\ \hvz$ we project the cumulants $\v{\kappa}_n$ onto the line of sight $\kappa_n= \kappa_{n}^{i_1\cdots i_n} \hat{z}_{i_1}\cdots \hat{z}_{i_n}$ \eqref{cumproj}. Expanding $W$ in \eqref{W2} up to second order in $\v{J}$ implies that all redshift space distortion induced clustering is encoded in the scale dependent mean and variance given by the pairwise velocity $v_{12}$ and its dispersion $\sigma_{12}^2$. As we will shortly see, this corresponds to the Gaussian streaming model (GSM). Since the GSM is known to be a good approximation, we will perform an expansion around this Gaussian
\begin{align*}
&\exp\left[\ln (1+ \xi_X(r,t))+ i \v{v}_{12} \cdot \v{J} - \frac{1}{2}\v{J}^{T} \v{\sigma}^2_{12} \v{J}  - \frac{i}{6} \Lambda_{12}^{ijk} J^{i}J^{j}J^k\right]\\ &\approx (1+ \xi_X(r,t)) \exp\left[i \v{v}_{12} \cdot \v{J} - \frac{1}{2}\v{J}^{T} \v{\sigma}^2_{12} \v{J} \right] \left[1- \frac{i}{6} \Lambda_{12}^{ijk} J^{i}J^{j}J^k\right] \,.
\end{align*}
This approach is similar to the idea behind Convolution Lagrangian perturbation theory (CLPT), see \cite{CRW13}. To obtain the Gaussian streaming model it is crucial to expand in cumulants and keep the pairwise velocity mean and dispersion in the exponent, corresponding to specific resummation of moments. Within the distribution function approach to redshift space distortions developed in \cite{SMc11,VSMOB12,VSOD13} a moment expansion without such an resummation was performed such that the connection to the Gaussian streaming model is not manifest and has not been discussed.

Later when testing the accuracy of this model, we will restrict ourselves to the leading order non-Gaussian term. However one can systematically expand the exponential of the non-Gaussian contributions to $Z(\v{J})$ in an Edgeworth expansion \cite{BK94,JWACB95} around a Gaussian pairwise velocity probability distribution. The Edgeworth series $E_n$ is an asymptotic expansion to approximate a probability distribution using its cumulants $\kappa_n$. With the Gaussian distribution as reference function it can be written as, see Eq.\,(43) in \cite{BM97},
\begin{subequations}
\highlight{
\begin{align}
\label{Edgeworth}
 E_n(x)=&\frac{1}{\sqrt{2\pi\kappa_2}}\exp\left(-\frac{(x-\kappa_1)^2}{2\kappa_2}\right)\\
\notag &\ \times \Bigg[1+ \sum_{s = 1}^n \sum_{r = 1}^s \frac{B_{s,r}(\lambda_3,..., \lambda_{s-r+3})}{s!}  H_{s+ 2 r}\left(\frac{x-\kappa_1}{\sqrt {\kappa_2}}\right) \Bigg] \,,
\end{align}
where $\lambda_n$ are the normalized and rescaled cumulants
\begin{equation}
\lambda_n \equiv \frac{\kappa_n}{n(n-1) \kappa_2^{n/2}} \,,
\end{equation}
}
$B_{s,r}$ the Bell polynomials 
\begin{align}
\label{Bell}
B_{1,1}(\lambda_3)=\lambda_3 \ , \
B_{2,1}(\lambda_3,\lambda_4)=\lambda_4 \ ,\
B_{2,2}(\lambda_3)=\lambda_3^2 \,,
\end{align}
and $H_n$ the probabilists' Hermite polynomials
\begin{align}
\label{Hermite}
H_3(x)&=x^3-3x\ ,\ H_4(x)=x^4-6x^2+3\,,\\
\notag H_6(x)&=x^6-15x^4+45x^2-15\,.
\end{align}
\end{subequations}
In the following we perform the Edgeworth expansion up to $n=1$ explicitly taking into account the first non-Gaussian correction given by the pairwise velocity skewness $\kappa_3=\Lambda_{12}$. Kurtosis $\kappa_4$ would arise in the next order but won't be considered in this paper.

We can now plug the Edgeworth expansion \eqref{Edgeworth} of $Z=\exp W$ according to Eq.\,\eqref{Wexpansion} into the correlation function Eq.\,\eqref{xizspaceexprgeneral}. In the course of the calculation we will use cylindrical coordinates 
$$\v{s} =  s_{\perp} [\cos (\phi)\hat{\v{x}} + \sin (\phi)\hat{\v{y}} ] + s_{||} \hvz  \,,$$
since $\xi_X(\v{s},t)$ does not depend on the angle $\phi$. Performing five of the six integrals in Eq.\,\eqref{xizspaceexprgeneral} we obtain the Gaussian streaming model (GSM) Eq.\,\eqref{GSM2} at second order in the cumulant expansion and the leading and up to order $n$ corrections of the Edgeworth streaming model (ESM) \eqref{ESM2}, \eqref{ESMnlo} at third and $n$-th order, respectively
\begin{subequations}
\label{ESM}
\begin{align} 
1+\xi_X(s_{||},s_\perp,t)&= \int^{\infty}_{-\infty} \frac{ \vol{}{r_{||}}}{\sqrt{2 \pi} \sigma_{12}(r,r_{||},t)}  (1+\xi_X(r,t)) \notag\\ 
&\!\!\!\!\!\!\!\!\! \!\!\!\! \! \!\times\exp\left[-\frac{\left(s_{||}- r_{||} - v_{12}(r,t) r_{||}/r\right)^2}{2 \sigma_{12}^2(r,r_{||},t)}\right] \label{GSM2}\\
&\!\!\!\!\!\!\!\!\! \!\!\!\! \! \!\times \Bigg[ 1+\frac{\Lambda_{12}}{6 \sigma_{12}^3} H_3\left(\frac{ \Delta_{srv} }{\sigma_{12}}\right) \label{ESM2} \\
 & \!\!\!\!\!\!\!\!\! \! \! \!+  \sum_{s = 2}^{n=\infty} \sum_{r = 1}^s \frac{B_{s,r}(\lambda_3,..., \lambda_{s-r+3})}{s!}  H_{s+ 2 r}\left(\frac{ \Delta_{srv} }{\sigma_{12}}\right)\Bigg]\,. \label{ESMnlo}
\end{align}
\end{subequations}

In more detail, the $k_z$ integral in Eq.\,\eqref{xizspaceexprgeneral} introduces the pairwise probability distribution multiplied by $(1+ \xi_X(r,t))$, while the trivial $k_x, k_y$ integrals enforce $\v{r}_\perp = \v{s}_\perp$. The $r_{\perp}$ integral ensures $r^2 = r_{||}^2 + s_\perp^2$, while the $\phi$-integral gives a factor of $2\pi$. 
We defined
\begin{subequations}
\label{cumproj}
\begin{align}
v_{12}(r,t) r_{||}/r &:= \v{v}_{12}(\v{r},t) \cdot \hvz = \kappa_1\,, \\
\notag \Delta_{srv}&:=s_{||}- r_{||} - v_{12}(r,t) r_{||}/r\,,\\
\notag \sigma^2_{12}(r,r_{||},t) &:= \hvz^{T} \v{\sigma}^2_{12}(\v{r},t)  \hvz = \kappa_2\\
&= \tilde\sigma_{12}^2(r,r_{||},t) -v_{12}(r,t)^2(r_{||}/r)^2  \\
 &= \sigma_{||}^2(r,t) (r_{||}/r)^2 +\sigma_\perp^2(r,t)\left[1-(r_{||}/r)^2\right] \,, \label{sigperpparsplit}\\
\Lambda_{12} &:= \Lambda_{12}^{ijk}   \hat{z}^i\hat{z}^j\hat{z}^k = \kappa_3 \label{lamperpparsplit} \\
\notag &= \left(\Lambda_{||} (r_{||}/r)^2 +\Lambda_\perp\left[1-(r_{||}/r)^2\right] \right) r_{||}/r\,.
\end{align}
\end{subequations}

In a previous study of the GSM \cite{RW11}, the following formula, inspired by the exact result from \cite{F94} for the case where both density and velocity fields are Gaussian and related to one another as in linear theory, was suggested to calculate Gaussian streaming redshift space distortions 
\begin{align} \label{GSM}
1+\xi_X(s_{||},s_\perp,t)=& \int^{\infty}_{-\infty} \frac{ \vol{}{r_{||}}}{\sqrt{2 \pi} \tilde{\sigma}_{12}(r,r_{||},t)}  (1+\xi_X(r,t)) \notag\\
&\ \times\exp\left[-\frac{\left( s_{||} - r_{||} - v_{12}(r,t) r_{||}/r\right)^2}{2 \tilde{\sigma}_{12}^2(r,r_{||},t)}\right]\,,
\end{align}
where our $s_{\|}$ and $r_{\|}$ corresponds to $r_{\|}$ and $y$ used in \cite{RW11,WRW14}, respectively. Note that, \eqref{GSM} corresponds to \eqref{GSM2} when the variance, given by the second pairwise velocity moment $\tilde \sigma_{12}^2$, is replaced by the pairwise velocity dispersion $\sigma_{12}^2$. The two quantities are related via Eq.\,\eqref{sigperpparsplit} such that $ \sigma_{||}^2 =\tilde \sigma_{||}^2 - v_{12}^2$ and $ \sigma_{\perp}^2 =\tilde \sigma_{\perp}^2$. By expanding $Z\simeq\exp(W_0+W_1+W_2)$ one obtains the GSM \eqref{GSM2} with the second cumulant $\sigma_{12}^2$ as variance whereas when expanding $Z\simeq Z_0(1+Z_1/Z_0+Z_2/Z_0)\simeq Z_0\exp(Z_1/Z_0+Z_2/Z_0)$ one obtains the GSM \eqref{GSM} with the second moment $\tilde \sigma_{12}^2$ as variance. When linearized, both expressions \eqref{GSM2} and \eqref{GSM} agree, because $v_{12}^2$ is second order, and correctly reproduce the Kaiser formula as shown in \cite{F94,RW11}. 

\begin{figure}[h!]
\includegraphics[width=0.49\textwidth]{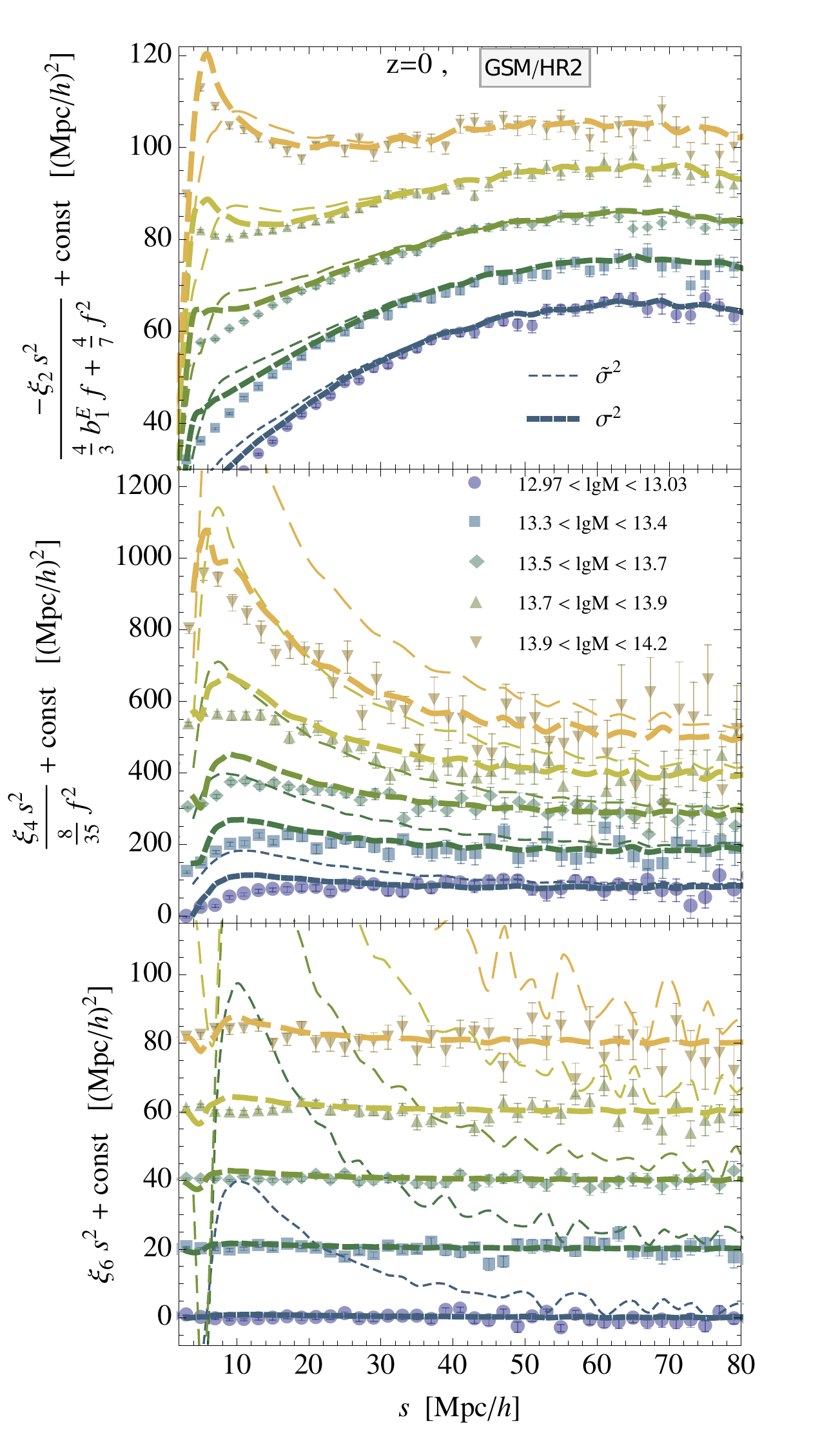}
\caption{Comparison between the Horizon Run 2 measurement ({\it data points}) of the multipoles of the redshift space correlation function defined in \eqref{ximultipoles} and
the GSM \eqref{GSM2} using $\sigma_{12}^2$ ({\it thick dashed}) or $\tilde\sigma_{12}^2$ ({\it thin dashed}) related via \eqref{sigperpparsplit}. {\it upper panel} The quadrupole $\xi_2$ shifted by $10 (i -1)(\Mpc)^2$ where $i$ labels the mass bin. {\it middle panel} The hexadecapole $\xi_4$ shifted by $100 (i -1) (\Mpc)^2$. {\it lower panel} The hexacontatetrapole (64-pole) $\xi_6$ shifted by $20 (i -1)(\Mpc)^2$.
 }
\label{fig:WvsZ}
\end{figure}

It is natural to follow an expansion in $W$ and to keep only the Gaussian part in the exponential in case the pairwise velocity distribution is close to a Gaussian. On the other hand the moment expansion of \cite{SMc11} is natural from a perturbation theory perspective, in which only moment spectra are kept that are nonzero up to certain order in perturbation theory. 

That the pairwise distribution function is indeed approximately Gaussian with a variance given by $\sigma_{12}$ rather than $\tilde\sigma_{12}$ becomes clear in Fig.\,\ref{fig:WvsZ}, where we compare the GSM with the second cumulant $\sigma_{12}^2$ \eqref{GSM2} to the GSM with the second moment $\tilde \sigma_{12}^2$ \eqref{GSM} as the variance of the Gaussian. The exact definition of the redshift space multipoles $\xi_n$ depicted in Fig.\,\ref{fig:WvsZ} and the reason for their normalization will be given in the next subsection. As we can clearly see the use of the second cumulant $\sigma_{12}^2$ significantly improves the agreement for the redshift space distribution function with the $N$-body simulation compared to the model where the second moment $\tilde\sigma_{12}^2$ is used. In \cite{RW11} it has been phenomenologically accounted for that difference by subtracting the square of the mean infall $v_{12}^2$ from $\tilde\sigma_{||}^2$ to get the dispersion about the mean. \highlight{We leave it for future work  to directly compare the ESM to the distribution function approach \cite{VSOD13}.}

\subsection{Accuracy of the GSM and ESM}
\label{accGSM}

\begin{figure*}[t!]
\includegraphics[width=0.335\textwidth]{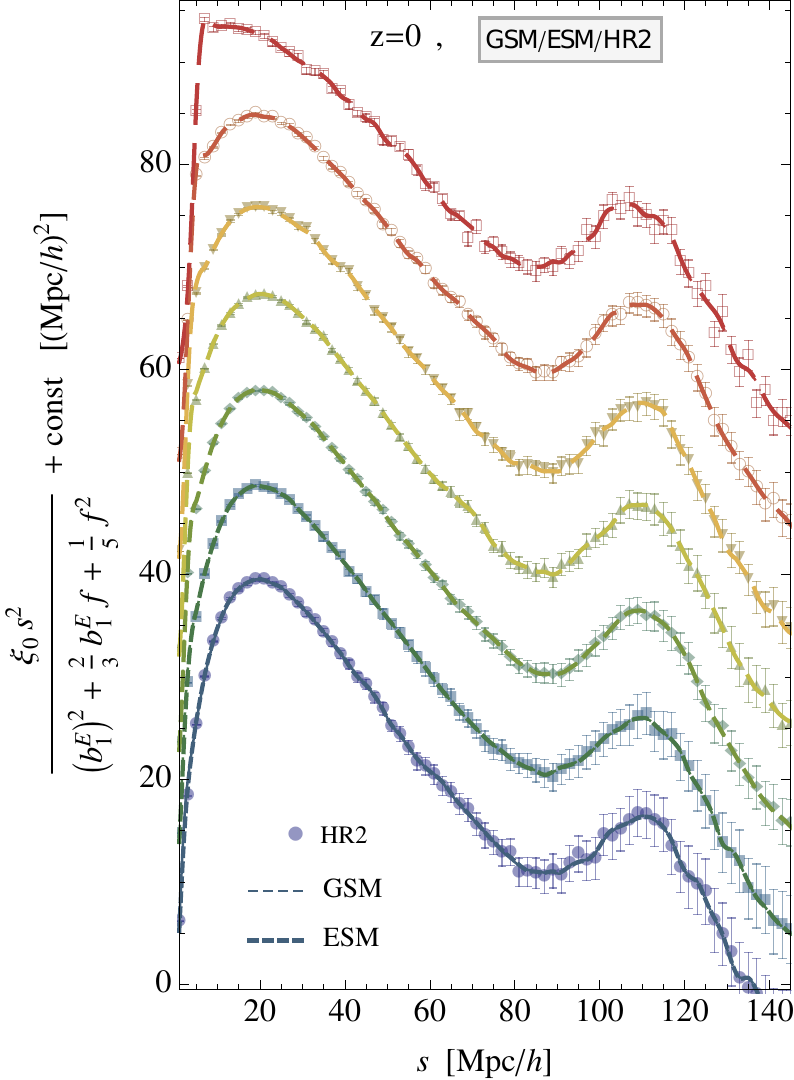}
\includegraphics[width=0.34\textwidth]{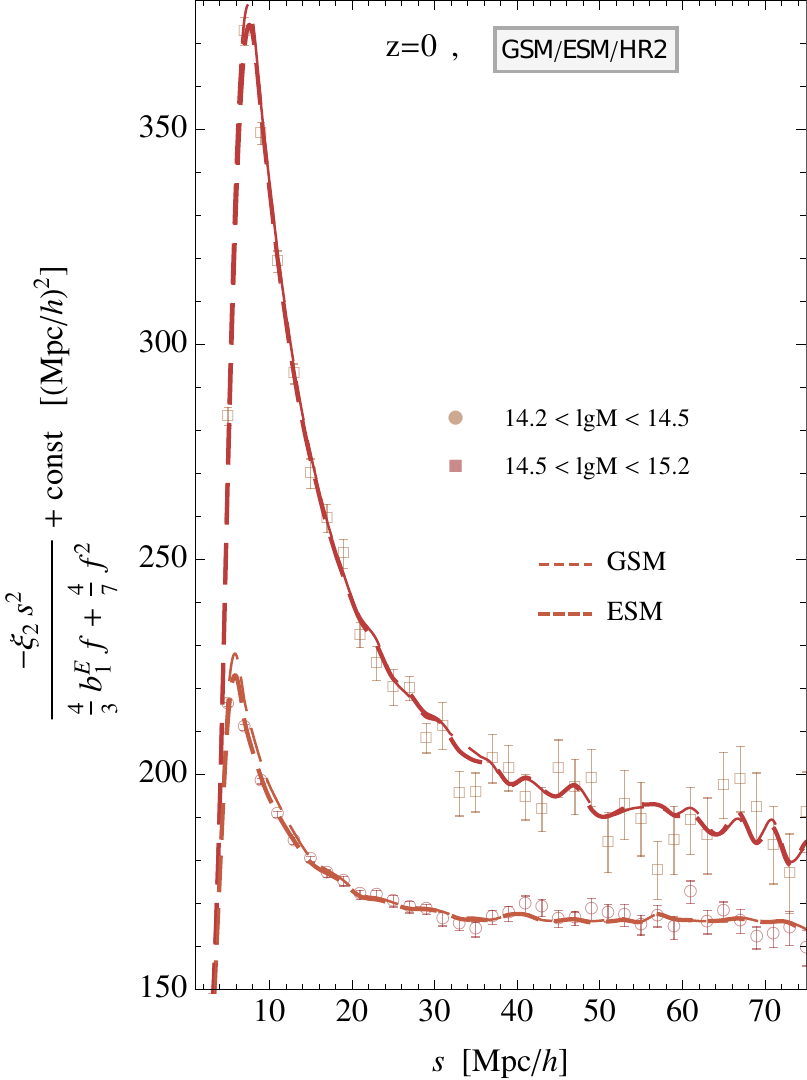}
\includegraphics[width=0.305\textwidth]{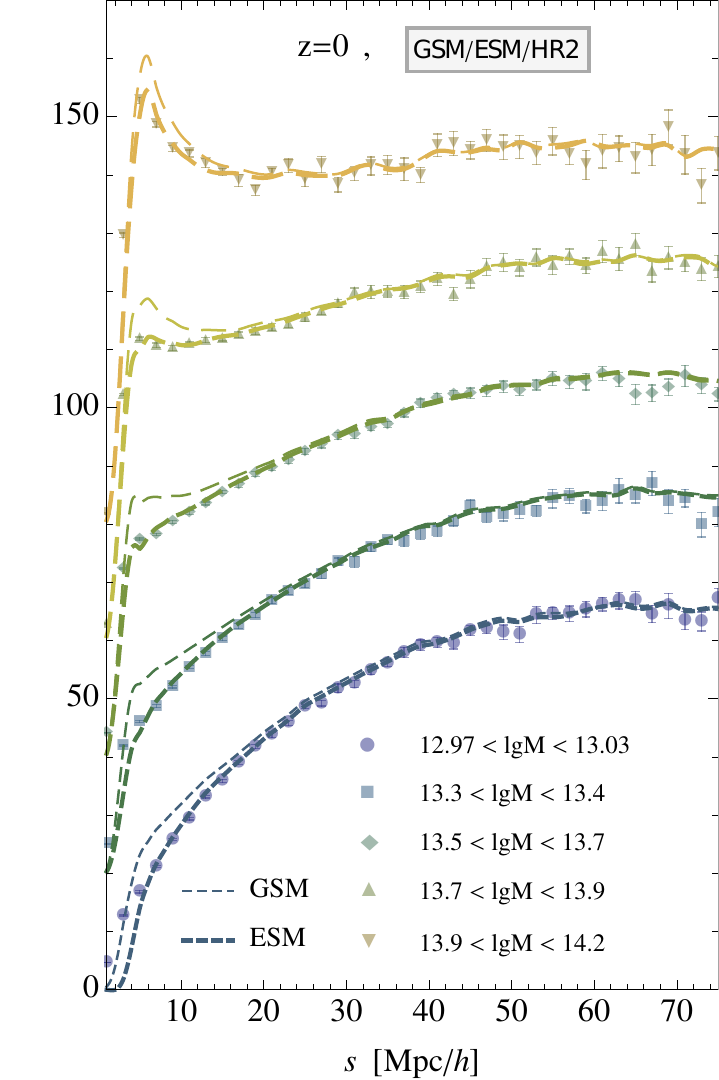}\\[12pt]
\includegraphics[width=0.50\textwidth]{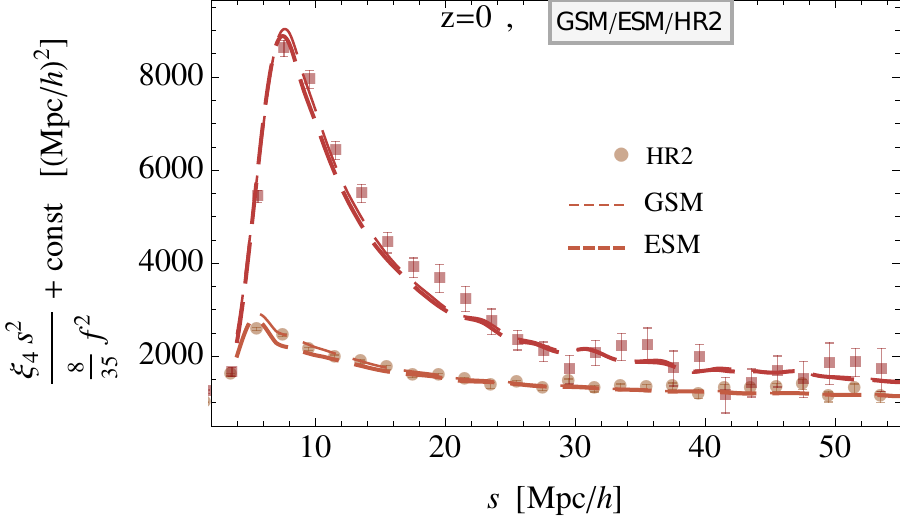}
\includegraphics[width=0.45\textwidth]{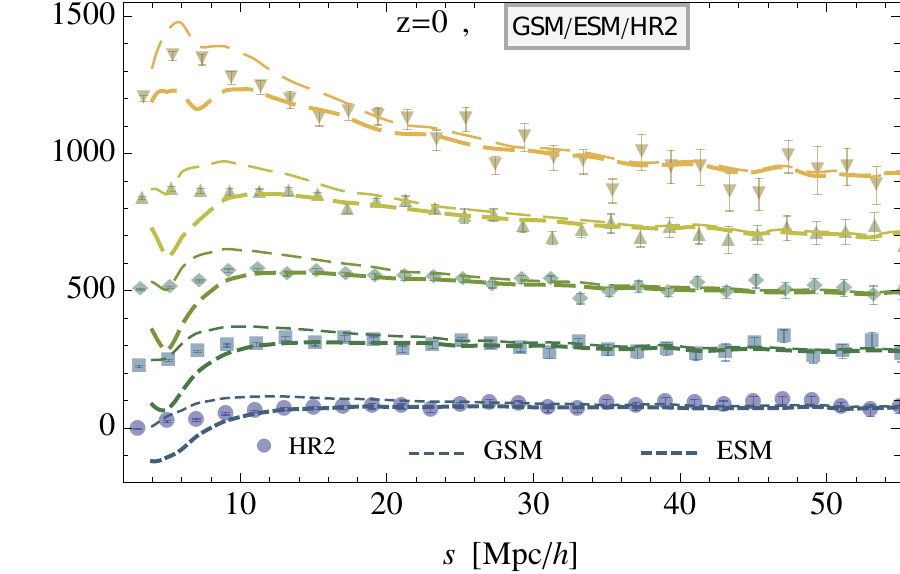}
\caption{The redshift-space multipoles $\xi_0$, $\xi_2$ and $\xi_4$ at $z=0$ predicted by the GSM \eqref{GSM2} ({\it thin dashed}) and ESM \eqref{ESM2} ({\it thick dashed}) using the HR2 data compared to the direct HR2 measurement ({\it data points}) normalized with respect to their bias factors for the different mass bins. Similar to previous plots we added a mass bin dependent constant to all curves for better visibility. 
{\it upper left panel} The monopole $\xi_0(s)$ shifted by $10(i-1)(\Mpc)^2$.
{\it upper middle and right panel} The quadrupole $\xi_2(s)$ shifted by $20(i-1)(\Mpc)^2$.
{\it lower panel} The hexadecapole $\xi_4(s)$ shifted by $200(i-1)(\Mpc)^2$.}
\label{fig:GSMcons02}
\end{figure*}

\begin{figure*}[t!]
\includegraphics[width=0.33\textwidth]{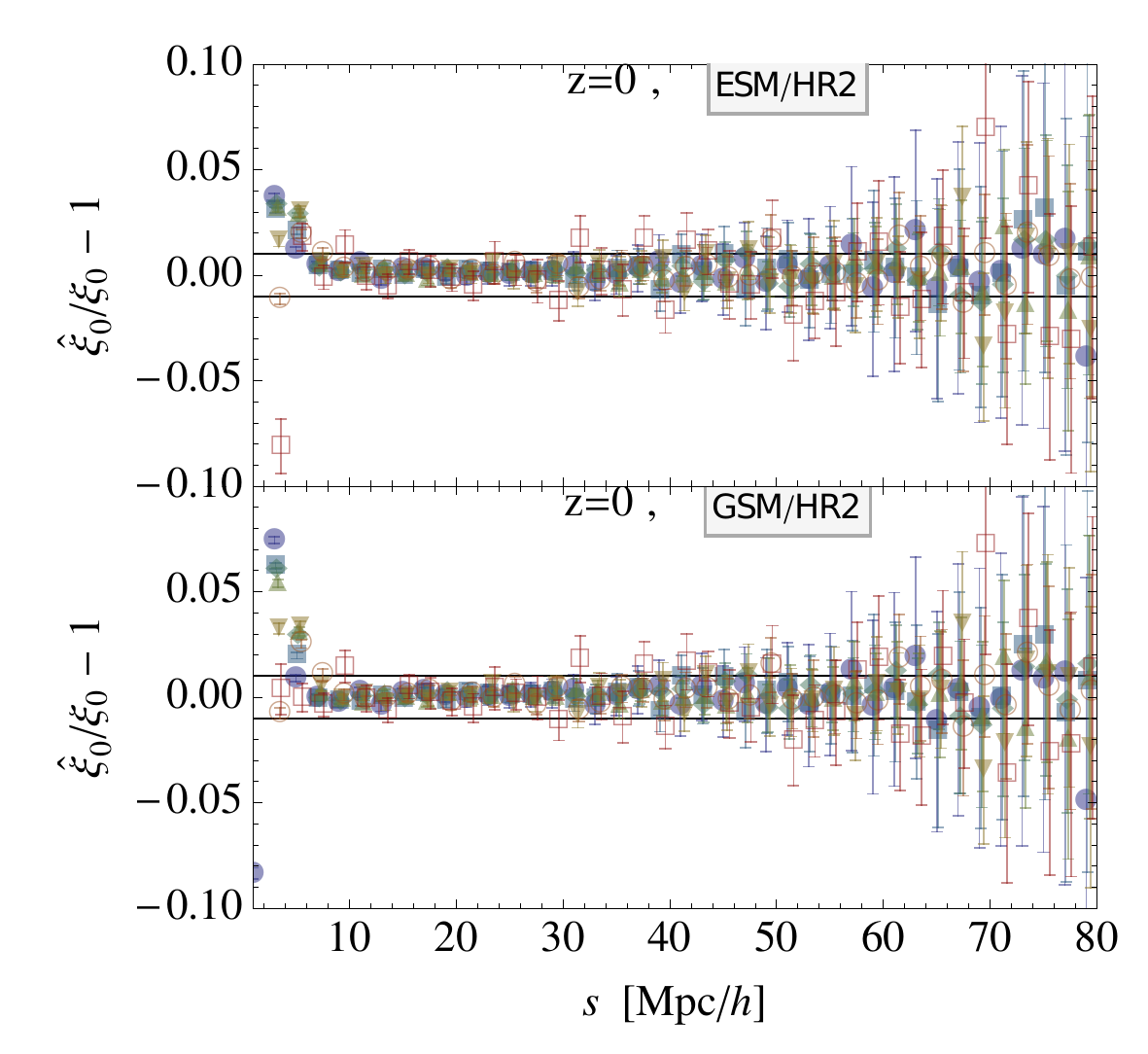}
\includegraphics[width=0.33\textwidth]{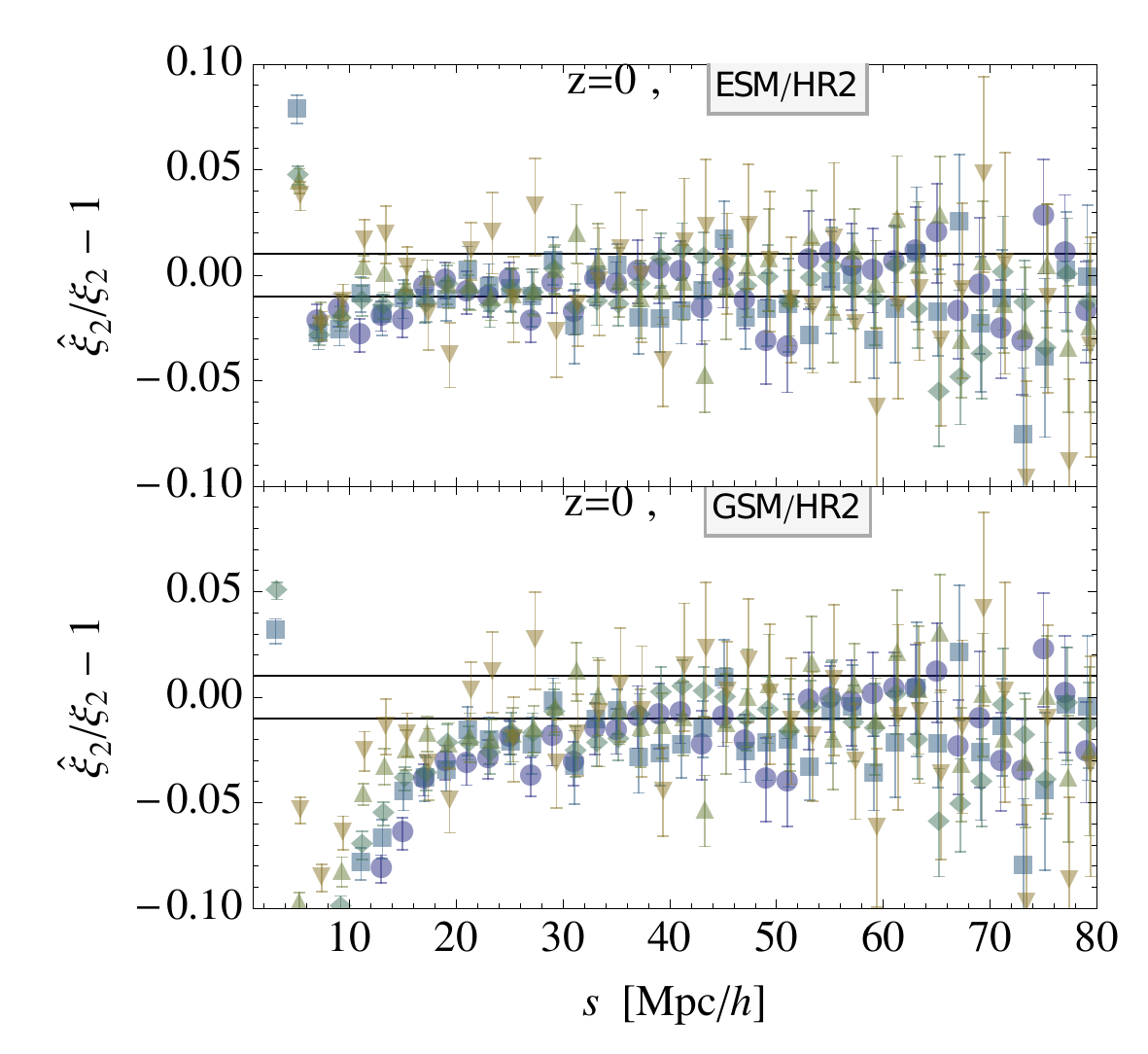}
\includegraphics[width=0.33\textwidth]{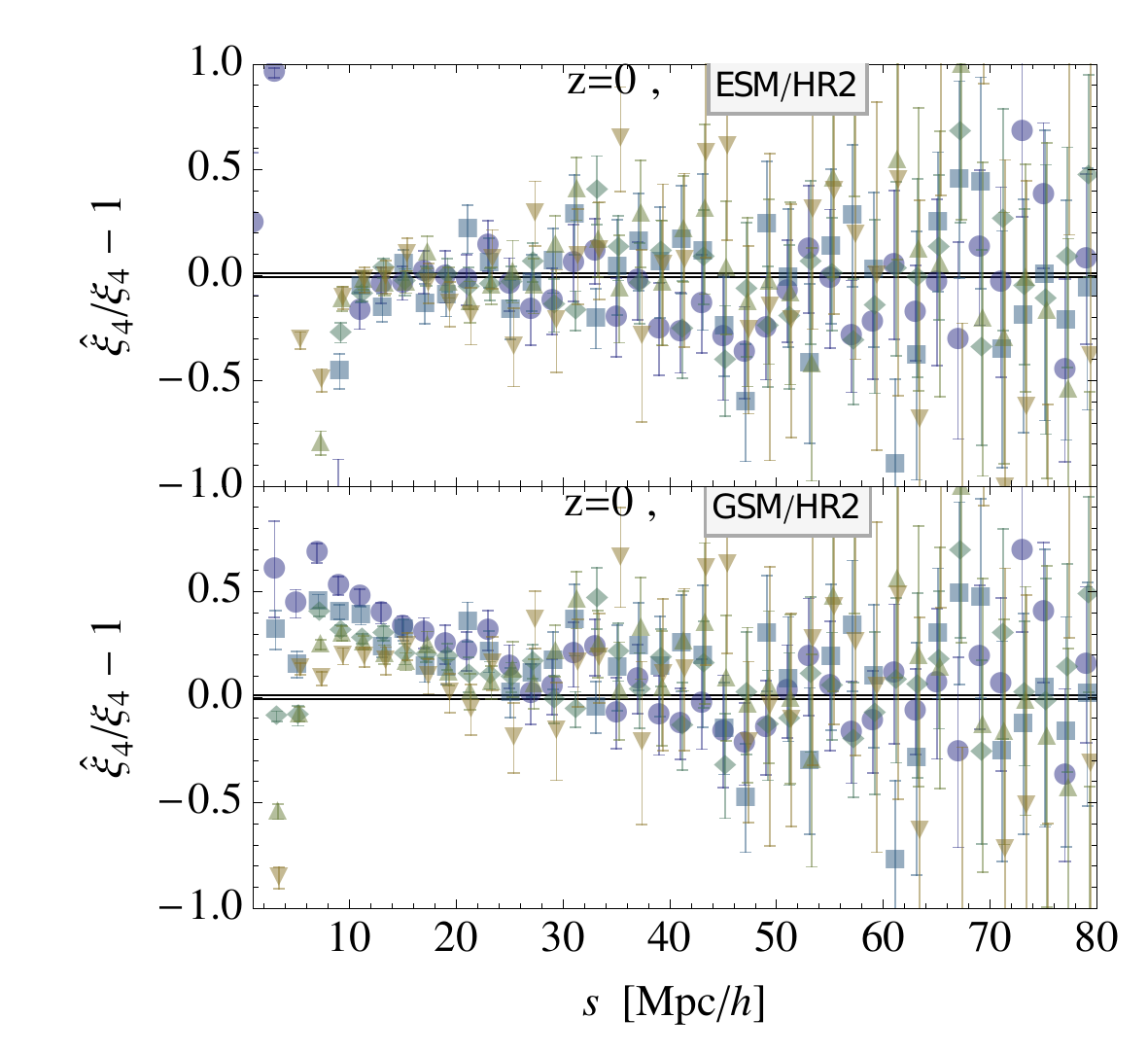}
\caption{The ratio of the GSM \eqref{GSM2} ({\it lower panel}) and ESM \eqref{ESM2} ({\it upper panel}) prediction using the HR2 data and the direct HR2 measurement ({\it data points}) for the lowest three redshift-space multipoles. {\it left panel} The monopole $\xi_0(s)$  {\it middle panel} The quadrupole $\xi_2(s)$. {\it right panel} The hexadecapole $\xi_4(s)$. }
\label{fig:GSMconsratio02}
\end{figure*}

In the following, we assess the accuracy of the GSM \eqref{GSM2} and the leading order of the ESM \eqref{ESM2} by comparing the results of the corresponding integrals \eqref{ESM} with the directly measured redshift space halo correlation function $\hat{\xi}(s, \mu,t )$. This is done by inserting the real space correlation $\xi(r)$, the pairwise velocity $v_{12}(r)$ and velocity dispersion $\sigma_{12}^2(r,\mu)$ measured in an $N$-body simulation into Eq.\,\eqref{GSM2} and additionally measuring the skewness $\Lambda_{12}(r,\mu)$ and plugging it into Eq.\, \eqref{ESM2}. 

The Horizon Run 2 (HR2) $N$-body simulation \cite{KPetal09, KPetal11} has an enormous size of 7200\,Mpc/$h$ and consists of $6000^3$ particles of mass $\lgM = 11.097$. For the mass units we use the notation $\lgM \equiv \log_{10}(M h/M_\odot)$. We measured halo correlation functions and velocity statistics from large galaxy-sized haloes $\lgM = 13.0$ to cluster-sized halos $\lgM = 15.2$ at the redshift $z=0$. In an accompanying work \cite{KUA15} we describe in detail how the correlation functions and Gaussian streaming ingredients haven been determined from the HR2 halo catalog. 
In order to evaluate and compare  $\xi_X(s_{||},s_\perp,t)$ from Eq.\,\eqref{ESM} to simulations it is useful to expand $\xi_X(s_{||},s_\perp)$ into Legendre polynomials $L_n(\mu)$ using $s^2 = s_{||}^2+s_\perp^2$ and $\mu = s_{||}/s$  
\begin{subequations}
\label{ximultipoles}
\begin{align} \label{legendremom}
\xi_X(s,\mu)&= \sum_{n=0}^\infty L_n(\mu) \xi_{X,n}(s)\,,\\
\label{xin}
\xi_{X,n}(s) &=\frac{1+2 n}{2}\, \int_{-1}^1 \xi_X(s,\mu,t) L_{n}(\mu) d \mu \,.
\end{align}
\end{subequations}

$\xi_n$ vanishes for all odd $n$. In linear perturbation theory, the only non-zero multipoles are the monopole $\xi_0$, quadrupole $\xi_2$ and hexadecapole $\xi_4$ and even in the nonlinear regime the magnitude of $\xi_n$ rapidly decreases with $n$. \highlight{The linear results go back to \cite{H92} and are given in \cite{RW11} for the case of Eulerian bias as}
\begin{align}
\notag \xi^L_0(s) &= 
\left( (b_1^E)^2 + \frac{2}{3} b_1^E f + \frac{1}{5} f^2 \right)  \frac{1}{2\pi} \int d k\, k^2 P_L(k) j_0(ks)  \\
 \xi^L_2(s)& = 
 - \left(  \frac{4}{3} b_1^E f + \frac{4}{7} f^2 \right)\frac{1}{2\pi}\int d k\, k^2 P_L(k) j_2(ks)  \\
\notag  \xi^L_4(s)& = 
  \frac{8}{35} f^2 \frac{1}{2\pi}\int d k\, k^2 P_L(k) j_4(ks)     \,,
   \end{align}
where $f$ is the linear growth rate and $j_n(x)$ are the spherical Bessel functions. We use their prefactors, given in terms of linear local Eulerian bias $b^{\rm E}_1 = 1+ b_1(\lgM_{\rm opt})$ determined from the best fitting mass for the real space correlation function, see Tab.\,\ref{biasfittab}, as a normalization when plotting multipoles. 
In Fig.\,\ref{fig:GSMcons02} we compare the redshift space halo correlation function predicted from the GSM \eqref{GSM2} and ESM \eqref{ESM2}, by measuring their ingredients from the HR2 data, to the direct measurements within HR2 for the redshift-space monopole $\xi_0$, quadrupole $\xi_2$ and hexadecapole $\xi_4$. 
We find that the ESM \eqref{ESM2} clearly improves the quadrupole $\xi_2$ and hexadecapole $\xi_4$ on small scales compared to the GSM \eqref{GSM2}. As evident from Fig.\,\ref{fig:GSMconsratio02} the quadrupole predicted by ESM \eqref{ESM2} is accurate to 2\% within statistical errors down to $10\Mpc$ in contrast to the GSM \eqref{GSM2} which breaks down below $30\Mpc$. A similar trend can be observed for the hexadecapole $\xi_4$ whereas the monopole $\xi_0$ is less sensitive to non-Gaussian terms. Apparently smaller halos are more sensitive to non-Gaussian corrections which is in line with the expectation that smaller objects are more affected by nonlinear dynamics. 

We conclude that the GSM is a very accurate model for the multipoles $\xi_n$, $n=0,2,4$ of the redshift space halo correlation function on scales larger than $30\Mpc$ while the ESM stays accurate down to $10\Mpc$. Our result is consistent with the previous finding that the GSM monopole is accurate on the percent level down to $10\Mpc$  and the quadrupole down to $30\Mpc$, compare Fig.\,6 in \cite{RW11}. This shows that the expansion of $Z$ around $\v{J}=0$ was justified and that halos over a wide range of masses can indeed be reasonably described by the GSM/ESM (\ref{GSM2}/\ref{ESM2}). 

Having established the range of validity of the streaming models GSM and ESM, we can use them as a basis for the theoretical modeling of redshift space halo correlation functions being aware of their limitations. As a next step, accurate theoretical predictions for the streaming model ingredients, $\xi(r)$, $v_{12}(r)$, $\sigma_{12}^2(r,\mu)$ and $\Lambda_{12}(r,\mu)$ are needed. 
In the following two sections we will combine the GSM/ESM with perturbation theory employing that halos can be treated as single-streaming objects when the fluid description is only applied on scales larger than their size, given by the Lagrangian radius. More precisely, we will calculate the streaming model ingredients from Convolution Lagrangian Perturbation Theory (CLPT) based on the dust model and extend it to include a coarse-graining scale chosen to be the Lagrangian radius.

\subsection{Pairwise generating and tracer cumulants}
By performing a cumulant expansion we can relate the term $\v{J}\cdot (\v{u}_2-\v{u}_1)$ contained in the exponential of the generating function $Z$ from \eqref{defZgeneral} to the cumulants $C_X^{(n)}$ of the tracer distribution function $f_X$. Therefore we introduce the moment generating functional 
\begin{align}
&G[\v{\tilde J}] := \int \vol{3}{u} \exp\left[i\v{\tilde J}\cdot\v{u}\right] f_X(\v{r},\v{u},t) \,, \label{genfun}
\end{align}
which allows to compute the cumulants $C_X^{(n)}$ of the distribution function $f_X$ according to
\begin{align}
&C^{(n)}_{X,i_1 \cdots i_n}:= (-i)^n \left.\frac{\del^n \ln G[\v{\tilde J}]}{\del \tilde J_{i_1} \ldots \del \tilde J_{i_n}} \right|_{\v{\tilde J}=0} \label{cumulants}\,.
\end{align}
This can be used to re-express $Z\left(\v{r},\v{J}\right) $ from \eqref{defZgeneral}
\begin{align}
Z &= \Bigg \langle \int \vol{3}{u_1}\!\!\!\int \vol{3}{u_2} f_{X,1} f_{X,2}  \exp\left[i \v{\tilde J}\cdot(\v{u}_2-\v{u}_1)\right] \Bigg \rangle  \notag\\
&= \Bigg \langle \int \vol{3}{u_2} f_{X,2}  \exp\left[i \v{\tilde J}\cdot\v{u}_2\right] \int \vol{3}{u_1} f_{X,1}  \exp\left[-i \v{\tilde J}\cdot\v{u}_1\right] \Bigg \rangle \notag \\ 
&= \Big \langle G[\v{\tilde J}] (\v{r}_2)G[-\v{\tilde J}](\v{r}_1) \Big \rangle \label{defZcum}  \\ 
&=\Bigg \langle \exp \left[ \sum_{N=0}^\infty \frac{i^N}{N!} \frac{J_{i_1}... J_{i_N}}{a^{2N}H^N} \left(C^{(N)}_{X,2,i_1 ... i_N} + (-1)^N C^{(N)}_{X,1,i_1 ... i_N}\right) \right] \Bigg \rangle \,, \notag
\end{align}
where $\v{\tilde J}=\v{J}/(a^2H)$,  $f_{X,1(2)}=f_X(\v{r}_{1(2)},\v{u}_{1(2)},t)$ and $C^{(N)}_{X,1(2)}=C^{(N)}_{X}(\v{r}_{1(2)})$.

\section{Determining Streaming model ingredients based on the dust model} 
\label{sec:GSM-dust}

In this Section we describe how the scale-dependent functions entering the streaming model can be determined once a phase-space distribution $f_X$ of the tracers is specified. 
To draw conclusions based on theoretical modeling it is due to connect the (proto-)halo distribution $f_X$ to an underlying dark matter distribution $f$ whose dynamics is known to be governed by the Vlasov-Poisson equation. Halos are biased tracers of dark matter, since according to spherical collapse and excursion set theory \cite{Bondetal, MW96, CLMP98}, the probability of forming a halo depends on the initial density field. Therefore, there are two steps for determining streaming model ingredients:
\begin{enumerate}
\item Choose a model for the distribution function $f$ of dark matter that reasonably approximates Vlasov dynamics.
\item Specify a bias model in order to relate the (proto-)halo cumulants $C_X^{(N)}$ to the ones of dark matter $C^{(N)}$.
\end{enumerate}

In the following we employ the pressureless fluid model as standard model for cold dark matter and discuss different possibilities to incorporate a coarse-graining in this fluid picture. For relating the halo to the dark matter density we use local Lagrangian bias with zero velocity bias and present two possibilities to generalize this notion to higher cumulants.

\subsection{The single-stream case: dust model}

In the context of analytical modellng CDM dynamics, usually the dark matter distribution is assumed to be described by the pressureless fluid (dust) model
\begin{align}
f_{\rm d}(\v{r},\v{u},t)= (1+\delta(\v{r},t) )\delta_{\rm D}(\v{u}-a \v{v}(\v{r},t)) \label{fdust}
\end{align}
which encodes all properties in terms of a number density $n(\v{r})=1+\delta(\v{r})$ and a single-streaming and curl-free velocity $\v{v}(\v{r})$ fulfilling the coupled continuity, Euler and Poisson equations \cite{B02}.  The cumulants of the dust model are
\begin{align}
\label{dustcumulant}
C^{(0)}&=\ln \left(1+\delta\right)\ , \quad C^{(1)}_{i} = av_{i} \ , \quad C^{(N\geq 2)}_{i_1 \cdots i_n}\equiv 0
\end{align}
which displays that the dust model is entirely described by density and velocity and all higher cumulants such as velocity dispersion vanish identically. Although the dust model is an exact solution of the Vlasov equation, its applicability is limited to the single-stream regime. It does not allow to describe the nonlinear stage of structure formation during which higher cumulants are sourced by the occurence of shell-crossing after which multiple streams form. For (proto-)halos this limitation is not as severe since they approximately behave as single-streaming objects even though a large fraction of dark matter particles resides in halos where it is multi-streaming and not accessible by the dust model. 
Hence, the proto-halos can also be described in terms of a single-streaming dust fluid
\begin{align}
f_{X}(\v{r},\v{u},t)= (1+\delta_X(\v{r},t) )\delta_{\rm D}\left(\v{u}-a \v{v}_X(\v{r},t)\right) \label{fXdust} \,.
\end{align}
To connect the density of halos to the dark matter density, we assume local Lagrangian bias 
\begin{align}
(1+\delta_X(\v{r},t))\, \varvol{3}{r} &= F[\delta_R(\v{q}),t]\,\varvol{3}{q}\,. \label{locLagBias}
\end{align}
This equality states that proto-halos identified in the linear initial conditions, depending only on the smoothed initial linear density field $\delta_R(\v{q})$, are conserved until they form a proper halo at time $t$. The proto-halo initial density field is assumed to be a local function $F[\delta_R(\v{q}),t]$ of the initial linear density field $\delta_{L}(\v{q})$ smoothed over some scale related to the Lagrangian size $R$ of the proto-halo by applying a window function $W$ in Fourier space
\begin{equation}
\delta_R(\v{q}) = \int \frac{\vol{3}{k}}{(2\pi)^3} W(k R) e^{i \v{k}\cdot \v{q}} \delta_{L}(\v{k}) \,.
\end{equation}
The choice of the appropriate smoothing scale $R$ will be elaborated in more detail in a forthcoming paper \cite{KUA15}. Note that for the computations in both iPT and CLPT this smoothing scale $R$ is effectively removed by setting the window function to unity. In \cite{M08} this is justified by claiming that the large-scale clustering of biased objects should not depend on the artificial choice of $R$ to define the background field and seconded
by the assertion that this is demanded by consistency with the approximation being valid only  on scales larger than the smoothing radius $R$. In \cite{WRW14} it is furthermore argued that $R$ naturally drops out in the final statistics of interest and is only necessary to keep intermediate quantities well-behaved. \highlight{ We will preserve the smoothing and see in Sec.\,\ref{sec:GSM-CLPT}, in particular Figs.\, \ref{fig:z0realAllmass} and \ref{fig:xicgCLPTvsTCLPT} how large the effect on the baryon acoustic peak is when a smoothing at the Lagrangian scale is performed compared to the case where the smoothing is dropped.}

The tracer velocity field $\v{v}_X = a {\v{\dot \varPsi}}_X$ displaces the proto-halos to their halo virialization sites $\v{r}=\v{q}+\v{\varPsi}_X(\v{q})$. We assume zero velocity bias such that  proto-halos move along dust fluid trajectories $\v{r}=\v{q}+\v{\varPsi}(\v{q})$ with the dust velocity $\v{v}_X=\v{v}=a {\v{\dot \varPsi}}$ 
\begin{subequations}
\label{CLPTmodel}
\begin{align} 
\hspace{-0.3cm}\boxed{\text{CLPT}:\ \ \  f_{X}(\v{r},\v{u},t)= (1+\delta_X(\v{r},t) )\delta_{\rm D}\left(\v{u}-a\v{v}(\v{r},t)\right)} \,. \label{fXdustzerovelbias}
\end{align}
Local Lagrangian bias \eqref{locLagBias} allows us to relate the densities in real and Lagrangian space in the following way
\begin{align}
1+\delta_X(\v{r},t) = &\int \vol{3}{q} F[\delta_R(\v{q}),t]\, \delta_{\rm D}\left(\v{r} -\v{q} - \v{\varPsi}(\v{q},t)\right) \label{deltarq} \,.
\end{align}
\end{subequations}
Due to the single-valuedness of the velocity we recover from inserting \eqref{fXdustzerovelbias} into \eqref{realtoredshift} a simpler relation between densities in real and redshift space
\begin{subequations} \label{deltasrandsq}
\begin{align} 
1+\delta_X(\v{s},t) = &\int \vol{3}{r} (1+\delta_X(\v{r},t))\, \delta_{\rm D}\left(\v{s} -\v{r} -  \frac{\v{v}(\v{r},t)\cdot\hvz}{aH} \hvz\right)  \,. \label{deltasr}
\end{align}
We could combine both relations in a single expression by expressing the velocity in terms of the displacement $\v{v} = a \dot{\v{\varPsi}}$ 
\begin{align} \label{deltasq}
1+\delta_X(\v{s},t) &= \int \vol{3}{q} F[\delta_R(\v{q}),t]\times \\
\notag &\qquad\qquad \times \delta_{\rm D}\left(\v{s} -\v{q} - \v{\varPsi}(\v{q},t) - \frac{\dot{\v{\varPsi}}(\v{q},t)  \cdot \hvz}{H}\hvz\right)\,.
\end{align}
\end{subequations}
\highlight{Note however that we will not use formulas \eqref{deltasrandsq} explicitly. Instead we will rely on the GSM \eqref{ESM} to go from real space to redshift space and Eq.\,\eqref{xizspace} below, to go from Lagrangian to Eulerian space.}

\paragraph*{Derivation}
We already showed in \eqref{GSM2} that one can obtain the GSM \eqref{GSM} from quite general assumptions, \highlight{in particular that no assumptions about tracer dynamics and bias are required}. Now, we will specialize $Z$ from \eqref{defZcum} to the dust ansatz \eqref{CLPTmodel} for the tracer phase-space distribution $f_X$ combined with local Lagrangian bias \eqref{locLagBias} as considered in \cite{WRW14}. 

First, we use the dust model cumulants \eqref{dustcumulant} applied to the proto-halo distribution from CLPT \eqref{CLPTmodel} and plug them into the general expression for $Z(\v{r}, \v{J})$ in terms of tracer cumulants \eqref{defZcum} obtaining
\begin{align*}
Z &=\Bigg \langle \left[1+\delta_X(\v{r}_1) \right] \left[1+\delta_X(\v{r}_2) \right] \exp\left[i \frac{\v{J}\cdot (\v{v}(\v{r}_2)-\v{v}(\v{r}_1))}{aH}\right] \Bigg\rangle\,.
\end{align*}
We then switch to Lagrangian space making use of local Lagrangian bias \eqref{deltarq} and express the bias function
\begin{equation}
F[\delta_R(\v{q})] = \int \frac{d \lambda}{2\pi} \tilde F(\lambda) e^{i \lambda \delta_R(\v{q}) }\,,
\end{equation}
as well as the delta function $\delta_{\rm D}\left(\v{r} -\v{q} - \v{\varPsi}(\v{q},t)\right)$ in Fourier space. Next we replace the single streaming velocity by the derivative of the displacement field $\v{v}(\v{r})=a\v{\dot\varPsi}(\v{q})$ and integrate over $\v{Q}=\v{q}_1+\v{q}_2$ to obtain
\begin{subequations} \label{xizspace}
\begin{align}
Z(\v{r}, \v{J}, t) &=  \int \vol{3}{q}\!\! \int \frac{\vol{3}{k}}{(2\pi)^3} e^{i \v{k}\cdot (\v{q} - \v{r}) }  \int \frac{\vol{}{\lambda_1} \vol{}{\lambda_2}}{(2\pi)^2} \notag\\
& \quad \times \tilde{F}(\lambda_1)\tilde{F}(\lambda_2)\langle e^{i X}\rangle \label{defZdust}\,,
\end{align}
with
\begin{align}
X&= \lambda_1 \delta_1+\lambda_2 \delta_2+\v{k}\cdot\v{\Delta} +\v{J}\cdot \frac{\dot{\v{\Delta}}}{H} \label{X}\,,
\end{align}
\end{subequations}
where $\vq=\vq_2-\vq_1$, $\delta_{1(2)}=\delta_R(\v{q}_{1(2)})$ and  $\v{\Delta} = \v{\varPsi}(\v{q}_2,t)- \v{\varPsi}(\v{q}_1,t)$.   

\paragraph*{Previous studies} 
Originally, in \cite{CRW13}, Eq.\,\eqref{deltasq} was used to derive an expression for the two-point correlation function 
\begin{align}\label{xizspaceOri}
1+\xi_X(\v{s},t) & =  \int \vol{3}{q}\!\! \int \frac{\vol{3}{k}}{(2\pi)^3} e^{i \v{k}\cdot (\v{q} - \v{s}) }   \\
\notag & \qquad\quad\int \frac{\vol{}{\lambda_1} \vol{}{\lambda_2}}{(2\pi)^2}  \tilde{F}(\lambda_1)\tilde{F}(\lambda_2)\langle e^{i X( \v{J} = (\v{k}\cdot\hvz)\ \hvz)}\rangle\,,
\end{align}
which was then evaluated within CLPT to obtain a Post-Zel'dovich approximation for biased tracers in redshift space. The formula \eqref{GSM} was suggested in \cite{RW11} to calculate Gaussian streaming redshift space distortions, following the idea of \cite{F94} to reconcile the streaming model \cite{P80} for nonlinear scales with linear theory \cite{K87} by considering a scale-dependent variance. In \cite{RW11}, the pairwise velocity mean $ v_{12}$ and second moment $ \tilde{\sigma}_{12}^2$ entering the streaming model \eqref{GSM} were calculated from SPT with linear bias while the real space correlation $\xi(r)$ was inferred from LPT with local Lagrangian bias \eqref{deltarq}. Later on, in \cite{WRW14}, the real space correlation and velocity statistics were treated on the same footing and determined within CLPT \cite{CRW13} together with local Lagrangian bias. Note that \eqref{xizspaceOri} involves a three-dimensional $\v{q}$-integral which needs to be evaluated numerically within CLPT \cite{CRW13}. Studying the expression \eqref{xizspace} for $Z$ in CLPT, the streaming model ingredients can be calculated according to \eqref{GSMparam} and involve at most two-dimensional numerical integrals \cite{WRW14}. This a practical reason to chose to perform an Edgeworth expansion of $Z$ to obtain the Gaussian streaming model \eqref{GSM2} and its non-gaussian generalization -- the Edgeworth streaming model -- whose numerical evaluation is more efficient than the full CLPT expression \eqref{xizspaceOri}.

\subsection{Beyond single-stream: coarse-graining the dust model}
\label{sec:multGSM}

In the following we compare several distinct approaches of coarse-graining a dust fluid, namely a coarse-graining in Eulerian space (cgCLPT) and a coarse-graining in Lagrangian space implemented by smoothing the initial power spectrum in the spirit of the truncated Zel’dovich approximation (TCLPT). A key question is how to generalize the biasing scheme employed for CLPT based on the dust model to the coarse-grained case. So far we assumed local Lagrangian bias for the density \eqref{deltarq} and zero velocity bias. This might be generalized by
{(\it a)} assuming zero velocity bias and that higher cumulants for the tracer vanish identically $C_X^{(N\geq 2)}\equiv 0$ motivated by the fact that proto-halos can be described well by single-stream physics such that, in analogy to the CLPT case,
\begin{subequations}
\label{velbiasassumpt}
\begin{equation} \label{velbiasassumpt-a}
f_X(\v{u},\v{r},t)=\left(1+\delta_X(\v{r},t)\right)  \delta_{\rm D}( \v{u} - a\v{v}(\v{r},t))\,.
\end{equation}
or {(\it b)} assuming that tracers and dark matter are only biased with respect to density such that all higher tracer cumulants are identical to those of dark matter $C_X^{(N\geq 1)}=C^{(N\geq 1)}$ and 
\begin{equation} \label{velbiasassumpt-b}
 \frac{f_X(\v{u},\v{r},t)}{1+\delta_X(\v{r},t)}  = \frac{f(\v{u},\v{r},t)}{1+\delta(\v{r},t)}\,.
\end{equation}
\end{subequations}

Note that in order to write the biasing in analogy to the dust case \eqref{defZdust} it is necessary that $f_X(\v{u},\v{r},t)/(1+\delta_X(\v{r},t))$ is independent of the bias function $F$ which is achieved by both relations \eqref{velbiasassumpt}. Then, the redshift space correlation takes the form
\begin{subequations} \label{fxizspaceanalogue}
\begin{align}
\ \tilde Z(\v{r}, \v{J}, t) &=  \int \vol{3}{q}\!\! \int \frac{\vol{3}{k}}{(2\pi)^3} e^{i \v{k}\cdot (\v{q} - \v{r}) } \int \frac{\vol{}{\lambda_1} \vol{}{\lambda_2}}{(2\pi)^2} \notag\\
& \quad\times \tilde{F}(\lambda_1)\tilde{F}(\lambda_2)\langle e^{i \tilde X}\rangle\,, \label{defZwithbias}
\end{align}
with
\begin{align}
\label{tildeX}
\tilde X &=   \lambda_1\delta_1 + \lambda_2\delta_2+ \vk\cdot \v{\Delta} \\
\notag &\ + \sum_{N=1}^\infty \frac{i^{N-1}}{N!} \frac{J_{i_1}...J_{i_N}}{a^{2N}H^N} \left[C^{(N)}_{X,2,i_1 ... i_N} + (-1)^N C^{(N)}_{X,1,i_1 ... i_N}\right] \,,
\end{align}
\end{subequations}
where $C^{(N)}_{X,1(2)}=C^{(N)}_{X}\left(\v{r}_{1(2)}(\vq_{1(2)})\right)$.
Hence, the ESM ingredients are still computed according to Eqs.\,\eqref{GSMparam} with $Z$ from \eqref{xizspace} replaced by $\tilde{Z}$ from \eqref{fxizspaceanalogue}. 
If we consider the GSM, expanding up to second order in $\v{J}$, we see that the first cumulant $C_X^{(1)}$ corresponds to the term $\dot{\v{\Delta}}/H$ that is also present in the single streaming Gaussian streaming model \eqref{X} and contributes both to the mean $v_{12}$ and variance $\sigma_{12}^2$ of the Gaussian. In contrast, the second cumulant $C_X^{(2)}$ is conceptually new and contributes only to the variance of the Gaussian, whereas all higher cumulants $ C_X^{(N\geq 3)}$ are irrelevant for the GSM but only contribute to the ESM.  

\subsubsection{Coarse-graining in Eulerian space (cgCLPT)}
Coarse-graining the dust model on a length scale $\sigx$ in Eulerian space and a velocity scale $\sigu$ gives rise to the so-called coarse-grained dust model as described in detail in \cite{UK14}. We shortly recap the main results that are of direct relevance here. 
The coarse-grained dust model is defined as a smoothing of the dust phase space distribution with a Gaussian filter of width $\sigx$ and $\sigu$ in $\vx$ and $\v{u}$ space, respectively
\begin{align}
\label{fcgdust}
 \bar f_\d&= \int \frac{\vol{3}{\tilde x}\vol{3}{\tilde u}}{(2 \pi\sigx\sigu)^3 } \exp\left[-\frac{(\vx-\tilde\vx)^2}{2\sigx^2}-\frac{(\v{u}-\v{\tilde u})^2}{2\sigu^2} \right] f_\d(\tilde\vx,\v{\tilde u}) \\
\notag &= \int \frac{\vol{3}{\tilde x}}{(2 \pi\sigx\sigu)^3} \exp\left[-\frac{(\vx-\tilde\vx)^2}{2\sigx^2}-\frac{(\v{u}-a\v{v}(\tilde\vx))^2}{2\sigu^2}\right]n(\tilde\vx)\,.
\end{align}
If $x_{\rm typ}$ and $u_{\rm typ}$ are the (minimal) scales of interest we have to ensure that $\sigx \ll x_{\text{typ}}$ and $\sigu \ll u_{\text{typ}}$ in order to be able to resolve these scales. The coarse-grained dust model features higher cumulants which are absent in the pressureless fluid case and given by
\begin{subequations} \label{cgdustcumulant}
\begin{align}
\bar C^{(0)}&=\ln \bar n\ , \quad \bar C^{(1)}_i = a\frac{\overline{nv_i}}{\bar n} =: a\bar v_i \label{cgdustcumulant01} \,, \\
\bar C^{(2)}_{ij} 
&= \sigu^2 \delta_{ij} + a^2 \left(\frac{\overline{nv_iv_j}}{\bar n} - \frac{\overline{nv_i}\ \overline{nv_j}}{\bar n^2} \right)\label{barC2} \,,\\
\bar C^{(3)}_{ijk} 
&= a^3\left(\frac{\overline{nv_iv_jv_k}}{\bar n} - \stackrel{+ \text{cyc. perm.}}{\frac{\overline{nv_iv_j}\ \overline{nv_k}}{\bar n^2}} +2 \frac{\overline{nv_i}\ \overline{nv_j}\ \overline{nv_k}}{\bar n^3} \right)\,,
\end{align}
\end{subequations}
see \cite{UK14}, where we also defined 
$$\bar g(\vx):=\int \frac{{\text{d}^3\tilde x}}{\left(\sqrt{2 \pi}\sigx\right)^3} \exp\left[-\frac{(\vx-\v{\tilde x})^2}{2\sigx^2}\right]g(\v{\tilde x})\,.$$ 
The coarse-grained velocity $\bar{\v{v}}$ is the mass-weighted dust velocity which is obtained by smoothing the momentum field $n v_i$ and then dividing by the smoothed density field $\bar{n}$. From a physical point of view $\bar{\v{v}}$ describes the center-of-mass velocity of the collection of particles inside a coarsening cell of diameter $\sigx$ around $\v{x}$. 
Whereas the contribution from the velocity smoothing scale $\sigu$ has no dynamical effect but only contributes to the velocity dispersion, the smoothing on a fixed Eulerian scale $\sigx$ affects the fluid dynamics \cite{D00,UKH14}. The modified fluid equations resulting from \eqref{fcgdust} and the dependence of the smoothing scale $\sigx$ have been studied perturbatively in Eulerian and Lagrangian space \cite{UK14}. 
The displacement field $\bar{\v{\varPsi}}$ was defined as the integral lines of $\v{\bar v}=:a\dot{\v{\bar\varPsi}}$ and can be determined perturbatively from the coarse-grained Eulerian quantities $\bar\delta$ and $\bar{\v{v}}$, as described in \cite{UK14}. 
Our Eulerian-coarse-grained displacement $\bar{\v{\varPsi}}$ should not be confused with the direct coarse-graining of ${\v{\varPsi}}$ in Lagrangian space considered in \cite{PSZ13}. 

If we combine the biasing scheme \eqref{velbiasassumpt-a} for single-streaming tracers with the coarse-grained dust model (scgCLPT), we effectively consider
\begin{subequations}
\label{cgCLPTmodel-a}
\begin{equation}
\hspace{-0.3cm}\boxed{\text{scgCLPT}:\quad \ \  f_X(\v{r},\v{u},t) = \left(1+\bar \delta_X\right) \delta_{\rm D}\left( \v{u} - a \bar{\v{v}}(\v{r})\right) }\,, \label{scgCLPT}
\end{equation} 
which is analogous to \eqref{CLPTmodel} but the velocity and tracer density are now expressed in terms of the coarse-grained displacement field,  $\bar{\v{v}}=a\dot{\bar{\v{\varPsi}}}$ and  
\begin{align}
1+\bar \delta_X(\v{r}) = &\int \vol{3}{q} \! F[\delta_{R=\sigx}(\v{q})]\, \delta_{\rm D}\left(\v{r} -\v{q} - \bar{\v{\varPsi}}(\v{q})\right) \label{deltarq2} \,.
\end{align}
Then, the evaluation of the streaming ingredients \eqref{GSMparam} from \eqref{fxizspaceanalogue}  within cgCLPT is affected by the coarse-graining only via modified expressions for the kernels involving coarse-grained displacements $\v{\bar\varPsi}$. We present the Lagrangian correlators for the coarse-grained dust model that are relevant for the CLPT evaluation in App.\,\ref{AppL}.
\end{subequations}

When instead choosing the biasing scheme \eqref{velbiasassumpt-b}, we assume that halos and dark matter are only biased with respect to density 
\begin{equation}
\hspace{-0.3cm}\boxed{\text{cgCLPT}:\quad \ \ \ \  f_X(\v{r},\v{u},t)  = \frac{1+\bar\delta_X(\v{r},t)}{1+\bar\delta(\v{r},t)} \bar f_d(\v{r},\v{u},t) }\,. \label{cgCLPTmodel-b}
\end{equation} 

Then, in addition to having modified displacement kernels, the computation of the streaming ingredients is also affected by the occurrence of velocity dispersion encoded in higher cumulants \eqref{cgdustcumulant}. This affects the variance $\sigma_{12}^2$ through $\bar C_X^{(2)}$, as computed in App.\,\ref{AppK2} and the leading-order non-Gaussian correction $\Lambda_{12}$ through $\bar C_X^{(2)}$ and $\bar C_X^{(3)}$. 

\subsubsection{Coarse-graining in Lagrangian space (TCLPT)}

Our coarse-graining in Lagrangian space is based on a fixed smoothing scale $\sigq$ in the Lagrangian or initial condition space $\v{q}$ corresponding to the scgCLPT model \eqref{cgCLPTmodel-a} 
\begin{subequations}
\begin{equation} \label{fXLagrange}
\text{TCLPT}: f_X(\v{r},\v{u},t) =\left(1+\bar\delta_{X, \sigq}\right)  \delta_{\rm D}\left( \v{u} - a\bar{\v{v}}_{\sigq}(\v{r}) \right) \,,
\end{equation}
but with the velocity and density given by the smoothed displacement field $\bar{\v{\varPsi}}_{\sigq}$ via $\bar{\v{v}}_\sigq= a\dot{\bar{\v{\varPsi}}}_{\sigq}$ and
\begin{align}
1+\bar \delta_{X, \sigq}(\v{r}) = &\int \vol{3}{q}\! F[\delta_{R=\sigq}(\v{q})]\, \delta_{\rm D}\left(\v{r} -\v{q} - \v{\bar\varPsi}_{\sigq}(\v{q})\right) \label{deltarq3} .
\end{align}
\end{subequations}
If one considers coarse-grained dust dynamics in Lagrangian space then one is lead to \cite{PSZ13}. For TCLPT, we instead perform a smoothing in the initial conditions while keeping the pressureless fluid dynamics unchanged. This is implemented by smoothing the initial linear density field $\delta_{L}(\v{q})$, or equivalently the power spectrum $P_{L}\rightarrow \bar P_L$ when calculating the Lagrangian correlators for CLPT \eqref{CLPTmodel} that are given in \cite{WRW14}
\begin{align}
\label{TCLPTmodel}
\hspace{-0.3cm}\boxed{\text{TCLPT: \quad \ \  CLPT with } P_L(k) \rightarrow \exp\left(-\sigq^2k^2\right)P_L(k)} \,.
\end{align}
Note that when evaluated in first order LPT in which case CLPT is identical to the Zel'dovich approximation), then also TCLPT \eqref{TCLPTmodel} and cgCLPT (\ref{cgCLPTmodel-a}/\ref{cgCLPTmodel-b}) are identical to the TZA and differences arise when nonlinearities in the displacement field are taken into account.
 In the following section we evaluate the streaming model ingredients within CLPT up to second order in the power spectrum.

\section{Evaluation of streaming model ingredients within Convolution Lagrangian Perturbation Theory} 
\label{sec:GSM-CLPT}

In this Section we evaluate the scale-dependent functions entering the streaming model within Convolution Lagrangian perturbation theory (CLPT), introduced in \cite{CRW13} based on the pressureless fluid model \eqref{CLPTmodel} and its coarse-grained versions scgCLPT \eqref{cgCLPTmodel-a}, cgCLPT \eqref{cgCLPTmodel-b} and TCLPT \eqref{TCLPTmodel} defined in the last section. First, we present the calculation of the real space correlation function and the pairwise velocity statistics for the coarse-grained case, which relies on results from ordinary CLPT presented in \cite{WRW14}. In the last subsection \ref{subsec:results} we synoptically compare the CLPT prediction for the streaming model ingredients to those of its coarse-grained generalizations scgCLPT, cgCLPT and TCLPT.

\subsection{CLPT formalism}
In analogy to \cite{CRW13,WRW14} we define
\begin{subequations}
 \begin{align}
K_{p,{i_1,...,i_p}}(\v{k},\v{q},\lambda_1,\lambda_2) &= \Bigg \langle \left(\frac{\partial}{i\partial  \v{J}_{i_k}} \right)^p e^{i \tilde X} \Bigg\rangle \Bigg|_{\v{J}=0} \,, \label{Kp}
\end{align}
which allows to compute the ingredients of the streaming models arising from $Z$ as given in \eqref{fxizspaceanalogue}
 \begin{align}
\frac{\partial^p Z}{(i\partial  \v{J}_{i_k})^p} &= \int \vol{3}{q}\!\! \int \frac{\vol{3}{k}}{(2\pi)^3} e^{i \v{k}\cdot (\v{q} - \v{r}) } \int \frac{\vol{}{\lambda_1} \vol{}{\lambda_2}}{(2\pi)^2} \notag\\
& \quad\times \tilde{F}(\lambda_1)\tilde{F}(\lambda_2) K_{p,{i_1,...,i_p}}(\v{k},\v{q},\lambda_1,\lambda_2)  \,.
\end{align}
By integrating over $\lambda$ we obtain the bias parameters, which are expectation values of derivatives of the Lagrangian halo density field $F[\delta_R(\vq)]$ \highlight{with respect to $\delta_R(\vq)$}, according to \cite{M08}
\begin{align}
\label{biasparam}
\int \frac{\text{d}{\lambda}}{2\pi} \tilde{F}(\lambda) (i\lambda)^n \exp\left(-\tfrac{1}{2}\lambda^2\sigma_R^2\right) =\langle F^{(n)}\rangle \,,
\end{align}
\end{subequations}
where $\sigma_R^2=\langle \delta_R^2(\v{q}) \rangle$. Furthermore also the integration over $\vk$ can be performed analytically and only two dimensions of the $\v{q}$ integration have to be done numerically, see \cite{CRW13,WRW14}. 

To numerically evaluate the real space correlation function \eqref{gsmXi} and the pairwise velocity statistics \eqref{gsmv}, \eqref{gsmsig} within CLPT we resort to the C++ code\footnote{https://github.com/wll745881210/CLPT GSRSD.git} that has been implemented by \cite{WRW14} for the dust model and which we extended to the coarse-grained dust case. 
The two quantities $K_0$ and $K_1$ do not depend on higher cumulants $C_X^{(N\geq 2)}$ and hence do not discriminate between the two bias versions of cgCLPT (\ref{cgCLPTmodel-a}/\ref{cgCLPTmodel-b}). These quantities can be computed straightforwardly in full analogy to CLPT \cite{WRW14} by simply replacing dust correlators with their coarse-grained counterparts, which of course differ for TCLPT and cgCLPT.
For $K_2$, additionally the effect of velocity dispersion $C_X^{(2)}$ becomes relevant such that its form depends on whether scgCLPT \eqref{cgCLPTmodel-a} corresponding to $C_X^{(N\geq 2)}\equiv 0$ or cgCLPT \eqref{cgCLPTmodel-b} corresponding to $C_X^{(N\geq 2)}\equiv \bar C^{(N\geq 2)}$ is employed. Both cases will be considered and compared to each other.

\subsection{Real space correlation function \boldmath $1+\xi_X$}
The real-space two-point correlation function $1+\xi_X(\v{r},t)$ is given by
\begin{subequations}
\label{gsmIngredients}
 \begin{align}
 1+\xi_X(\v{r},t) &= \int \vol{3}{q}\!\! \int \frac{\vol{3}{k}}{(2\pi)^3} e^{i \v{k}\cdot (\v{q} - \v{r}) } \int \frac{\vol{}{\lambda_1} \vol{}{\lambda_2}}{(2\pi)^2} \notag\\
& \quad\times \tilde{F}(\lambda_1)\tilde{F}(\lambda_2) K_0(\v{k},\v{q},\lambda_1,\lambda_2)  \,, \label{gsmXi}\,.
\end{align}
where $K_0$ has to be evaluated according to the cumulant expansion theorem \cite{K62}
\begin{align}
\label{cumexpK0}
\bar K_0 &=\langle e^{i\tilde X_{\v{J}=0}} \rangle \,|_{\mathcal{O}(P_L^2)} = \exp\left[\sum_{N=1}^\infty \frac{i^N}{N!} \langle \tilde X^N_{\v{J}=0}\rangle_c\right] \Bigg|_{\mathcal{O}(P_L^2)} \,.
\end{align}
First we expand the exponent up to second order in the linear power spectrum $P_L$
\begin{align}
\label{sumiX}
\notag &\sum_{N=0}^\infty \frac{i^N}{N!} \Big\langle \tilde X^N_{\v{J}=0} \Big\rangle_c \Bigg|_{\mathcal{O}(P_L^2)} =1 - \frac{1}{2} (\lambda_1^2+\lambda_2^2)\bar \sigma_R^2 - \lambda_1\lambda_2\bar \xi_L\\
&-(\lambda_1+\lambda_2)\bar U_ik_i-\frac{1}{2}\bar A_{ij}k_ik_j  -\frac{i}{2}(\lambda_1^2+\lambda_2^2)\bar U_i^{20(2)}k_i\\
\notag & -i\lambda_1\lambda_2\bar U_i^{11(2)}k_i  - \frac{i}{2}(\lambda_1+\lambda_2)\bar A_{ij}^{10}k_ik_j-\frac{i}{6}\bar W_{ijk}k_ik_jk_k\,,
\end{align}
where the Lagrangian correlators are defined as 
\begin{align}
\label{defCor}
\sigma_R^2 &:= \langle \delta_1^2 \rangle_c = \langle \delta_2^2 \rangle_c \  , \ 
\xi_L := \langle \delta_1 \delta_2 \rangle_c \  , \notag \\
U_i^{mn(p)} &:= \langle \delta_1^m \delta_2^n \Delta_i^{(p)} \rangle_c \  , \ 
A_{ij}^{mn(pq)} := \langle \delta_1^m \delta_2^n \Delta_i^{(p)} \Delta_j^{(q)}\rangle_c \  , \\ 
W_{ijk}^{mn(pqr)} &:= \langle \delta_1^m \delta_2^n \Delta_i^{(p)} \Delta_j^{(q)} \Delta_k^{(r)} \rangle_c \,, \notag
\end{align}
and we adopt the shorthand notation $U_i = U_i^{10}$, $A_{ij} = A_{ij}^{00}$ and $W_{ijk}=W_{ijk}^{00}$ introduced in \cite{CRW13,WRW14}. Whenever indices in brackets are omitted they have been summed over to the appropriate order of perturbation theory, for example $U_i=U_i^{(1)}+U_i^{(3)}$ and $A_{ij}=A_{ij}^{(11)} + A_{ij}^{(22)} + A_{ij}^{(13)}+ A_{ij}^{(31)} $. Note that this notation for the correlators replaces that in Eq.\,\eqref{defC} which was defined according to \cite{M08}. In the following calculation we will keep this notation, such that whenever $A$, $U$ or $W$ occur they refer to the usual kernels given in \cite{CRW13,WRW14}. In contrast, we will use $\bar A$, $\bar U$ or $\bar W$ for correlators arising from smoothed quantities
\begin{align}
\label{defCorcg}
\bar \sigma_R^2 &:= \langle \bar \delta_1^2 \rangle_c = \langle \bar \delta_2^2 \rangle_c \  , \ 
\bar \xi_L := \langle \bar \delta_1 \bar \delta_2 \rangle_c \  , \ 
\bar U_i^{mn(k)} := \langle \bar \delta_1^m\bar  \delta_2^n \bar \Delta_i^{(k)} \rangle_c \  , \notag \\
\bar A_{ij}^{mn} &:= \langle \bar \delta_1^m \bar \delta_2^n \bar \Delta_i \bar \Delta_j\rangle_c \  , \ 
\bar W_{ijk}^{mn} := \langle \bar \delta_1^m \bar \delta_2^n \bar \Delta_i \bar \Delta_j \bar \Delta_k \rangle_c \,.
\end{align}
Plugging the exponential \eqref{sumiX} into the expression \eqref{cumexpK0} and keeping only the two terms exponentiated which are linear in the power spectrum and have non-zero limits as $|\vq| \rightarrow \infty$ gives
\begin{align}
\label{eiX}
\notag \bar K_0 &= e^{-\tfrac{1}{2}\bar A_{ij}k_ik_j} e^{-\tfrac{1}{2} (\lambda_1^2+\lambda_2^2)\bar \sigma_R^2}\\\
\notag &\ \ \times \Big\{1 - \lambda_1\lambda_2\bar \xi_L -(\lambda_1+\lambda_2)\bar U_ik_i +\frac{1}{2} \lambda_1^2\lambda_2^2\bar \xi_L^2\\
&\quad\  +\frac{1}{2}(\lambda_1+\lambda_2)^2\bar U_ik_i\bar U_jk_j  +\lambda_1\lambda_2(\lambda_1+\lambda_2)\bar\xi_L\bar U_ik_i  \\
 \notag & \quad\ -\frac{i}{2}(\lambda_1^2+\lambda_2^2)\bar U_i^{20(2)}k_i -i\lambda_1\lambda_2\bar U_i^{11(2)}k_i \\
\notag &\quad\ - \frac{i}{2}(\lambda_1+\lambda_2)\bar A_{ij}^{10}k_ik_j-\frac{i}{6}\bar W_{ijk}k_ik_jk_k \Big\} \,.
\end{align}
\end{subequations}
Note that this corresponds to Eq.\,(18) in \cite{WRW14} and resembles Eq.\,(72) in \cite{CRW13} except for the typo regarding the sign of the $U_i^{20(2)}$ term.

For the inference of the halo correlation function from the matter correlation function we used local Lagrangian bias \eqref{locLagBias} and fit the mass $\lgM$ determining the two bias parameters $b_1(\lgM_{\rm opt})=\langle F'\rangle$ and $b_2(\lgM_{\rm opt})=\langle F''\rangle$ from \eqref{biasparam} that are listed in Tab.\,\ref{biasfittab} for CLPT, cgCLPT and TCLPT. \highlight{This procedure can be interpreted as fitting only $b_1$ while predicting $b_2(b_1)$. We will give more details on the employed bias model and the fitting procedure in a forthcoming paper \cite{KUA15}.}

\begin{table}[h!]
  \begin{tabular}{| l | c | c | c | c | c | c | c |}
\hline
$\overline \lgM$               & 13.00 & 13.35 & 13.59 & 13.79 & 13.99 &14.25  & 14.67  \\\hline
$R(M)\,[\Mpc]$ & 3.21 & 4.20 & 5.05 & 5.89 & 6.86 & 8.38 & 11.57 \\\hline
\hline
\multicolumn{8}{|c|}{CLPT \eqref{CLPTmodel}} \\ \hline
$\lgM_{\rm opt}$  & 12.93 & 13.35 & 13.63 & 13.85 &  14.07 & 14.34 & 14.79    \\
\hline
$b_1(\lgM_{\rm opt})$            & -0.01 & 0.26 & 0.51 & 0.77 &  1.11 & 1.68 & 3.17 \\
$b_2(\lgM_{\rm opt})$     & -0.74 & -0.83 & -0.79 & -0.61 &  -0.18 & 1.06 & 7.41    \\
\hline

\multicolumn{8}{|c|}{cgCLPT (\ref{cgCLPTmodel-a}/\ref{cgCLPTmodel-b}) with $\sigx=R(M)$} \\ \hline
$\lgM_{\rm opt}$  & 12.93 & 13.35 & 13.62 & 13.84 &  14.05 & 14.30 & 14.66    \\
\hline
$b_1(\lgM_{\rm opt})$            & 0.00 & 0.26 & 0.50 & 0.76 &  1.08 & 1.58 & 2.65  \\
$b_2(\lgM_{\rm opt})$     & -0.74 & -0.83 & -0.79 & -0.62 &  -0.23 & 0.82 & 4.70    \\
\hline

\multicolumn{8}{|c|}{TCLPT \eqref{TCLPTmodel} with $\sigq=R(M)$} \\ \hline
$\lgM_{\rm opt}$  & 12.91 & 13.33 & 13.60 & 13.83 &  14.01 & 14.26 & 14.64    \\
\hline
$b_1(\lgM_{\rm opt})$            & -0.02 & 0.24 & 0.48 & 0.71 &  1.01 & 1.50 & 2.58  \\
$b_2(\lgM_{\rm opt})$     & -0.74 & -0.83 & -0.80 & -0.66 &  -0.32 & 0.61 & 4.35    \\
\hline
\end{tabular}
\caption{Best fit mass $\lgM_{\rm opt}$ for the mass bins of average mass $\overline \lgM$ for $z=0$ with the corresponding bias parameters evaluated at $\lgM_{\rm opt}$.}
\label{biasfittab}
\end{table}

\begin{figure}[h!]
\includegraphics[width=0.48\textwidth]{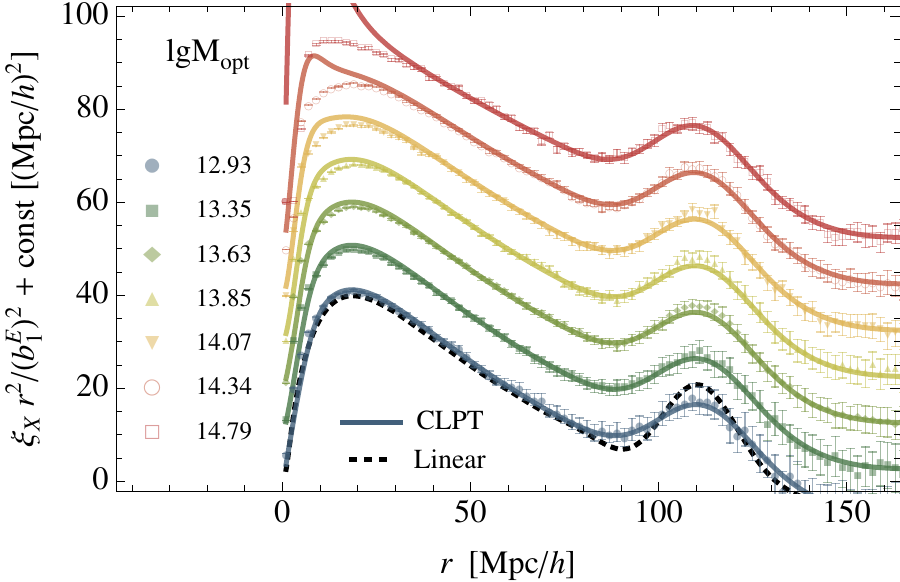}
\caption{Normalized real space halo correlation function $\xi_X$ times $r^2$ for all 7 mass bins measured in HR2 ({\it data points}) and predicted from linear theory ({\it thick black dashed}) and CLPT \eqref{CLPTmodel} {\it (thick solid)} colored according to the mass bin. To achieve a clear representation for all masses we divided $\xi_X r^2$ by the corresponding linear Eulerian bias $b_1^E(\lgM_{\mathrm{opt},i})$ and shifted all values by a constant $10(i-1)(\Mpc)^2$ according to the $i$-th mass bin.  }
\label{fig:z0realAllmass}
\end{figure}

\begin{figure}[h!]
\includegraphics[width=0.46\textwidth]{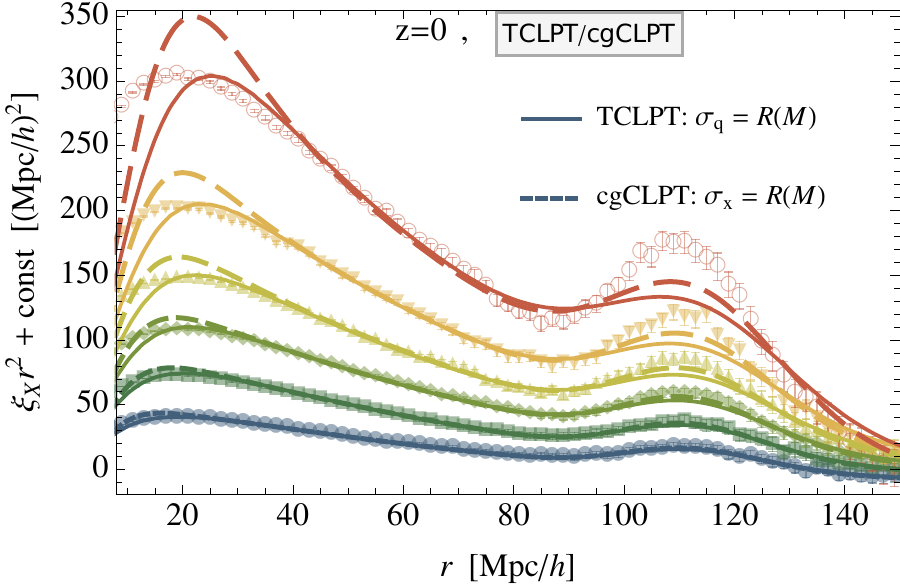}
\caption{Comparison between $\xi_X r^2$ predictions of cgCLPT (\ref{cgCLPTmodel-a}/\ref{cgCLPTmodel-b}) {\it (thick dashed)} and TCLPT \eqref{TCLPTmodel} {\it (thin solid)} smoothed on the Lagrangian radius $R(M)$ with the measurement from HR2 ({\it data points}) for the lowest 6 mass bins. For better visibility we normalized all functions by $b_1^E(\lgM_{\mathrm{opt},i})$ and shifted all values by a constant $10(i-1)(\Mpc)^2$ according to the $i$-th mass bin. }
\label{fig:xicgCLPTvsTCLPT}
\end{figure}

Fig.\,\ref{fig:z0realAllmass} shows the real space halo correlation function for all mass bins measured from the HR2 simulation in comparison to the linear and CLPT \eqref{CLPTmodel} predictions. In Fig.\,\ref{fig:xicgCLPTvsTCLPT} we compare the smoothing effect on $\xi(r)$ caused by cgCLPT (\ref{cgCLPTmodel-a}/\ref{cgCLPTmodel-b}) corresponding to a coarse-graining in Eulerian space and our coarse-graining in Lagrangian space TCLPT \eqref{TCLPTmodel} computed with a smoothed input power spectrum \highlight{with a smoothing scale given by the Lagrangian size of the halo. That this smoothing scale should be relevant for halos will be discussed in \cite{KUA15}.}

As can be seen in Fig.\,\ref{fig:z0realAllmass}, CLPT provides a quite accurate fit to the data points over a vast range of halo masses while only failing at small scales for the highest masses. As evident from Fig.\,\ref{fig:xicgCLPTvsTCLPT} smoothing on the Lagrangian scale considerable flattens out the BAO peak around $r\approx 110\Mpc$ for both, cgCLPT and TCLPT, spoiling the agreement with the $N$-body data. \highlight{It is therefore clear that keeping the smoothing scale and using a natural value for it, significantly affects the result even on scales that are naively much larger than the filter size. }

\subsection{Mean pairwise velocity \boldmath $v_{12}$}
\begin{subequations}
The expression for the mean pairwise velocity $\v{v}_{12}(\v{r},t)$ is
\label{gsmV}
 \begin{align}
[(1+\xi_X) v_{12,i}](\v{r},t) &= \int \vol{3}{q}\!\! \int \frac{\vol{3}{k}}{(2\pi)^3} e^{i \v{k}\cdot (\v{q} - \v{r}) } \int \frac{\vol{}{\lambda_1} \vol{}{\lambda_2}}{(2\pi)^2} \notag\\
& \quad\times \tilde{F}(\lambda_1)\tilde{F}(\lambda_2) K_{1,i}(\v{k},\v{q},\lambda_1,\lambda_2)  \label{gsmv}\,,
\end{align}
where $K_1$ is to be evaluated according to 
\begin{align}
\label{cumexpK1}
 \bar K_{1,i} &=\exp\left[\sum_{N=1}^\infty \frac{i^N}{N!} \langle \tilde X^N_{\v{J}=0}\rangle_c\right] \left[\sum_{N=0}^\infty \frac{i^N}{N!} \Bigg\langle \dot{\bar \Delta}_i\tilde X^N_{\v{J}=0}\Bigg\rangle_c \right]\Bigg|_{\mathcal{O}(P_L^2)} \,.
\end{align}
Plugging the exponential \eqref{sumiX} into the expression \eqref{cumexpK1} and expanding the second term gives
\begin{align}
\notag  \bar K_{1,i} &= e^{-\tfrac{1}{2}\bar A_{ij}k_ik_j} e^{-\tfrac{1}{2} (\lambda_1^2+\lambda_2^2)\bar \sigma_R^2}\\
\notag  &\ \ \times \Big\{i(\lambda_1+\lambda_2)\dot{\bar U}_i+ik_j\dot{\bar A}_{ji}-\frac{1}{2}(\lambda_1^2+\lambda_2^2)\dot{\bar U}^{20}_i \\
  &\quad  -\lambda_1\lambda_2\dot{\bar U}^{11}_i -\frac{1}{2}k_jk_k\dot{\bar W}_{jki} -(\lambda_1+\lambda_2)k_j\dot{\bar A}^{10}_{ji} \\
\notag  &\quad -i\lambda_1\lambda_2(\lambda_1+\lambda_2)\bar\xi_L\dot{\bar U}_i -i(\lambda_1+\lambda_2)^2k_j\bar U_j\dot{\bar U}_i \\
 \notag &\quad -i\lambda_1\lambda_2\bar\xi_Lk_j\dot{\bar A}_{ji}-i(\lambda_1+\lambda_2)k_jk_k\bar U_j\dot{\bar A}_{ki} \Big\} \,,
\end{align}
where in addition to \eqref{defCorcg} we defined
\begin{align}
\label{defCorcgDot}
\dot{\bar U}_i^{mn(k)} &:= \langle \bar \delta_1^m \bar \delta_2^n \dot{\bar \Delta}_i^{(k)} \rangle_c \  , \ 
\dot{\bar A}_{ij}^{mn} := \langle \bar \delta_1^m \bar \delta_2^n \bar \Delta_i \dot{\bar \Delta}_j\rangle_c \  , \notag\\
\dot{\bar W}_{ijk}^{mn} &:= \langle \bar \delta_1^m \bar \delta_2^n \bar \Delta_i \bar \Delta_j \dot{\bar \Delta}_k \rangle_c \,.
\end{align}
\end{subequations}
This result is analogous to Eq.\,(29) in \cite{WRW14} except for the use of coarse-grained instead of dust correlators. From this the pairwise velocity $v_{12}(r)$ defined as $\v{v}_{12}\cdot\hat\vz = v_{12}(r) r_{\|}/r$ is computed according to \eqref{gsmV}. 

\begin{figure}[h!]
\includegraphics[width=0.46\textwidth]{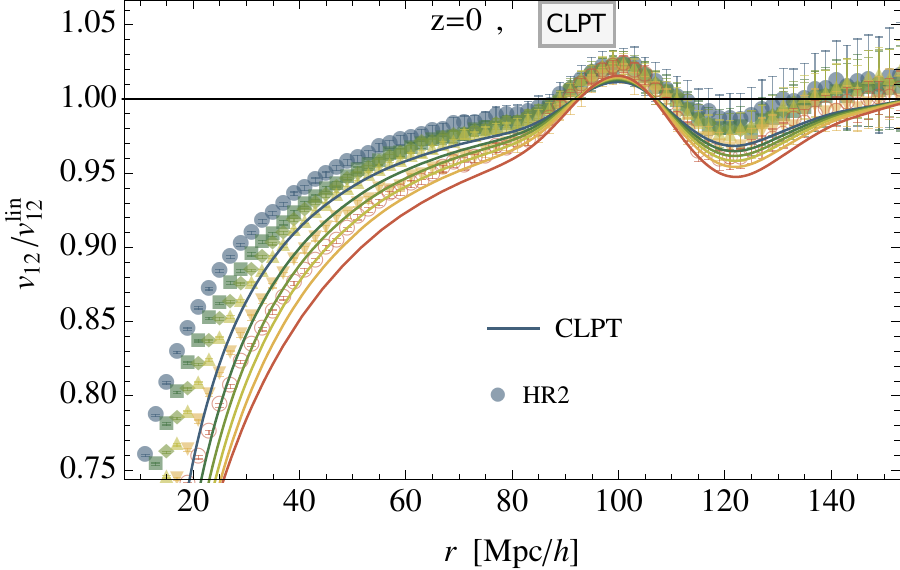}\\[8pt]
\caption{Mean pairwise velocity $v_{12}$ from \eqref{gsmV} compared to linear theory $v_{12}^{\text{lin}}$ \eqref{linv12} for the 6 lowest mass bins for CLPT \eqref{CLPTmodel} ({\it thin solid lines}) and measurements from HR2 ({\it thin dashed}). }
\label{fig:v12comp}
\end{figure}

In Fig.\,\ref{fig:v12comp} we show the CLPT \eqref{CLPTmodel} prediction for the pairwise velocity together with the HR2 data points normalized by the linear result, see \cite{RW11}, which is given by
\begin{align}
v_{12,L} &= -2\sH f b_1^{\rm E}\ \frac{1}{2\pi^2} \int_0^\infty d k\, P_L(k) j_1(k r) \,. \label{linv12}
\end{align}
In Fig.\,\ref{fig:v12comp} we see that the CLPT prediction for the pairwise velocity is relatively inaccurate, especially compared to the excellent agreement found in Fig.\,\ref{fig:z0realAllmass} for the real space correlation function. On the largest scales CLPT correctly describes the $N$-body data and reproduces the linear theory result \eqref{linv12}, \highlight{however around the BAO scale at $120\Mpc$, there is systematic offset.}

\begin{figure}[b!]
\includegraphics[width=0.46\textwidth]{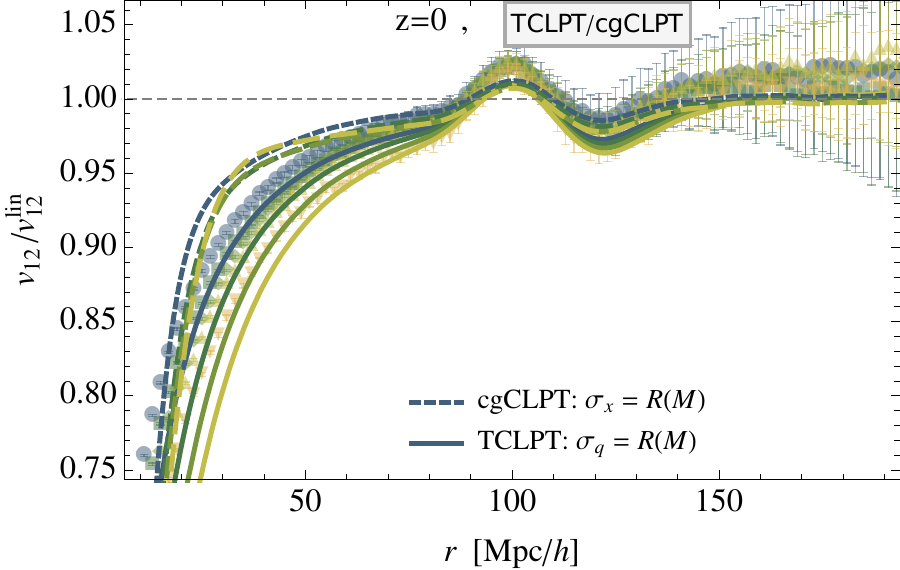}\\[8pt]
\caption{Comparison between predictions for the mean pairwise velocity $v_{12}$ compared to linear theory for cgCLPT (\ref{cgCLPTmodel-a}/\ref{cgCLPTmodel-b}) {\it (thick dashed)} and TCLPT \eqref{TCLPTmodel} {\it (thick solid)} smoothed on the Lagrangian radius $R(M)$ with the measurement from HR2 ({\it thin dashed}) for the lowest 2 mass bins.
}
\label{fig:v12cgCLPTvsTCLPT}
\end{figure} 

In Fig.\,\ref{fig:v12cgCLPTvsTCLPT} we show a comparison of cgCLPT (\ref{cgCLPTmodel-a}/\ref{cgCLPTmodel-b}) and TCLPT \eqref{TCLPTmodel}, both smoothed on the Lagrangian radius $R(M)$, together with the HR2 measurements. 

\begin{figure}[h!]
\includegraphics[width=0.47\textwidth]{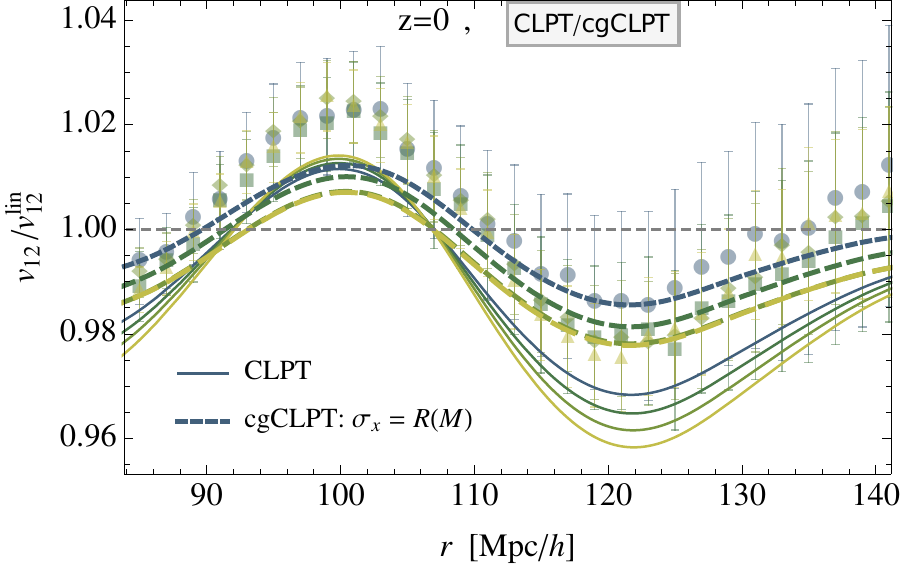}
\caption{Detailed view on large scales of the mean pairwise velocity $v_{12}$ compared to linear theory for CLPT \eqref{CLPTmodel} ({\it thin solid}) and cgCLPT (\ref{cgCLPTmodel-a}/\ref{cgCLPTmodel-b}) {\it (thick dashed)} smoothed on the Lagrangian radius $R(M)$ with the HR2 measurement ({\it data points}) for the lowest 4 masses.
}
\label{fig:v12CLPTvscgCLPT}
\end{figure}

\highlight{From Fig.\,\ref{fig:v12cgCLPTvsTCLPT} and Fig.\,\ref{fig:v12CLPTvscgCLPT} we deduce that cgCLPT systematically increases the absolute value of the pairwise velocity compared to CLPT on small and on BAO scales. It shows only a small mass-dependence if the only the four smallest masses are considered. While cgCLPT improves the CLPT prediction on very large scales, TCLPT generally behaves better below $90\Mpc$. TCLPT exhibits a stronger mass-dependence than cgCLPT.}

\subsection{Mean pairwise velocity dispersion \boldmath $\sigma_{12}^2$}
In order to evaluate the pairwise velocity dispersion $\v{\sigma}_{12}^2=\v{\tilde \sigma}_{12}^2-\v{v}_{12}\v{v}_{12}$ we have to determine
\begin{subequations}
\label{gsmSig} 
 \begin{align}
[(1+\xi_X)\tilde\sigma^2_{12,ij}](\v{r},t) &= \int \vol{3}{q}\!\! \int \frac{\vol{3}{k}}{(2\pi)^3} e^{i \v{k}\cdot (\v{q} - \v{r}) } \int \frac{\vol{}{\lambda_1} \vol{}{\lambda_2}}{(2\pi)^2} \notag\\
& \ \times \tilde{F}(\lambda_1)\tilde{F}(\lambda_2) K_{2,{ij}}(\v{k},\v{q},\lambda_1,\lambda_2) \label{gsmsig} \,,
\end{align}
where $K_2$ is computed using Eq.\,\eqref{fxizspaceanalogue}. 
As mentioned before, $K_2$ depends on the tracer's velocity dispersion which vanishes identically for the models CLPT \eqref{CLPTmodel}, scgCLPT \eqref{cgCLPTmodel-a} and TCLPT \eqref{TCLPTmodel} 
but is relevant for cgCLPT \eqref{cgCLPTmodel-b}. Therefore we split $K_2$ into one contribution $\bar K_2$ from the dust model and another $K^{\sigx,\sigu}_2$ from velocity dispersion which is for cgCLPT \eqref{cgCLPTmodel-b} controlled by the smoothing scales $\sigx$ and $\sigu$
\begin{align}
\label{K2}
K_{2,ij} = \bar K_{2,ij} + K^{\sigx,\sigu}_{2,ij} \,.
\end{align}
\end{subequations}

The standard contribution to $K_{2,ij}$ for the dustlike model is identical to Eq.\,(34-35) in \cite{WRW14}
\begin{subequations}
\label{K2bar}
\begin{align}
\bar K_{2,ij} &= \exp\left[\sum_{N=1}^\infty \frac{i^N}{N!} \langle \tilde X^N_{\tilde J=0}\rangle_c\right] \Bigg[ \sum_{N=0}^\infty \frac{i^N}{N!} \Big\langle \dot{\bar\Delta}_i\dot{\bar\Delta}_j\tilde X^N_{\v{J}=0}\Big\rangle_c \\
\notag &\qquad\qquad\qquad + \sum_{N,M=0}^\infty \frac{i^{N+M}}{N!M!} \Big\langle \dot{\bar\Delta}_i\tilde X^N_{\v{J}=0}\Big\rangle_c  \Big\langle \dot{\bar\Delta}_j\tilde X^M_{\v{J}=0} \Big\rangle_c \Bigg] \Bigg|_{\mathcal{O}(P_L^2)}\label{standardK2} \,,
\end{align}
but evaluated with the kernels for the smoothed quantities
\begin{align}
\notag \bar K_{2,ij}&= e^{-\tfrac{1}{2}\bar A_{ij}k_ik_j} e^{-\tfrac{1}{2} (\lambda_1^2+\lambda_2^2)\bar \sigma_R^2} \\
\notag &\ \ \times \Big\{ (\lambda_1+\lambda_2)^2\dot{\bar U}_i\dot{\bar U}_j -(\lambda_1+\lambda_2)(\dot{\bar A}_{ki}k_k\dot{\bar U}_j +\dot{\bar A}_{kj}k_k\dot{\bar U}_i) \\
\notag &\quad -\dot{\bar A}_{ki}k_k \dot{\bar A}_{lj}k_l +[1-\lambda_1\lambda_2\bar\xi_L-(\lambda_1+\lambda_2)\bar U_k k_k]\ddot{\bar A}_{ij} \\
 & \quad + i(\lambda_1+\lambda_2)\ddot{\bar A}^{10}_{ij}+i\ddot{\bar W}_{kij}k_k\Big\} \,, 
\end{align}
where in addition to \eqref{defCorcg} and \eqref{defCorcgDot} we defined
\begin{align}
\label{defCorcgDDot}
\ddot{\bar A}_{ij}^{mn} := \langle \bar \delta_1^m \bar \delta_2^n \dot{\bar \Delta}_i \dot{\bar \Delta}_j\rangle_c \  , \ 
\ddot{\bar W}_{ijk}^{mn} := \langle \bar \delta_1^m \bar \delta_2^n \bar \Delta_i \dot{\bar \Delta}_j \dot{\bar \Delta}_k \rangle_c \,.
\end{align}
\end{subequations}

\begin{figure}[b!]
\includegraphics[width=0.46\textwidth]{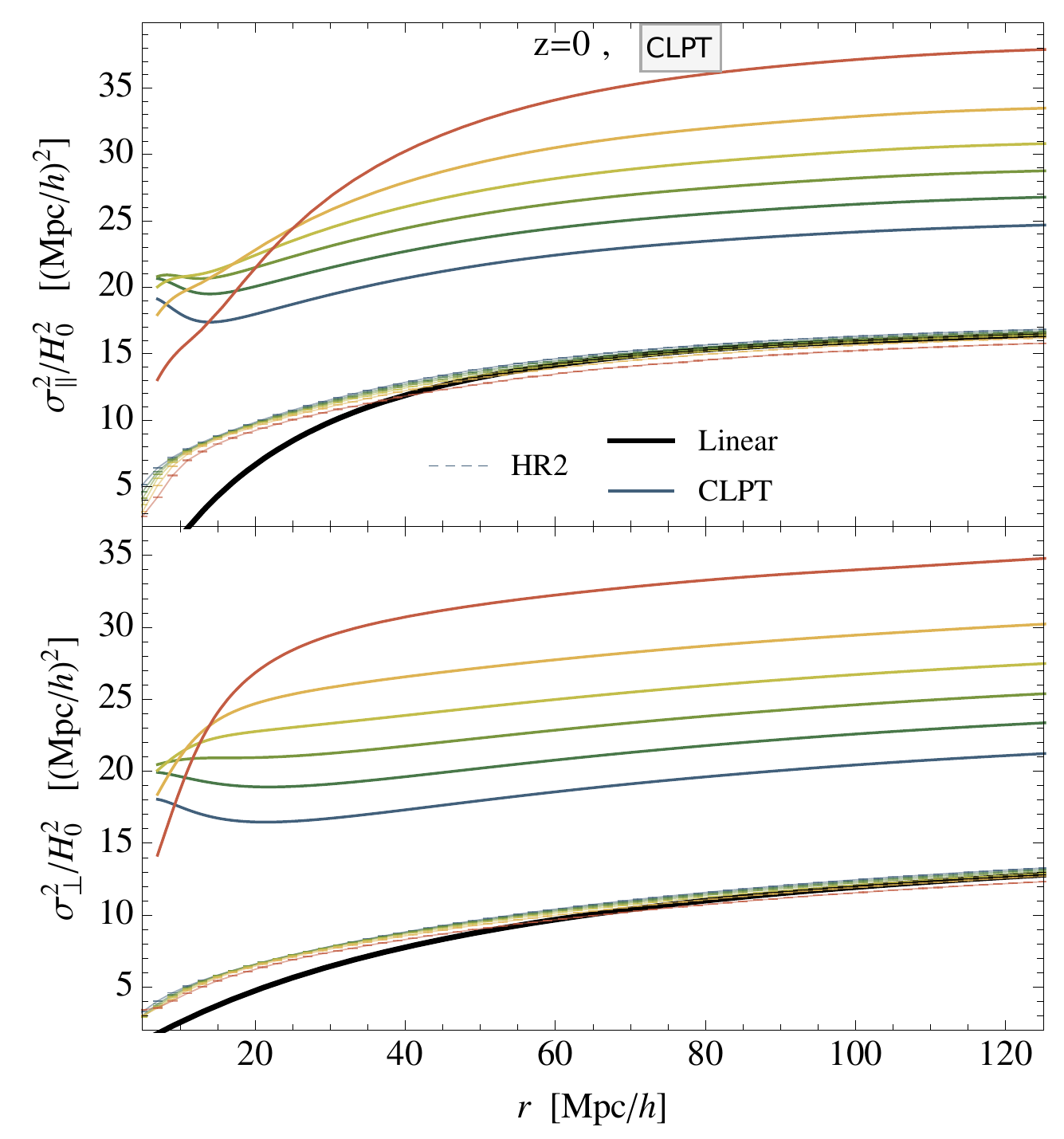}
\caption{Pairwise velocity dispersion $\sigma_{12}^2$ from \,\eqref{K2bar} split according to \eqref{sigperpparsplit} into parts parallel $\sigma_{||}^2$ and perpendicular $\sigma_{\perp}^2$ to the pair separation. Shown are the HR2 measurements ({\it thin dashed}) lying close to each other for all masses, the mass-independent linear theory prediction \eqref{sig12lin} ({\it thick black}) and the CLPT prediction \eqref{CLPTmodel} ({\it thin solid}).}
\label{fig:sigcomp}
\end{figure}
In Fig.\,\ref{fig:sigcomp} the pairwise velocity dispersion $\sigma^2_{12}$ from CLPT \eqref{CLPTmodel} is shown together with the $N$-body measurements and the mass-independent result of linear theory, see \cite{RW11},
\begin{align}
\label{sig12lin}
\notag  \sigma^2_{||,L}&=  2\sH^2 f^2 \left(R_{\rm NL}^2 -\frac{1}{2\pi^2} \int_0^\infty d k\, P_L(k) \left( j_0(k r)- \frac{2 j_1(k r)}{k r}\right) \right)\\
 \sigma^2_{\perp,L}& =  2\sH^2 f^2 \left(R_{\rm NL}^2 - \frac{1}{2\pi^2} \int_0^\infty d k\, P_L(k) \frac{ j_1(k r)}{k r}\right)\\
\notag R_{\rm NL}^2&= \frac{1}{6 \pi^2} \int_0^\infty dk\, P_L(k) \,.
\end{align}
We can clearly see that CLPT significantly overestimates the amplitude of the pairwise velocity dispersion even on large scales and exhibits a strong mass dependence. The data points from HR2 lie close to each other for all masses which is captured by the linear theory result on large scales. The comparatively excellent performance of linear perturbations must be considered as accidental: the pairwise dispersion $\sigma_{12}^2$ contains contributions of the one-point velocity variance $\langle v^i(\v{x}) v^j(\v{x}) \rangle$, which we denoted by $R_{\rm NL}^2$ in the linear perturbation theory \eqref{sig12lin}. This term is obtained as the limit $r\rightarrow 0$ of $\langle v^i(\v{x}+\v{r}) v^j(\v{x}) \rangle$ and thus is sensitive to smallest scales. Therefore a large scale-independent offset error of $\sigma_{12}^2$ is expected. In \cite{RW11,WRW14} it has been accounted for that error by shifting the CLPT predictions to agree with the $N$-body measurements on large scales.

\begin{figure}[b!]
\includegraphics[width=0.46\textwidth]{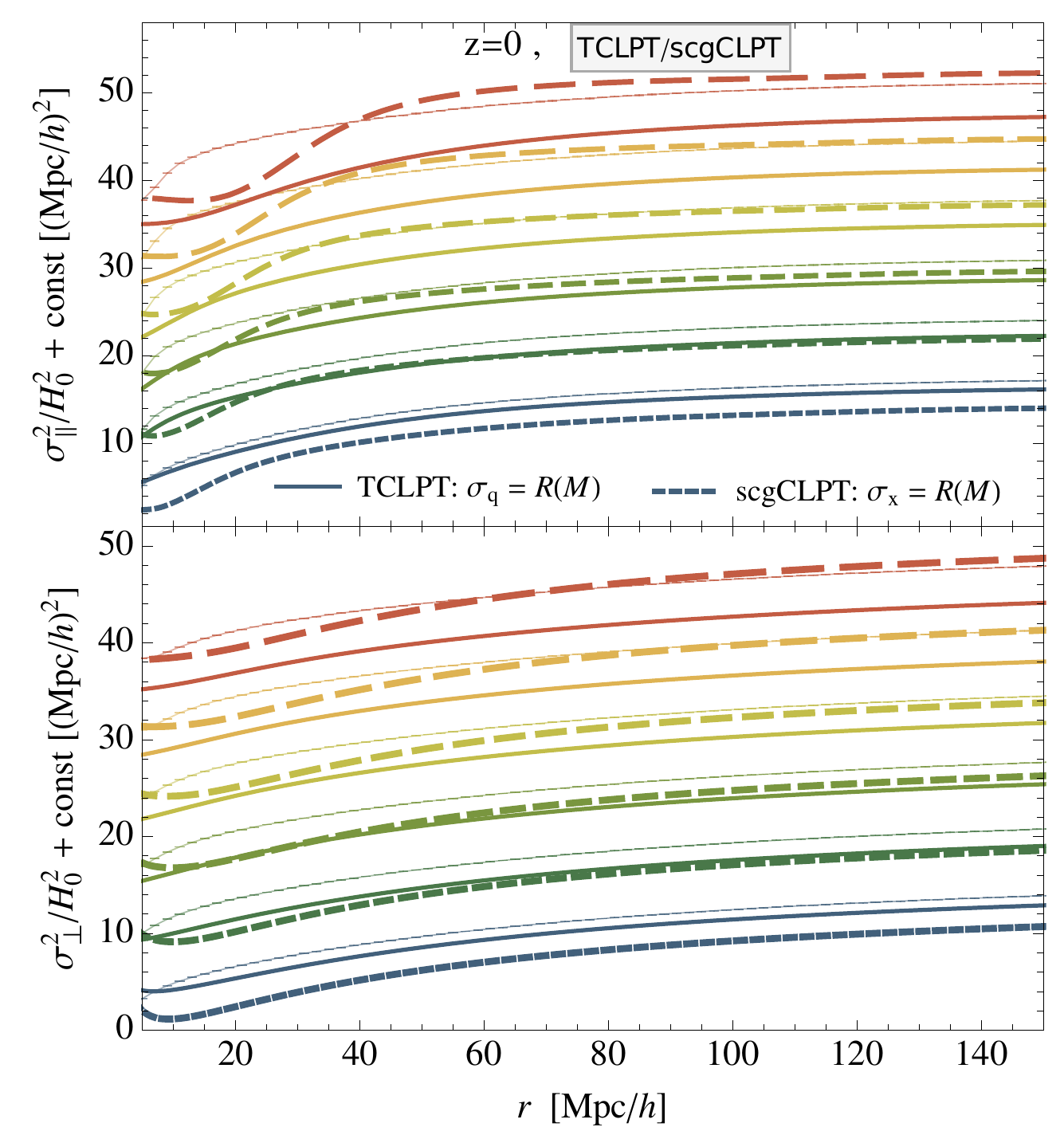}
\caption{Comparison between $\sigma_{12}^2$ predictions of scgCLPT \eqref{cgCLPTmodel-a} {\it (thick dashed)} and TCLPT \eqref{TCLPTmodel} {\it (thick solid)} smoothed on the Lagrangian radius $R(M)$ with the measurement from HR2 ({\it data points}) for the lowest 6 mass bins. For better visibility we shifted all values by a constant $7(i-1) (\Mpc)^2$ according to the $i$-th mass bin.}
\label{fig:sigxcgCLPTvsTCLPT}
\end{figure}

Fig.\,\ref{fig:sigxcgCLPTvsTCLPT} compares scgCLPT \eqref{cgCLPTmodel-a} containing modified fluid dynamics to TCLPT \eqref{TCLPTmodel} based on smoothing the input power spectrum. The smoothing on the Lagrangian scale $R(M)$ significantly reduces the amplitude of $\sigma_{12}$ for all masses and narrows down the mass dependence compared to CLPT, see Fig.\ref{fig:sigcomp}, bringing both models to better agreement with the data points. \highlight{We therefore suggest that  one should not adjust by hand the off-set of $\sigma_{12}$, but should take this as an indication that a smoothing around the Lagrangian size of the halo should be applied.}\\

\paragraph*{Contribution from the velocity dispersion of the tracer}
For the cgCLPT model \eqref{cgCLPTmodel-b} we have to consider the conceptually new contribution to $K_2$ that arises from the velocity dispersion of the tracers, encoded in $C_X^{(2)}=\bar C^{(2)}$, according to 
\begin{align}
K^{\sigx,\sigu}_{2,ij} &= K^{\sigx}_{2,ij} + K^{\sigu}_{2,ij} = \exp\left[\sum_{N=1}^\infty \frac{i^N}{N!} \langle \tilde X^N_{\tilde J=0}\rangle_c\right] \label{sigmaxpK2}\\
\notag &\times \Bigg[  \sum_{N=0}^\infty \frac{i^N}{N!} \Big\langle \tilde X^N_{\v{J}=0}  \left\{ \bar C^{(2)}_{ij}(\vx_1(\vq_1))+\bar C^{(2)}_{ij}(\vx_2(\vq_2)) \right\}\Big\rangle_c  \Bigg] \Bigg|_{\mathcal{O}(P_L^2)} \,.
\end{align}

\begin{figure}[b!]
\includegraphics[width=0.475\textwidth]{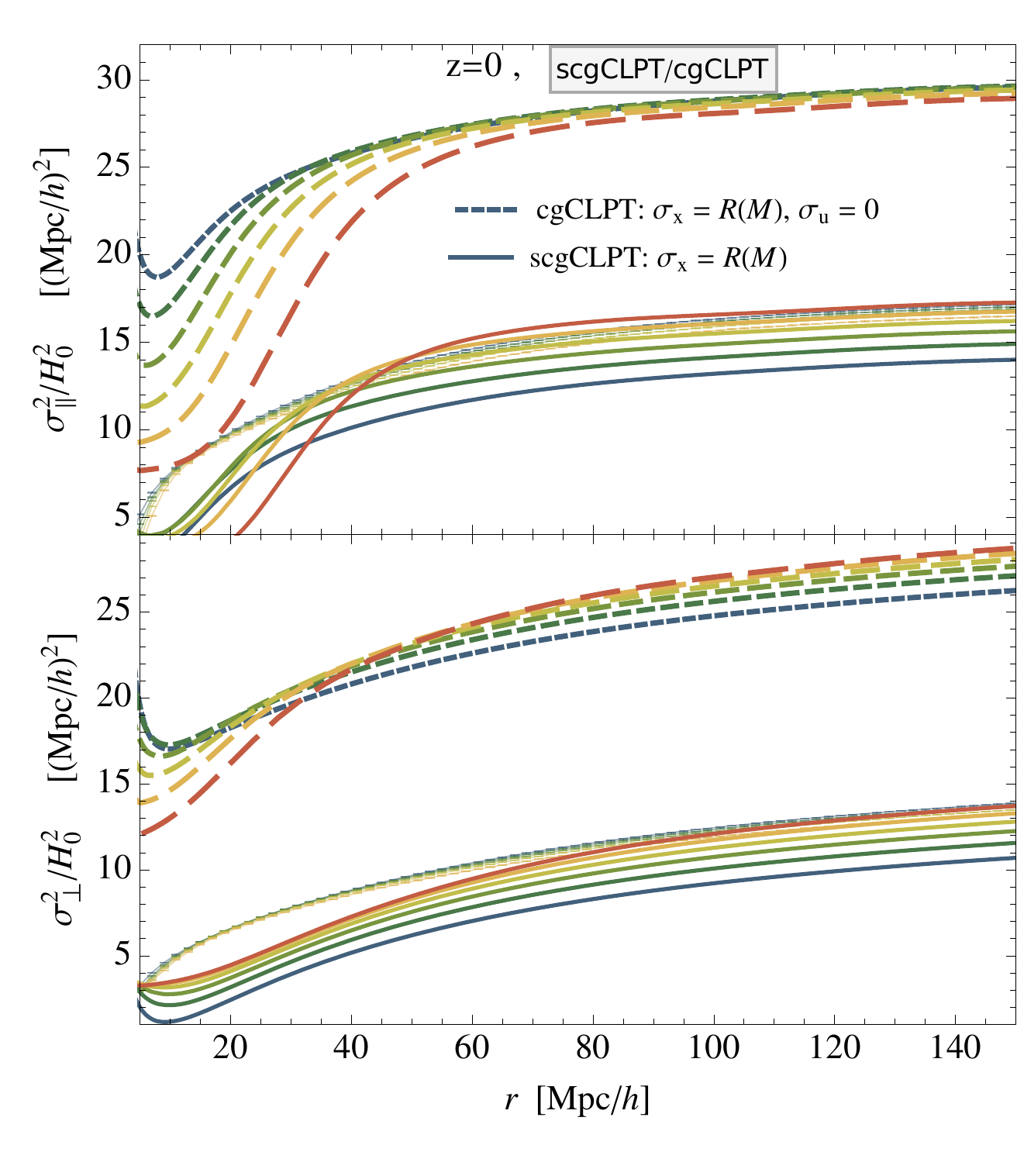}\\[2pt]
\includegraphics[width=0.475\textwidth]{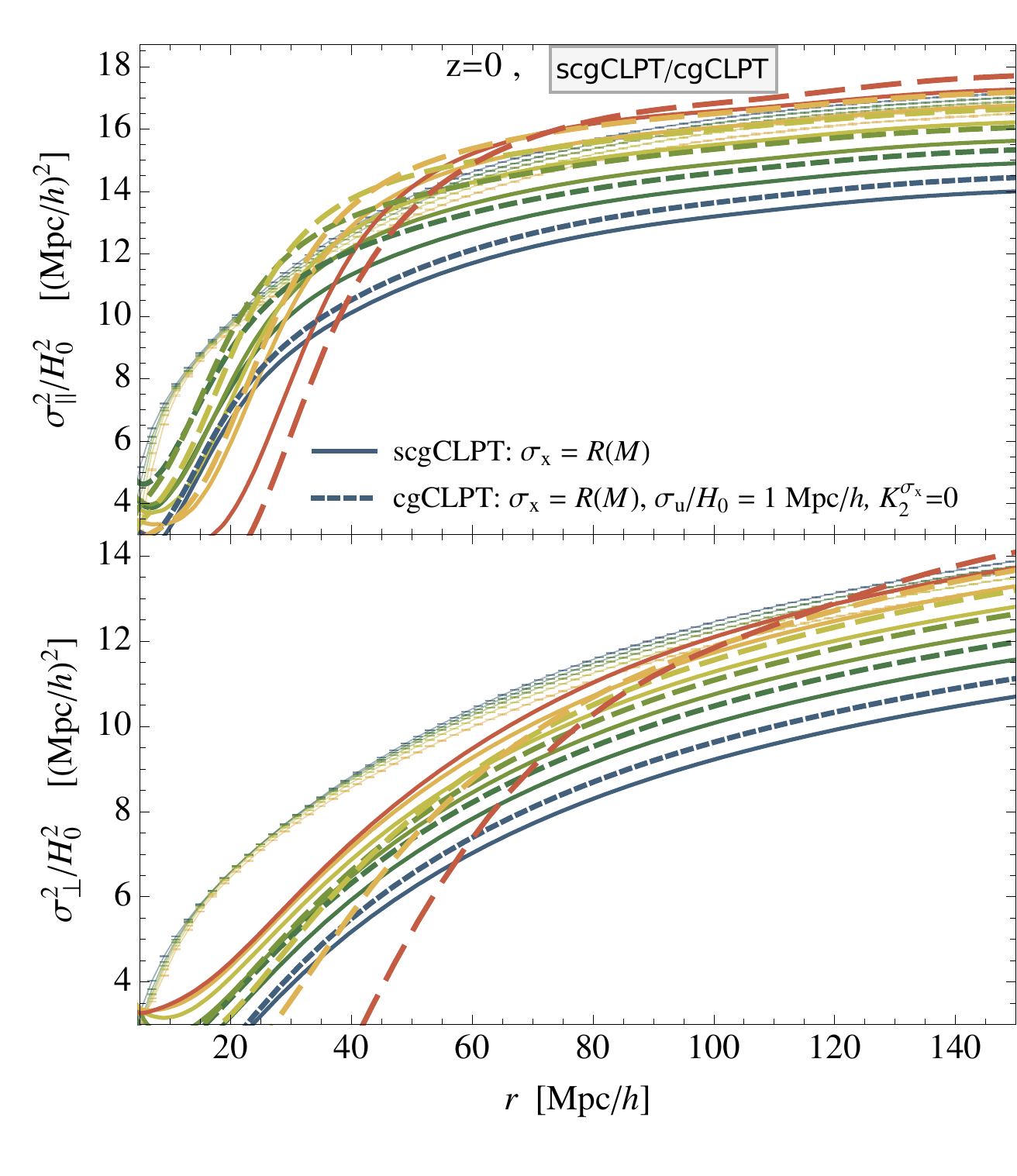}
\caption{Comparison between $\sigma_{12}^2$ predictions from 
scgCLPT \eqref{cgCLPTmodel-a} {\it (thin solid)}, based only on modified fluid dynamics encoded in $\bar K_2$ \eqref{K2bar}, and
cgCLPT \eqref{cgCLPTmodel-b} {\it (thick dashed)}, including higher tracer cumulants, smoothed on the Lagrangian radius $R(M)$ together with the HR2 measurement ({\it data points}). {\it upper panel} The contribution $K_2^\sigx$ \eqref{K2sigx} from the spatial smoothing $\sigx$. {\it lower panel}  The contribution $K_2^\sigu$ \eqref{K2sigu} from the velocity smoothing $\sigu/H_0=1\Mpc$. }
\label{fig:sigxpcgCLPTvsTCLPTonlyKbar}
\end{figure}
$\,$\bigskip\\
It is affected by both the spatial smoothing scale $\sigx$ and the velocity smoothing scale $\sigu$. Their contributions labeled $K_2^\sigx$ and $K_2^\sigu$ will be evaluated separately in the following.

The corrections to the pairwise velocity dispersion $\sigma_{12}^2$ connected to the spatial coarse-graining $\sigx$ are given by
\begin{align}
\notag K^{\sigma_x}_{2,ij} &:= \exp\left[\sum_{N=1}^\infty \frac{i^N}{N!} \langle \tilde X^N_{\tilde J=0}\rangle_c\right]  \\
&\quad \times \sum_{N=0}^\infty \frac{i^N}{N!} \Bigg\langle \tilde X^N_{\v{J}=0}\left[\left( \frac{\overline{(1+\delta)v_iv_j}}{1+\bar \delta}-\bar v_i\bar v_j \right)(\vx_1(\vq_1))\right. \label{K2sigx}\\
\notag &\qquad\qquad\qquad\quad \left.+\left(\frac{\overline{(1+\delta)v_iv_j}}{1+\bar \delta}-\bar v_i\bar v_j\right)(\vx_2(\vq_2))\right] \Bigg\rangle_c \ \Bigg|_{\mathcal O(P_L^2)} \,,
\end{align}
and have to be calculated by explicitly evaluating the corresponding correlators as done in App.\,\ref{AppK2}.  

The correction term due to the coarse-graining $\sigu$ with respect to velocity
\begin{subequations}
\label{K2sigu}
\begin{align}
K^{\sigu}_{2,ij}  &:=2\sigu^2 \delta_{ij} \exp\left[\sum_{N=1}^\infty \frac{i^N}{N!} \langle \tilde X^N_{\v{J}=0}\rangle_c\right] \left[\sum_{N=0}^\infty \frac{i^N}{N!} \Big\langle \tilde X^N_{\v{J}=0} \Big\rangle_c \right] \Bigg|_{\mathcal{O}(P_L^2)} \,, \label{K2sigp}
\end{align}
can be obtained easily by combining \eqref{sumiX} and \eqref{eiX}
\begin{align}
\notag K^{\sigu}_{2} &= 2\sigu^2 \ e^{-\tfrac{1}{2}\bar A_{ij}k_ik_j} e^{-\tfrac{1}{2} (\lambda_1^2+\lambda_2^2)\bar \sigma_R^2}\\
&\notag \times  \Big\{1 - 2\lambda_1\lambda_2\bar \xi_L -2(\lambda_1+\lambda_2)\bar U_ik_i - \frac{1}{2} (\lambda_1^2+\lambda_2^2)\bar \sigma_R^2  \\
\notag & -\frac{1}{2}\bar A_{ij}k_ik_j+\frac{3}{2} \lambda_1^2\lambda_2^2\bar \xi_L^2 + \frac{1}{2} \lambda_1\lambda_2(\lambda_1^2+\lambda_2^2) \bar \xi_L\bar \sigma_R^2
\\
 & +3\lambda_1\lambda_2(\lambda_1+\lambda_2)\bar \xi_L \bar U_ik_i +\frac{1}{2}\lambda_1\lambda_2\bar \xi_L \bar A_{ij}k_ik_j \\
\notag & +\frac{3}{2}(\lambda_1+\lambda_2)^2\bar U_ik_i\bar U_jk_j +\frac{1}{2} (\lambda_1+\lambda_2)(\lambda_1^2+\lambda_2^2) \bar \sigma_R^2\bar U_ik_i \\
\notag &  + \frac{1}{2} (\lambda_1+\lambda_2) \bar A_{ij}k_ik_j \bar U_kk_k  -i(\lambda_1^2+\lambda_2^2)\bar U_i^{20(2)}k_i\\
\notag&  -2i\lambda_1\lambda_2\bar U_i^{11(2)}k_i - i(\lambda_1+\lambda_2)\bar A_{ij}^{10}k_ik_j-\frac{i}{3}\bar W_{ijk}k_ik_jk_k \Big\}  \,.
\end{align}
\end{subequations}

In Fig.\,\ref{fig:sigxpcgCLPTvsTCLPTonlyKbar} a comparison is shown between the predictions of scgCLPT \eqref{cgCLPTmodel-a}, containing only modified dynamics corresponding to $\bar K_2$, and cgCLPT \eqref{cgCLPTmodel-b}, also including velocity dispersion encoded in $K_2^\sigx$ and $K_2^\sigu$. 
The inclusion of higher cumulants, which were assumed to be identical to those of dark matter, significantly increases the amplitude of the pairwise velocity dispersion thereby spoiling the agreement with the data. This effect is due to the contribution of the spatial coarse-graining scale $\sigx$ while the velocity coarse-graining $\sigu$ has hardly any effect for reasonable values of $\sigma_u$.

\subsection{Mean pairwise velocity skewness \boldmath $\Lambda_{12}$}
\begin{subequations}
The first non-Gaussian correction $\v{\tilde{\Lambda}}_{12}$ in the ESM \eqref{ESM2} is, in analogy to $\v{v}_{12}$ and $\v{\tilde{\sigma}}_{12}^2$ defined as
 \begin{align}
[(1+\xi_X)\tilde\Lambda_{12,ijk}](\v{r},t) &= \int \vol{3}{q}\!\! \int \frac{\vol{3}{k}}{(2\pi)^3} e^{i \v{k}\cdot (\v{q} - \v{r}) } \int \frac{\vol{}{\lambda_1} \vol{}{\lambda_2}}{(2\pi)^2} \notag\\
& \ \times \tilde{F}(\lambda_1)\tilde{F}(\lambda_2) K_{3,{ijk}}(\v{k},\v{q},\lambda_1,\lambda_2) \label{gsmLam} \,.
\end{align}
As it was the case for $K_2$, the explicit expression for $K_3$ comprises two parts
\begin{align}
K_{3,ijk} &= \bar K_{3,ijk} + K_{3,ijk}^{\sigx,\sigu} \,.
\end{align}
The first $\bar K_{3,ijk}$ is the contribution from the dust model evaluated with the kernels for the smoothed quantities, while the second one $K_{3,ijk}^{\sigx,\sigu}$ only appears if higher cumulants of the tracer are present. To isolate the effect of non-Gaussianities in the pairwise velocity distribution we explicitly calculate only the dust part 
\begin{align}
\bar K_{3,ijk} &= \notag \exp\left[\sum_{N=1}^\infty \frac{i^N}{N!} \langle \tilde X^N_{\tilde J=0}\rangle_c\right] \times \left[ \sum_{N=0}^\infty \frac{i^N}{N!} \Big\langle \dot{\bar\Delta}_i\dot{\bar\Delta}_j\dot{\bar\Delta}_k\tilde X^N_{\v{J}=0}\Big\rangle_c \right. \\
&+ \stackrel{+\text{ cyc. perm.}}{\sum_{N,M=0}^\infty \frac{i^{N+M}}{N!M!} \Big\langle \dot{\bar\Delta}_i\dot{\bar\Delta}_j\tilde X^N_{\v{J}=0}\Big\rangle_c \Big\langle \dot{\bar\Delta}_k\tilde X^M_{\v{J}=0} \Big\rangle_c}\label{K3standard}\\
&+\!\!\!\!  \left.\sum_{N,M,K=0}^\infty \frac{i^{N+M+K}}{N!M!K!} \Big\langle \dot{\bar\Delta}_i\tilde X^N_{\v{J}=0}\Big\rangle_c\Big\langle \dot{\bar\Delta}_j\tilde X^M_{\v{J}=0} \Big\rangle_c  \Big\langle \dot{\bar\Delta}_k\tilde X^K_{\v{J}=0} \Big\rangle_c \right] \Bigg|_{\mathcal{O}(P_L^2)} \!\! \,.\notag
%
\end{align}

\begin{figure}[h!]
\includegraphics[width=0.48\textwidth]{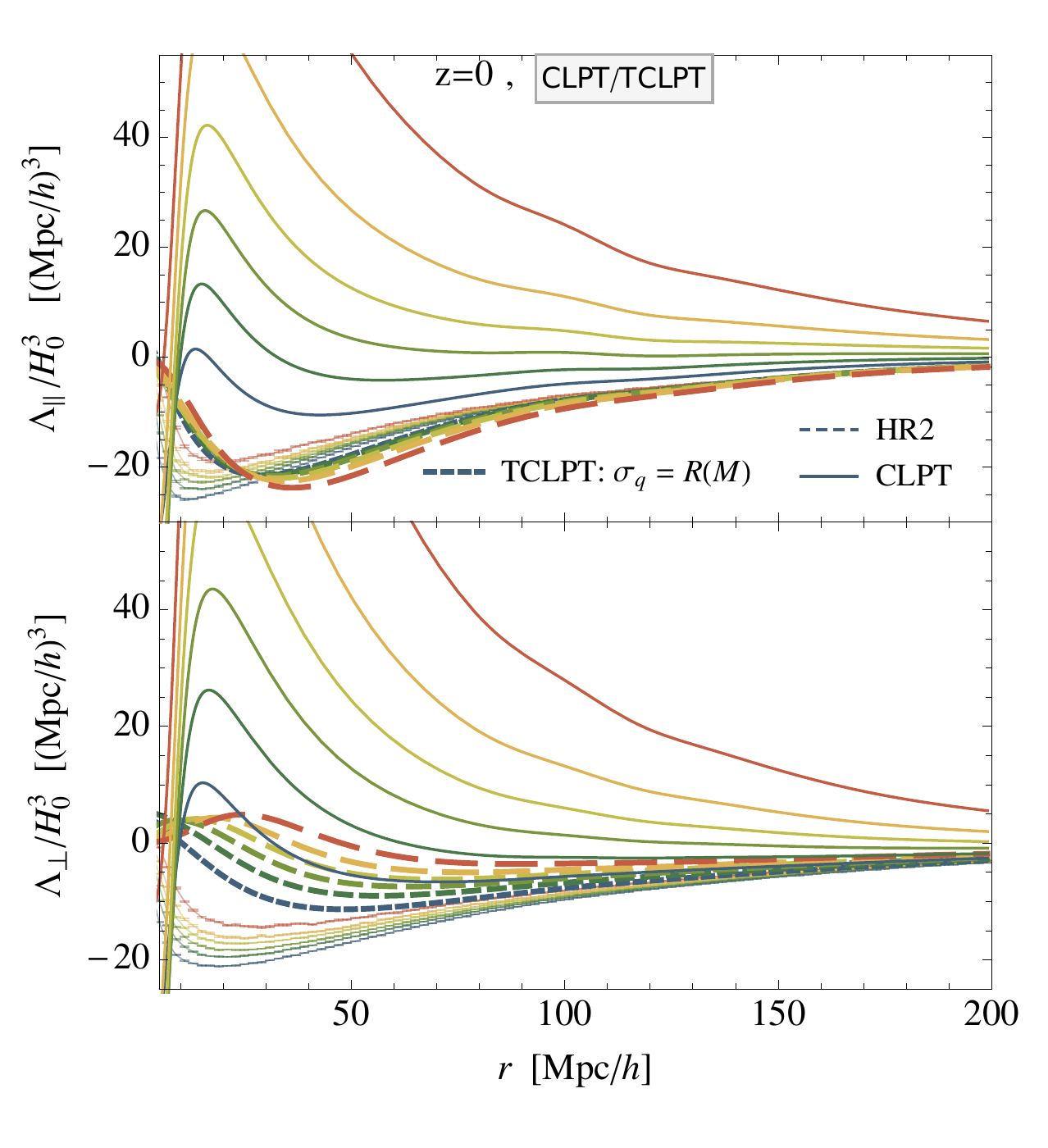}
\caption{Pairwise velocity skewness $\Lambda_{12}$ from \,\eqref{gsmLam} split according to \eqref{lamperpparsplit} into parts parallel $\Lambda_{||}$ and perpendicular $\Lambda_{\perp}$ to the pair separation vector. Shown are the measurements from HR2 ({\it data points}) which lie close to each other for all masses, the CLPT prediction \eqref{CLPTmodel} ({\it thin solid}) and the TCLPT prediction \eqref{TCLPTmodel} ({\it thick dashed}) smoothed on the Lagrangian radius $R=R(M)$.}
\label{fig:lamcomp}
\end{figure}
\end{subequations}

The CLPT result, when re-expressed in terms of the CLPT correlators defined in \eqref{defCorcg}, reads 
\begin{align}
\bar K_{3,ijk}&= (faH)^3 \times e^{-\tfrac{1}{2}\bar A_{ij}k_ik_j} e^{-\tfrac{1}{2} (\lambda_1^2+\lambda_2^2)\bar \sigma_R^2} \notag\\
&\quad \times \Big\{ 2 \bar W_{ijk} + i \bar A^{(11)}_{(ij}\left[(\lambda_1+\lambda_2) \bar U^{(1)}_{k)} + \bar A^{(11)}_{k)m}k^m\right]\Big\} \,.
\end{align}

In Fig.\,\ref{fig:lamcomp} we show the effect of the pairwise skewness $\Lambda_{12}$ which is the leading order non-Gaussian correction in the ESM \eqref{ESM2}. Similarly as done for the pairwise velocity dispersion $\sigma_{12}^2$, the skewness $\Lambda_{12}$ is also split into a part parallel and perpendicular to the line of sight according to \eqref{lamperpparsplit}. 
The result shows that both CLPT and TCLPT based on the dust model fail to capture the non-Gaussian effects that are encoded in the pairwise skewness basically for all scales below $r=100\Mpc$. \highlight{However, it should be noticed that a smoothing on the Lagrangian scale does again bring the theory closer to the $N$-body simulation, while reducing the spread in mass.}

\subsection{Results}
\label{subsec:results}

The comparison of the streaming ingredients and the $N$-body data for large galaxy to cluster sized halos shows that the accuracy of the CLPT prediction is not sufficient to warrent the use of the ESM, see Fig.\,\ref{fig:lamcomp}. Hence, the GSM is sufficient when applied to CLPT based on the dust model and the ESM requires an improved theoretical modeling of the ESM ingredients with percent precision on $10-30\Mpc$ scales. 

We investigated how the real space halo correlation function and the velocity statistics are affected by a smoothing on the \highlight{Lagrangian radius, the length scale corresponding to the physical size of a halo. In a forthcoming paper \cite{KUA15} we illustrate that a smoothing on the Lagrangian size indeed optimizes the LPT prediction of the displacement field on the level of realizations compared to an $N$-body simulation.} Within our base model, GSM combined with CLPT, we observed a distinct behavior depending on whether the coarse-graining is performed in the initial conditions (TCLPT \eqref{TCLPTmodel}) or in Eulerian space resulting only in modified dynamics (scgCLPT \eqref{cgCLPTmodel-a}) or also including higher cumulants of the tracer (cgCLPT \eqref{cgCLPTmodel-b}), assumed to be identical to those of dark matter. \highlight{Since all results have to be viewed in the light of fundamental limitations of perturbation theory to capture the correct small-scale behavior, we mainly focus on the impact of the smoothing on large scales for the model comparison.}

Fig.\,\ref{fig:xicgCLPTvsTCLPT} illustrates the real space halo correlation function and that the dominant effect of coarse-graining on the Lagrangian radius is to damp the BAO feature around $r\approx 110\Mpc$ compared to CLPT. As expected, the impact of this modification becomes greater for massive halos since the Lagrangian size of the halo grows with mass. Given the already quite good agreement between the CLPT prediction and the data points evident from Fig.\,\ref{fig:z0realAllmass} and the strong damping, the Lagrangian radius cannot be considered an optimal smoothing scale when aiming at optimizing the agreement with the $N$-body data for the real space correlation function. \highlight{In the case considered in \cite{KUA15}, where only the proto-halo centers in the realization are displaced, no damping around the BAO scale is observed. This means that the apparent disagreement  most likely originates from the underlying local Lagrangian biasing scheme being not suitable to account for the fact that proto-halo centers are special rather than random points in the initial density field. We will address this issue in a future paper in the context of peak bias \cite{D08}, where the correlation of density peaks is considered and spatial derivatives acting on the linear density correlation are included. We have preliminary results suggesting that the usage of peak bias indeed alleviates the issues present in the current model based on local Lagrangian bias.}

The pairwise velocity depicted in Fig.\,\ref{fig:v12cgCLPTvsTCLPT} shows that the two smoothing procedures either implemented in the initial conditions (TCLPT) or Eulerian space (cgCLPT) lead to different predictions. Smoothing only the initial conditions leaving the dynamics unaffected as done for TCLPT \eqref{TCLPTmodel} increases the absolute value of the pairwise velocity compared to CLPT in agreement with the data. Smoothing in Eulerian space as done for cgCLPT (\ref{cgCLPTmodel-a}/\ref{cgCLPTmodel-b}) leads to a reduced mass-dependence and systematically increases the absolute value of the mean pairwise velocity compared to CLPT, more prominently on small scales and around the BAO. cgCLPT improves the agreement with the data on large scales compared to CLPT for small and intermediate masses, see  Fig.\,\ref{fig:v12CLPTvscgCLPT}.

For the pairwise velocity dispersion the main consequence of the coarse-graining on the Lagrangian scale is to downsize the global amplitude and narrow down the mass dependence of the CLPT prediction as evident from Figs.\,\ref{fig:sigxcgCLPTvsTCLPT} and \ref{fig:sigxpcgCLPTvsTCLPTonlyKbar}. This effect is mainly caused by the mass dependence of the smoothing scale and there is only a minor difference between the modified dynamics contained in scgCLPT \eqref{cgCLPTmodel-a} and a smoothing of the initial conditions implemented by TCLPT \eqref{TCLPTmodel}. Furthermore, we showed in Fig.\,\ref{fig:sigxpcgCLPTvsTCLPTonlyKbar} that the cgCLPT model \eqref{cgCLPTmodel-b} including higher tracer cumulants significantly increases the amplitude of the velocity dispersion compared to the scgCLPT model \eqref{cgCLPTmodel-a} that follows the same modified dynamics but does not include velocity dispersion. Most notably, the measurements from the $N$-body simulation suggest that the bias assumption \eqref{velbiasassumpt-a} behind scgCLPT \eqref{cgCLPTmodel-a}, neglecting higher cumulants of the tracers themselves, is more appropriate than cgCLPT \eqref{cgCLPTmodel-b} to describe dark matter halos. This is expected since proto-halos behave significantly more like a single-streaming fluid than the underlying dark matter distribution.\\

\section{Conclusion and Outlook}
\label{sec:concl}
\label{sec:outlook}

In the first part of the paper we derived the Edgeworth streaming model (ESM \eqref{ESM}) for redshift space distortions  which reduces to the Gaussian streaming model (GSM) when neglecting non-Gaussianities, starting from a general distribution function for dark matter tracers without making any assumption about the dynamics of those tracers and their relation to the dark matter field. We then studied the accuracy of both the GSM and ESM on the basis of the Horizon Run 2 $N$-body simulation halo catalog with masses ranging from large galaxy to cluster sized halos finding excellent agreement of the GSM for the quadrupole and hexadecapole on scales $s\gtrsim 30\,\text{Mpc}/h$ and an improvement of the leading order ESM over GSM on scales $s\lesssim 30\,\text{Mpc}/h$, see Fig.\,\ref{fig:GSMconsratio02}. 

In the second part of the paper we describe how to infer the streaming model ingredients for the GSM/ESM from 
perturbation theory based on the dust model to determine the redshift space halo correlation function. Since halos carry an intrinsic length scale given by their Lagrangian radius, we considered a coarse-grained dust fluid to capture the basic properties of the proto-halo fluid being comprised of extended objects. In Sec.\,\ref{sec:GSM-dust} we presented two different ways to incorporate a smoothing into the dust fluid description, implemented either in Eulerian or Lagrangian space, and proposed two possible biasing schemes to relate dark matter and halo phase space distribution functions. In Sec.\,\ref{sec:GSM-CLPT} we evaluated the streaming model ingredients for the coarse-grained dust model within the Post-Zel'dovich approximation in its CLPT incarnation \cite{CRW13} which outperforms standard perturbation theory and even the Zel'dovich approximation in the nonlinear regime. We studied the impact of the smoothing on the streaming model ingredients and compared the coarse-graining in Eulerian space (scgCLPT \eqref{cgCLPTmodel-a} and cgCLPT \eqref{cgCLPTmodel-b}) to an approximate Lagrangian coarse-graining implemented by smoothing the initial power spectrum (TCLPT \eqref{TCLPTmodel}). We found that halos can be described well as single-streaming tracers of the underlying dark matter distribution. Incorporating a smoothing on the Lagrangian size of the halo turned out to have a strong effect on the BAO peak of the real space correlation function and hence as not suitable for improving the CLPT prediction when combined with local Lagrangian bias. By contrast, the smoothing improved the prediction for the pairwise velocity statistics, namely the mean pairwise velocity for small masses and most notably the overall magnitude of the pairwise velocity dispersion irrespective of whether scgCLPT or TCLPT are used. We expect that when combining the smoothing on the Lagrangian radius with the peak bias formalism \cite{D08} will considerably improve the agreement for the real space correlation function and the pairwise velocity statistics.

In a forthcoming work \cite{KUA15} we investigate the appropriate choice of the smoothing scale to predict proto-halo displacement fields using the Zel'dovich approximation on the basis of $N$-body simulations. We then employ the fact that TCLPT can be implemented straightforwardly in CLPT by simply smoothing the input power spectrum and suggest a pragmatic optimizing scheme to accurately predict the ingredients of the GSM and hence the redshift space correlation function from the GSM whose individual accuracy on scales $s\gtrsim 30\Mpc$ has been established. 

\section*{Acknowledgement}
The work of MK \& CU was supported by the DFG cluster of excellence ``Origin and Structure of the Universe''. \highlight{We would like to thank Tobias Baldauf, Luigi Guzzo, Eiichiro Komatsu, Roman Scoccimarro, Uros Seljak, Ravi Sheth and Atsushi Taruya for their input and interesting discussions. We also want to thank the anonymous referee for helpful comments and suggestions that helped to improve the manuscript.}

\newcommand{\apjl}{Astrophys. J. Letters}
\newcommand{\apjs}{Astrophys. J. Suppl. Ser.}
\newcommand{\mnras}{Mon. Not. R. Astron. Soc.}
\newcommand{\pasj}{Publ. Astron. Soc. Japan}
\newcommand{\apss}{Astrophys. Space Sci.}
\newcommand{\aap}{Astron. Astrophys.}
\newcommand{\physrep}{Phys. Rep.}
\newcommand{\mpla}{Mod. Phys. Lett. A}
\newcommand{\jcap}{J. Cosmol. Astropart. Phys.}
\bibliography{StreamingModelCLPTbib}

\begin{thebibliography}{54}
\expandafter\ifx\csname natexlab\endcsname\relax\def\natexlab#1{#1}\fi
\expandafter\ifx\csname bibnamefont\endcsname\relax
  \def\bibnamefont#1{#1}\fi
\expandafter\ifx\csname bibfnamefont\endcsname\relax
  \def\bibfnamefont#1{#1}\fi
\expandafter\ifx\csname citenamefont\endcsname\relax
  \def\citenamefont#1{#1}\fi
\expandafter\ifx\csname url\endcsname\relax
  \def\url#1{\texttt{#1}}\fi
\expandafter\ifx\csname urlprefix\endcsname\relax\def\urlprefix{URL }\fi
\providecommand{\bibinfo}[2]{#2}
\providecommand{\eprint}[2][]{\url{#2}}

\bibitem[{\citenamefont{{Sargent} and {Turner}}(1977)}]{ST77}
\bibinfo{author}{\bibfnamefont{W.~L.~W.} \bibnamefont{{Sargent}}}
  \bibnamefont{and} \bibinfo{author}{\bibfnamefont{E.~L.}
  \bibnamefont{{Turner}}}, \bibinfo{journal}{\apjl}
  \textbf{\bibinfo{volume}{212}}, \bibinfo{pages}{L3} (\bibinfo{year}{1977}).

\bibitem[{\citenamefont{{Kaiser}}(1987)}]{K87}
\bibinfo{author}{\bibfnamefont{N.}~\bibnamefont{{Kaiser}}},
  \bibinfo{journal}{\mnras} \textbf{\bibinfo{volume}{227}}, \bibinfo{pages}{1}
  (\bibinfo{year}{1987}).

\bibitem[{\citenamefont{{Tully} and {Fisher}}(1978)}]{TF78}
\bibinfo{author}{\bibfnamefont{R.~B.} \bibnamefont{{Tully}}} \bibnamefont{and}
  \bibinfo{author}{\bibfnamefont{J.~R.} \bibnamefont{{Fisher}}}, in
  \emph{\bibinfo{booktitle}{Large Scale Structures in the Universe}}, edited by
  \bibinfo{editor}{\bibfnamefont{M.~S.} \bibnamefont{{Longair}}}
  \bibnamefont{and} \bibinfo{editor}{\bibfnamefont{J.}~\bibnamefont{{Einasto}}}
  (\bibinfo{year}{1978}), vol.~\bibinfo{volume}{79} of
  \emph{\bibinfo{series}{IAU Symposium}}, pp. \bibinfo{pages}{31--45}.

\bibitem[{\citenamefont{{Jackson}}(1972)}]{J72}
\bibinfo{author}{\bibfnamefont{J.~C.} \bibnamefont{{Jackson}}},
  \bibinfo{journal}{\mnras} \textbf{\bibinfo{volume}{156}}, \bibinfo{pages}{1P}
  (\bibinfo{year}{1972}).

\bibitem[{\citenamefont{{Peebles}}(1980)}]{P80}
\bibinfo{author}{\bibfnamefont{P.~J.~E.} \bibnamefont{{Peebles}}},
  \emph{\bibinfo{title}{{The large-scale structure of the universe}}}
  (\bibinfo{year}{1980}).

\bibitem[{\citenamefont{{Park} et~al.}(1994)\citenamefont{{Park}, {Vogeley},
  {Geller}, and {Huchra}}}]{PVGH94}
\bibinfo{author}{\bibfnamefont{C.}~\bibnamefont{{Park}}},
  \bibinfo{author}{\bibfnamefont{M.~S.} \bibnamefont{{Vogeley}}},
  \bibinfo{author}{\bibfnamefont{M.~J.} \bibnamefont{{Geller}}},
  \bibnamefont{and} \bibinfo{author}{\bibfnamefont{J.~P.}
  \bibnamefont{{Huchra}}}, \bibinfo{journal}{\apj}
  \textbf{\bibinfo{volume}{431}}, \bibinfo{pages}{569} (\bibinfo{year}{1994}).

\bibitem[{\citenamefont{{Peacock} and {Dodds}}(1994)}]{PD94}
\bibinfo{author}{\bibfnamefont{J.~A.} \bibnamefont{{Peacock}}}
  \bibnamefont{and} \bibinfo{author}{\bibfnamefont{S.~J.}
  \bibnamefont{{Dodds}}}, \bibinfo{journal}{\mnras}
  \textbf{\bibinfo{volume}{267}}, \bibinfo{pages}{1020} (\bibinfo{year}{1994}),
  \eprint{astro-ph/9311057}.

\bibitem[{\citenamefont{{Hatton} and {Cole}}(1998)}]{HC98}
\bibinfo{author}{\bibfnamefont{S.}~\bibnamefont{{Hatton}}} \bibnamefont{and}
  \bibinfo{author}{\bibfnamefont{S.}~\bibnamefont{{Cole}}},
  \bibinfo{journal}{\mnras} \textbf{\bibinfo{volume}{296}}, \bibinfo{pages}{10}
  (\bibinfo{year}{1998}), \eprint{astro-ph/9707186}.

\bibitem[{\citenamefont{{Nishimichi} and {Taruya}}(2011)}]{TN11}
\bibinfo{author}{\bibfnamefont{T.}~\bibnamefont{{Nishimichi}}}
  \bibnamefont{and} \bibinfo{author}{\bibfnamefont{A.}~\bibnamefont{{Taruya}}},
  \bibinfo{journal}{\prd} \textbf{\bibinfo{volume}{84}}, \bibinfo{eid}{043526}
  (\bibinfo{year}{2011}), \eprint{1106.4562}.

\bibitem[{\citenamefont{Fisher}(1995)}]{F94}
\bibinfo{author}{\bibfnamefont{K.~B.} \bibnamefont{Fisher}},
  \bibinfo{journal}{Astrophys.J.} \textbf{\bibinfo{volume}{448}},
  \bibinfo{pages}{494} (\bibinfo{year}{1995}), \eprint{astro-ph/9412081}.

\bibitem[{\citenamefont{{Scoccimarro}}(2004)}]{S04}
\bibinfo{author}{\bibfnamefont{R.}~\bibnamefont{{Scoccimarro}}},
  \bibinfo{journal}{\prd} \textbf{\bibinfo{volume}{70}}, \bibinfo{eid}{083007}
  (\bibinfo{year}{2004}), \eprint{astro-ph/0407214}.

\bibitem[{\citenamefont{{Bianchi} et~al.}(2015)\citenamefont{{Bianchi},
  {Chiesa}, and {Guzzo}}}]{BCG15}
\bibinfo{author}{\bibfnamefont{D.}~\bibnamefont{{Bianchi}}},
  \bibinfo{author}{\bibfnamefont{M.}~\bibnamefont{{Chiesa}}}, \bibnamefont{and}
  \bibinfo{author}{\bibfnamefont{L.}~\bibnamefont{{Guzzo}}},
  \bibinfo{journal}{\mnras} \textbf{\bibinfo{volume}{446}}, \bibinfo{pages}{75}
  (\bibinfo{year}{2015}), \eprint{1407.4753}.

\bibitem[{\citenamefont{{Vlah} et~al.}(2012)\citenamefont{{Vlah}, {Seljak},
  {McDonald}, {Okumura}, and {Baldauf}}}]{VSMOB12}
\bibinfo{author}{\bibfnamefont{Z.}~\bibnamefont{{Vlah}}},
  \bibinfo{author}{\bibfnamefont{U.}~\bibnamefont{{Seljak}}},
  \bibinfo{author}{\bibfnamefont{P.}~\bibnamefont{{McDonald}}},
  \bibinfo{author}{\bibfnamefont{T.}~\bibnamefont{{Okumura}}},
  \bibnamefont{and}
  \bibinfo{author}{\bibfnamefont{T.}~\bibnamefont{{Baldauf}}},
  \bibinfo{journal}{\jcap} \textbf{\bibinfo{volume}{11}}, \bibinfo{eid}{009}
  (\bibinfo{year}{2012}), \eprint{1207.0839}.

\bibitem[{\citenamefont{Vlah et~al.}(2013)\citenamefont{Vlah, Seljak, Okumura,
  and Desjacques}}]{VSOD13}
\bibinfo{author}{\bibfnamefont{Z.}~\bibnamefont{Vlah}},
  \bibinfo{author}{\bibfnamefont{U.}~\bibnamefont{Seljak}},
  \bibinfo{author}{\bibfnamefont{T.}~\bibnamefont{Okumura}}, \bibnamefont{and}
  \bibinfo{author}{\bibfnamefont{V.}~\bibnamefont{Desjacques}},
  \bibinfo{journal}{JCAP} \textbf{\bibinfo{volume}{1310}}, \bibinfo{pages}{053}
  (\bibinfo{year}{2013}), \eprint{1308.6294}.

\bibitem[{\citenamefont{{Angulo} et~al.}(2014)\citenamefont{{Angulo}, {White},
  {Springel}, and {Henriques}}}]{AWSH13}
\bibinfo{author}{\bibfnamefont{R.~E.} \bibnamefont{{Angulo}}},
  \bibinfo{author}{\bibfnamefont{S.~D.~M.} \bibnamefont{{White}}},
  \bibinfo{author}{\bibfnamefont{V.}~\bibnamefont{{Springel}}},
  \bibnamefont{and}
  \bibinfo{author}{\bibfnamefont{B.}~\bibnamefont{{Henriques}}},
  \bibinfo{journal}{\mnras} \textbf{\bibinfo{volume}{442}},
  \bibinfo{pages}{2131} (\bibinfo{year}{2014}), \eprint{1311.7100}.

\bibitem[{\citenamefont{{Taruya} et~al.}(2013)\citenamefont{{Taruya},
  {Nishimichi}, and {Bernardeau}}}]{TNB13}
\bibinfo{author}{\bibfnamefont{A.}~\bibnamefont{{Taruya}}},
  \bibinfo{author}{\bibfnamefont{T.}~\bibnamefont{{Nishimichi}}},
  \bibnamefont{and}
  \bibinfo{author}{\bibfnamefont{F.}~\bibnamefont{{Bernardeau}}},
  \bibinfo{journal}{\prd} \textbf{\bibinfo{volume}{87}}, \bibinfo{eid}{083509}
  (\bibinfo{year}{2013}), \eprint{1301.3624}.

\bibitem[{\citenamefont{Reid and White}(2011)}]{RW11}
\bibinfo{author}{\bibfnamefont{B.~A.} \bibnamefont{Reid}} \bibnamefont{and}
  \bibinfo{author}{\bibfnamefont{M.}~\bibnamefont{White}},
  \bibinfo{journal}{Mon.Not.Roy.Astron.Soc.} \textbf{\bibinfo{volume}{417}},
  \bibinfo{pages}{1913} (\bibinfo{year}{2011}), \eprint{1105.4165}.

\bibitem[{\citenamefont{{Wang} et~al.}(2014)\citenamefont{{Wang}, {Reid}, and
  {White}}}]{WRW14}
\bibinfo{author}{\bibfnamefont{L.}~\bibnamefont{{Wang}}},
  \bibinfo{author}{\bibfnamefont{B.}~\bibnamefont{{Reid}}}, \bibnamefont{and}
  \bibinfo{author}{\bibfnamefont{M.}~\bibnamefont{{White}}},
  \bibinfo{journal}{\mnras} \textbf{\bibinfo{volume}{437}},
  \bibinfo{pages}{588} (\bibinfo{year}{2014}), \eprint{1306.1804}.

\bibitem[{\citenamefont{{White}}(2014)}]{W14}
\bibinfo{author}{\bibfnamefont{M.}~\bibnamefont{{White}}},
  \bibinfo{journal}{\mnras} \textbf{\bibinfo{volume}{439}},
  \bibinfo{pages}{3630} (\bibinfo{year}{2014}), \eprint{1401.5466}.

\bibitem[{\citenamefont{White et~al.}(2015)\citenamefont{White, Reid, Chuang,
  Tinker, McBride et~al.}}]{WR14}
\bibinfo{author}{\bibfnamefont{M.}~\bibnamefont{White}},
  \bibinfo{author}{\bibfnamefont{B.}~\bibnamefont{Reid}},
  \bibinfo{author}{\bibfnamefont{C.-H.} \bibnamefont{Chuang}},
  \bibinfo{author}{\bibfnamefont{J.~L.} \bibnamefont{Tinker}},
  \bibinfo{author}{\bibfnamefont{C.~K.} \bibnamefont{McBride}},
  \bibnamefont{et~al.}, \bibinfo{journal}{Mon.Not.Roy.Astron.Soc.}
  \textbf{\bibinfo{volume}{447}}, \bibinfo{pages}{234} (\bibinfo{year}{2015}),
  \eprint{1408.5435}.

\bibitem[{\citenamefont{Bernardeau et~al.}(2002)\citenamefont{Bernardeau,
  Colombi, Gaztanaga, and Scoccimarro}}]{B01}
\bibinfo{author}{\bibfnamefont{F.}~\bibnamefont{Bernardeau}},
  \bibinfo{author}{\bibfnamefont{S.}~\bibnamefont{Colombi}},
  \bibinfo{author}{\bibfnamefont{E.}~\bibnamefont{Gaztanaga}},
  \bibnamefont{and}
  \bibinfo{author}{\bibfnamefont{R.}~\bibnamefont{Scoccimarro}},
  \bibinfo{journal}{Phys.Rept.} \textbf{\bibinfo{volume}{367}},
  \bibinfo{pages}{1} (\bibinfo{year}{2002}), \eprint{astro-ph/0112551}.

\bibitem[{\citenamefont{{Buchert}}(1992)}]{B92}
\bibinfo{author}{\bibfnamefont{T.}~\bibnamefont{{Buchert}}},
  \bibinfo{journal}{\mnras} \textbf{\bibinfo{volume}{254}},
  \bibinfo{pages}{729} (\bibinfo{year}{1992}).

\bibitem[{\citenamefont{{Bond} and {Myers}}(1996)}]{BM96}
\bibinfo{author}{\bibfnamefont{J.~R.} \bibnamefont{{Bond}}} \bibnamefont{and}
  \bibinfo{author}{\bibfnamefont{S.~T.} \bibnamefont{{Myers}}},
  \bibinfo{journal}{\apjs} \textbf{\bibinfo{volume}{103}}, \bibinfo{pages}{1}
  (\bibinfo{year}{1996}).

\bibitem[{\citenamefont{{Porto} et~al.}(2014)\citenamefont{{Porto}, {Senatore},
  and {Zaldarriaga}}}]{PSZ13}
\bibinfo{author}{\bibfnamefont{R.~A.} \bibnamefont{{Porto}}},
  \bibinfo{author}{\bibfnamefont{L.}~\bibnamefont{{Senatore}}},
  \bibnamefont{and}
  \bibinfo{author}{\bibfnamefont{M.}~\bibnamefont{{Zaldarriaga}}},
  \bibinfo{journal}{\jcap} \textbf{\bibinfo{volume}{5}}, \bibinfo{eid}{022}
  (\bibinfo{year}{2014}), \eprint{1311.2168}.

\bibitem[{\citenamefont{{Cooray} and {Sheth}}(2002)}]{CS02}
\bibinfo{author}{\bibfnamefont{A.}~\bibnamefont{{Cooray}}} \bibnamefont{and}
  \bibinfo{author}{\bibfnamefont{R.}~\bibnamefont{{Sheth}}},
  \bibinfo{journal}{\physrep} \textbf{\bibinfo{volume}{372}},
  \bibinfo{pages}{1} (\bibinfo{year}{2002}), \eprint{astro-ph/0206508}.

\bibitem[{\citenamefont{{Bond} et~al.}(1991)\citenamefont{{Bond}, {Cole},
  {Efstathiou}, and {Kaiser}}}]{Bondetal}
\bibinfo{author}{\bibfnamefont{J.~R.} \bibnamefont{{Bond}}},
  \bibinfo{author}{\bibfnamefont{S.}~\bibnamefont{{Cole}}},
  \bibinfo{author}{\bibfnamefont{G.}~\bibnamefont{{Efstathiou}}},
  \bibnamefont{and} \bibinfo{author}{\bibfnamefont{N.}~\bibnamefont{{Kaiser}}},
  \bibinfo{journal}{\apj} \textbf{\bibinfo{volume}{379}}, \bibinfo{pages}{440}
  (\bibinfo{year}{1991}).

\bibitem[{\citenamefont{{Lacey} and {Cole}}(1993)}]{Laceyetal}
\bibinfo{author}{\bibfnamefont{C.}~\bibnamefont{{Lacey}}} \bibnamefont{and}
  \bibinfo{author}{\bibfnamefont{S.}~\bibnamefont{{Cole}}},
  \bibinfo{journal}{\mnras} \textbf{\bibinfo{volume}{262}},
  \bibinfo{pages}{627} (\bibinfo{year}{1993}).

\bibitem[{\citenamefont{{Tassev}}(2014)}]{Ta14}
\bibinfo{author}{\bibfnamefont{S.}~\bibnamefont{{Tassev}}},
  \bibinfo{journal}{\jcap} \textbf{\bibinfo{volume}{6}}, \bibinfo{eid}{008}
  (\bibinfo{year}{2014}), \eprint{1311.4884}.

\bibitem[{\citenamefont{{Zel'dovich}}(1970)}]{Z70}
\bibinfo{author}{\bibfnamefont{Y.~B.} \bibnamefont{{Zel'dovich}}},
  \bibinfo{journal}{\aap} \textbf{\bibinfo{volume}{5}}, \bibinfo{pages}{84}
  (\bibinfo{year}{1970}).

\bibitem[{\citenamefont{{Coles} et~al.}(1993)\citenamefont{{Coles}, {Melott},
  and {Shandarin}}}]{C93}
\bibinfo{author}{\bibfnamefont{P.}~\bibnamefont{{Coles}}},
  \bibinfo{author}{\bibfnamefont{A.~L.} \bibnamefont{{Melott}}},
  \bibnamefont{and} \bibinfo{author}{\bibfnamefont{S.~F.}
  \bibnamefont{{Shandarin}}}, \bibinfo{journal}{\mnras}
  \textbf{\bibinfo{volume}{260}}, \bibinfo{pages}{765} (\bibinfo{year}{1993}).

\bibitem[{\citenamefont{{Melott} et~al.}(1994)\citenamefont{{Melott},
  {Pellman}, and {Shandarin}}}]{M94}
\bibinfo{author}{\bibfnamefont{A.~L.} \bibnamefont{{Melott}}},
  \bibinfo{author}{\bibfnamefont{T.~F.} \bibnamefont{{Pellman}}},
  \bibnamefont{and} \bibinfo{author}{\bibfnamefont{S.~F.}
  \bibnamefont{{Shandarin}}}, \bibinfo{journal}{\mnras}
  \textbf{\bibinfo{volume}{269}}, \bibinfo{pages}{626} (\bibinfo{year}{1994}),
  \eprint{astro-ph/9312044}.

\bibitem[{\citenamefont{Pietroni et~al.}(2012)\citenamefont{Pietroni, Mangano,
  Saviano, and Viel}}]{P11}
\bibinfo{author}{\bibfnamefont{M.}~\bibnamefont{Pietroni}},
  \bibinfo{author}{\bibfnamefont{G.}~\bibnamefont{Mangano}},
  \bibinfo{author}{\bibfnamefont{N.}~\bibnamefont{Saviano}}, \bibnamefont{and}
  \bibinfo{author}{\bibfnamefont{M.}~\bibnamefont{Viel}},
  \bibinfo{journal}{JCAP} \textbf{\bibinfo{volume}{1201}}, \bibinfo{pages}{019}
  (\bibinfo{year}{2012}), \eprint{1108.5203}.

\bibitem[{\citenamefont{Buchert et~al.}(1994)\citenamefont{Buchert, Melott, and
  Weiss}}]{BMW94}
\bibinfo{author}{\bibfnamefont{T.}~\bibnamefont{Buchert}},
  \bibinfo{author}{\bibfnamefont{A.}~\bibnamefont{Melott}}, \bibnamefont{and}
  \bibinfo{author}{\bibfnamefont{A.}~\bibnamefont{Weiss}}
  (\bibinfo{year}{1994}), \eprint{astro-ph/9412075}.

\bibitem[{\citenamefont{Weiss et~al.}(1996)\citenamefont{Weiss, Gottlober, and
  Buchert}}]{WGB95}
\bibinfo{author}{\bibfnamefont{A.~G.} \bibnamefont{Weiss}},
  \bibinfo{author}{\bibfnamefont{S.}~\bibnamefont{Gottlober}},
  \bibnamefont{and} \bibinfo{author}{\bibfnamefont{T.}~\bibnamefont{Buchert}},
  \bibinfo{journal}{Mon.Not.Roy.Astron.Soc.} \textbf{\bibinfo{volume}{278}},
  \bibinfo{pages}{953} (\bibinfo{year}{1996}), \eprint{astro-ph/9505113}.

\bibitem[{\citenamefont{{Carlson} et~al.}(2013)\citenamefont{{Carlson}, {Reid},
  and {White}}}]{CRW13}
\bibinfo{author}{\bibfnamefont{J.}~\bibnamefont{{Carlson}}},
  \bibinfo{author}{\bibfnamefont{B.}~\bibnamefont{{Reid}}}, \bibnamefont{and}
  \bibinfo{author}{\bibfnamefont{M.}~\bibnamefont{{White}}},
  \bibinfo{journal}{\mnras} \textbf{\bibinfo{volume}{429}},
  \bibinfo{pages}{1674} (\bibinfo{year}{2013}), \eprint{1209.0780}.

\bibitem[{\citenamefont{Matsubara}(2008)}]{M08}
\bibinfo{author}{\bibfnamefont{T.}~\bibnamefont{Matsubara}},
  \bibinfo{journal}{Phys.Rev.} \textbf{\bibinfo{volume}{D78}},
  \bibinfo{pages}{083519} (\bibinfo{year}{2008}), \eprint{0807.1733}.

\bibitem[{\citenamefont{{Uhlemann} and {Kopp}}(2015)}]{UK14}
\bibinfo{author}{\bibfnamefont{C.}~\bibnamefont{{Uhlemann}}} \bibnamefont{and}
  \bibinfo{author}{\bibfnamefont{M.}~\bibnamefont{{Kopp}}},
  \bibinfo{journal}{\prd} \textbf{\bibinfo{volume}{91}}, \bibinfo{eid}{084010}
  (\bibinfo{year}{2015}), \eprint{1407.4810}.

\bibitem[{\citenamefont{{Kim} et~al.}(2009)\citenamefont{{Kim}, {Park}, {Gott},
  and {Dubinski}}}]{KPetal09}
\bibinfo{author}{\bibfnamefont{J.}~\bibnamefont{{Kim}}},
  \bibinfo{author}{\bibfnamefont{C.}~\bibnamefont{{Park}}},
  \bibinfo{author}{\bibfnamefont{J.~R.} \bibnamefont{{Gott}},
  \bibfnamefont{III}}, \bibnamefont{and}
  \bibinfo{author}{\bibfnamefont{J.}~\bibnamefont{{Dubinski}}},
  \bibinfo{journal}{\apj} \textbf{\bibinfo{volume}{701}}, \bibinfo{eid}{1547}
  (\bibinfo{year}{2009}), \eprint{0812.1392}.

\bibitem[{\citenamefont{{Kim} et~al.}(2011)\citenamefont{{Kim}, {Park},
  {Rossi}, {Lee}, and {Gott}}}]{KPetal11}
\bibinfo{author}{\bibfnamefont{J.}~\bibnamefont{{Kim}}},
  \bibinfo{author}{\bibfnamefont{C.}~\bibnamefont{{Park}}},
  \bibinfo{author}{\bibfnamefont{G.}~\bibnamefont{{Rossi}}},
  \bibinfo{author}{\bibfnamefont{S.~M.} \bibnamefont{{Lee}}}, \bibnamefont{and}
  \bibinfo{author}{\bibfnamefont{J.~R.} \bibnamefont{{Gott}},
  \bibfnamefont{III}}, \bibinfo{journal}{Journal of Korean Astronomical
  Society} \textbf{\bibinfo{volume}{44}}, \bibinfo{pages}{217}
  (\bibinfo{year}{2011}), \eprint{1112.1754}.

\bibitem[{\citenamefont{{Samushia} et~al.}(2012)\citenamefont{{Samushia},
  {Percival}, and {Raccanelli}}}]{SPR12}
\bibinfo{author}{\bibfnamefont{L.}~\bibnamefont{{Samushia}}},
  \bibinfo{author}{\bibfnamefont{W.~J.} \bibnamefont{{Percival}}},
  \bibnamefont{and}
  \bibinfo{author}{\bibfnamefont{A.}~\bibnamefont{{Raccanelli}}},
  \bibinfo{journal}{\mnras} \textbf{\bibinfo{volume}{420}},
  \bibinfo{pages}{2102} (\bibinfo{year}{2012}), \eprint{1102.1014}.

\bibitem[{\citenamefont{{Matsubara}}(2000)}]{M00}
\bibinfo{author}{\bibfnamefont{T.}~\bibnamefont{{Matsubara}}},
  \bibinfo{journal}{\apj} \textbf{\bibinfo{volume}{535}}, \bibinfo{pages}{1}
  (\bibinfo{year}{2000}), \eprint{astro-ph/9908056}.

\bibitem[{\citenamefont{{Seljak} and {McDonald}}(2011)}]{SMc11}
\bibinfo{author}{\bibfnamefont{U.}~\bibnamefont{{Seljak}}} \bibnamefont{and}
  \bibinfo{author}{\bibfnamefont{P.}~\bibnamefont{{McDonald}}},
  \bibinfo{journal}{\jcap} \textbf{\bibinfo{volume}{11}}, \bibinfo{eid}{039}
  (\bibinfo{year}{2011}), \eprint{1109.1888}.

\bibitem[{\citenamefont{Bernardeau and Kofman}(1995)}]{BK94}
\bibinfo{author}{\bibfnamefont{F.}~\bibnamefont{Bernardeau}} \bibnamefont{and}
  \bibinfo{author}{\bibfnamefont{L.}~\bibnamefont{Kofman}},
  \bibinfo{journal}{Astrophys.J.} \textbf{\bibinfo{volume}{443}},
  \bibinfo{pages}{479} (\bibinfo{year}{1995}), \eprint{astro-ph/9403028}.

\bibitem[{\citenamefont{{Juszkiewicz} et~al.}(1995)\citenamefont{{Juszkiewicz},
  {Weinberg}, {Amsterdamski}, {Chodorowski}, and {Bouchet}}}]{JWACB95}
\bibinfo{author}{\bibfnamefont{R.}~\bibnamefont{{Juszkiewicz}}},
  \bibinfo{author}{\bibfnamefont{D.~H.} \bibnamefont{{Weinberg}}},
  \bibinfo{author}{\bibfnamefont{P.}~\bibnamefont{{Amsterdamski}}},
  \bibinfo{author}{\bibfnamefont{M.}~\bibnamefont{{Chodorowski}}},
  \bibnamefont{and}
  \bibinfo{author}{\bibfnamefont{F.}~\bibnamefont{{Bouchet}}},
  \bibinfo{journal}{\apj} \textbf{\bibinfo{volume}{442}}, \bibinfo{pages}{39}
  (\bibinfo{year}{1995}), \eprint{astro-ph/9308012}.

\bibitem[{\citenamefont{Blinnikov and Moessner}(1998)}]{BM97}
\bibinfo{author}{\bibfnamefont{S.}~\bibnamefont{Blinnikov}} \bibnamefont{and}
  \bibinfo{author}{\bibfnamefont{R.}~\bibnamefont{Moessner}},
  \bibinfo{journal}{Astron.Astrophys.Suppl.Ser.}
  \textbf{\bibinfo{volume}{130}}, \bibinfo{pages}{193} (\bibinfo{year}{1998}),
  \eprint{astro-ph/9711239}.

\bibitem[{\citenamefont{Kopp et~al.}(2015)\citenamefont{Kopp, Uhlemann, and
  Achitouv}}]{KUA15}
\bibinfo{author}{\bibfnamefont{M.}~\bibnamefont{Kopp}},
  \bibinfo{author}{\bibfnamefont{C.}~\bibnamefont{Uhlemann}}, \bibnamefont{and}
  \bibinfo{author}{\bibfnamefont{I.}~\bibnamefont{Achitouv}}
  (\bibinfo{year}{2015}), \eprint{{\it in preparation}}.

\bibitem[{\citenamefont{{Hamilton}}(1992)}]{H92}
\bibinfo{author}{\bibfnamefont{A.~J.~S.} \bibnamefont{{Hamilton}}},
  \bibinfo{journal}{\apjl} \textbf{\bibinfo{volume}{385}}, \bibinfo{pages}{L5}
  (\bibinfo{year}{1992}).

\bibitem[{\citenamefont{{Mo} and {White}}(1996)}]{MW96}
\bibinfo{author}{\bibfnamefont{H.~J.} \bibnamefont{{Mo}}} \bibnamefont{and}
  \bibinfo{author}{\bibfnamefont{S.~D.~M.} \bibnamefont{{White}}},
  \bibinfo{journal}{\mnras} \textbf{\bibinfo{volume}{282}},
  \bibinfo{pages}{347} (\bibinfo{year}{1996}), \eprint{astro-ph/9512127}.

\bibitem[{\citenamefont{{Catelan} et~al.}(1998)\citenamefont{{Catelan},
  {Lucchin}, {Matarrese}, and {Porciani}}}]{CLMP98}
\bibinfo{author}{\bibfnamefont{P.}~\bibnamefont{{Catelan}}},
  \bibinfo{author}{\bibfnamefont{F.}~\bibnamefont{{Lucchin}}},
  \bibinfo{author}{\bibfnamefont{S.}~\bibnamefont{{Matarrese}}},
  \bibnamefont{and}
  \bibinfo{author}{\bibfnamefont{C.}~\bibnamefont{{Porciani}}},
  \bibinfo{journal}{\mnras} \textbf{\bibinfo{volume}{297}},
  \bibinfo{pages}{692} (\bibinfo{year}{1998}), \eprint{arXiv:astro-ph/9708067}.

\bibitem[{\citenamefont{{Bernardeau} et~al.}(2002)\citenamefont{{Bernardeau},
  {Colombi}, {Gazta{\~n}aga}, and {Scoccimarro}}}]{B02}
\bibinfo{author}{\bibfnamefont{F.}~\bibnamefont{{Bernardeau}}},
  \bibinfo{author}{\bibfnamefont{S.}~\bibnamefont{{Colombi}}},
  \bibinfo{author}{\bibfnamefont{E.}~\bibnamefont{{Gazta{\~n}aga}}},
  \bibnamefont{and}
  \bibinfo{author}{\bibfnamefont{R.}~\bibnamefont{{Scoccimarro}}},
  \bibinfo{journal}{\physrep} \textbf{\bibinfo{volume}{367}},
  \bibinfo{pages}{1} (\bibinfo{year}{2002}), \eprint{astro-ph/0112551}.

\bibitem[{\citenamefont{Dominguez}(2000)}]{D00}
\bibinfo{author}{\bibfnamefont{A.}~\bibnamefont{Dominguez}},
  \bibinfo{journal}{Phys.Rev.} \textbf{\bibinfo{volume}{D62}},
  \bibinfo{pages}{103501} (\bibinfo{year}{2000}).

\bibitem[{\citenamefont{{Uhlemann} et~al.}(2014)\citenamefont{{Uhlemann},
  {Kopp}, and {Haugg}}}]{UKH14}
\bibinfo{author}{\bibfnamefont{C.}~\bibnamefont{{Uhlemann}}},
  \bibinfo{author}{\bibfnamefont{M.}~\bibnamefont{{Kopp}}}, \bibnamefont{and}
  \bibinfo{author}{\bibfnamefont{T.}~\bibnamefont{{Haugg}}},
  \bibinfo{journal}{\prd} \textbf{\bibinfo{volume}{90}}, \bibinfo{eid}{023517}
  (\bibinfo{year}{2014}), \eprint{1403.5567}.

\bibitem[{\citenamefont{{Kubo}}(1962)}]{K62}
\bibinfo{author}{\bibfnamefont{R.}~\bibnamefont{{Kubo}}},
  \bibinfo{journal}{Journal of the Physical Society of Japan}
  \textbf{\bibinfo{volume}{17}}, \bibinfo{pages}{1100} (\bibinfo{year}{1962}).

\bibitem[{\citenamefont{{Desjacques}}(2008)}]{D08}
\bibinfo{author}{\bibfnamefont{V.}~\bibnamefont{{Desjacques}}},
  \bibinfo{journal}{\prd} \textbf{\bibinfo{volume}{78}}, \bibinfo{eid}{103503}
  (\bibinfo{year}{2008}), \eprint{0806.0007}.

\end{thebibliography}

\begin{widetext}

\appendix

\section{Abbreviations}
\label{AppAbb}
\begin{tabular}{llc}
\textbf{abbr.} & \textbf{full expression} & \textbf{reference }\\ \hline
HR2 & Horizon Run 2 halo catalog & \cite{KPetal09, KPetal11}\\\hline
ZA & Zel'dovich approximation & \cite{Z70}\\
TZA & truncated Zel'dovich approximation computed with smoothed input power spectrum & \cite{M94} \\
PZA & Post-Zel'dovich approximation (higher order perturbation theory)& \cite{B92}\\
TPZA & truncated Post-Zel'dovich approximation computed with smoothed input power spectrum &  \cite{BMW94,WGB95}\\ \hline
LPT & Lagrangian perturbation theory & \cite{B92}\\
CLPT & Convolution Lagrangian perturbation theory (dust model), approximation to PZA & \eqref{fXdust}, \cite{CRW13} \\
cgCLPT &  coarse-grained Convolution Lagrangian perturbation theory (coarse-grained dust model) & (\ref{cgCLPTmodel-a}/\ref{cgCLPTmodel-b}), \cite{UK14} \\
TCLPT & truncated  Convolution Lagrangian perturbation theory (dust model) & \eqref{TCLPTmodel}\\ \hline
GSM & Gaussian Streaming model & \eqref{GSM2}\\
ESM & Edgeworth Streaming model & \eqref{ESM2} \\
\end{tabular}

\section{Lagrangian correlators}
\label{AppL}

\subsection{Lagrangian framework}
In Lagrangian perturbation theory the exact displacement field $\v{\varPsi}(\tau,\vq)$ is expanded in a series with spatial parts $\v{\varPsi}^{(n)}(\vq)$ and temporal coefficients given by the scale factor $a(\tau)$ in an Einstein-de Sitter universe
\begin{align}
\label{pertExpPsi}
\v{\varPsi}(\tau,\vq) &= \sum_{n=1}^\infty a^n(\tau)\v{\varPsi}^{(n)}(\vq) \,.
\end{align}
The orders $\v{\varPsi}^{(n)}$ are expressed in Fourier space with the help of perturbative kernels $\v{L}^{(n)}$ in terms of the linear density field $\delta_L$
\begin{align}
\label{defL}
\v{\varPsi}^{(n)}(\vk)&= i \int \frac{\vol{3}{p_1}\ldots\ \vol{3}{p_n}}{(2\pi)^{3(n-1)}} \delta_{\rm D}(\vk-\vp_{1\cdots n})  \v{L}^{(n)}(\vp_1,\ldots,\vp_n) \delta_L(\vp_1) \cdots \delta_L(\vp_n) \,.
\end{align}
Note that we employ here a different notation for $\v{L}^{(n)}$ compared to Eq.\,(A2) in \cite{M08} such that when translating the results an additional prefactor $n!$ has to be taken into account. The vector valued kernels $\v{L}^{(n)}$ can be split into a longitudinal component $\v{S}^{(n)}$ and a transverse part $\v{T}^{(n)}$ according to
\begin{align}
\label{decompL}
\v{L}^{(n)} &= \v{S}^{(n)}+ \v{T}^{(n)} \ , \
\vk \times  \v{S}^{(n)}(\vp_1,\ldots,\vp_n) = 0\ , \
\vk\cdot \v{T}^{(n)}(\vp_1,\ldots,\vp_n)=0 \text{ with } \vk:=\vp_1+\ldots+\vp_n\,.
\end{align}

In addition to those definitions, it is useful to define the following mixed polyspectra of the linear density field and the displacement field in the same way as done in \cite{M08}
\begin{align}
\label{defC}
\Big\langle \delta_L(\vk_1)\cdots\delta_L(\vk_l)\varPsi_{i_1}^{(n_1)}(\vp_1)\cdots\varPsi_{i_m}^{(n_m)}(\vp_m) \Big\rangle_c = (2\pi)^3 \delta_{\mathrm D}(\vk_1+\ldots+\vk_l+\vp_1+\ldots+\vp_m) (-i)^m C_{i_1\cdots i_m}^{(n_1\cdots n_m)} (\vk_1,\ldots,\vk_l;\vp_1,\ldots,\vp_m) \,,
\end{align}
where an angle bracket with index $c$ denote cumulants (connected correlators).
For computations up to 1-loop level we only have to consider terms up to $\mathcal O(P_L^2)$ which implies $l+n_1+n_2+n_3\leq 4$ since due the properties of the cumulants only terms with $l+m \leq 3$ are relevant. Furthermore only even $l+n_1+n_2+n_3 \in 2\mathbb N$ contribute because the initial density field is assumed to be a random Gaussian field. For $l+m=2$ we adopt the simplified notation
\begin{align}
C(\vk):= C(\vk,-\vk) \quad , \quad C_i(\vk):= C_i(\vk;-\vk) \quad , \quad C_{ij}(\vk):= C_{ij}(\vk,-\vk) \,.
\end{align}
The $C$ as defined in \eqref{defC} should not be confused with cumulants of the phase space distribution function. We will also encounter mixed polyspectra with some $\v{\Psi}$ replaced by $\bar{\v{\Psi}}$ or by time derivatives. When combining the Gaussian streaming model with Convolution Lagrangian perturbation theory (CLPT) we will adopt another notation for the correlators, see Eq.\,\eqref{defCor} in Sec.\,\ref{sec:GSM-CLPT} and \cite{CRW13,WRW14}.
For the correlators involving time derivatives of $\v{\Psi}$ we simply use that $\dot{\v{\Psi}}^{(n)} = nf\sH \v{\Psi}^{(n)}$ and similarly for $\bar{\v{\Psi}}$.
The specific index structure enforced by translation symmetry allows to describe Lagrangian correlators in real space entirely in terms of scalar functions of $q=|\vq| = |\vq_2-\vq_1|$, since any tensor can be decomposed in terms of $\delta_{ij}$ and $\hat q_i$. For example, any rank-1 tensor can be written as $T_i(\vq)=T(q)\hat q_i$ and similarly   any rank-2 tensor can be decomposed according to $T_{ij}(\vq)=T_{\delta q}(q) \delta_{ij}+T_{qq}(q) \hat q_i \hat q_j$. The $q$-dependence of the functions can be expressed using spherical Bessel functions, namely
\begin{align}
\int_{-1}^1 d\mu\ \cos(x\mu) &= 2j_0(x) \qquad \int_{-1}^1 d\mu\ \mu^2 \cos(x\mu) = 2\left(j_0(x) - 2\frac{j_1(x)}{x}\right)\,.\label{Bessel}
\end{align}

We maintain the notation used in \cite{M08} and \cite{CRW13}, in which the $R$ and $Q$-functions, defined as 
\begin{align}
Q_n(k) &= \frac{k^3}{4\pi^2} \int_0^\infty dr P_L(kr) \int_{-1}^1 dx \ P_L(k\sqrt{1-2rx+r^2})\ \tilde Q_n(k,r,x) \,,\\
R_n(k) &= \frac{k^3}{4\pi^2} P_L(k) \int_0^\infty dr P_L(kr)\ \tilde R_n(k,r) \,,
\end{align}
have been computed for the standard fluid case. 

\subsection{Lagrangian correlators for the coarse-grained dust model (cgCLPT)}

In the following we state the results for the Lagrangian correlators obtained from the coarse-grained dust model \eqref{cgCLPTmodel-b}. Since for the CLPT computation, see Eqs. (B20-30) and (B41-46) in \cite{CRW13} only $Q_{1,2,5,8}$ are relevant, they are the only ones which will be listed here. 

\begin{subequations}
\label{tildeQ}
\begin{align}
\notag \tilde{\bar{Q}}_1(k,r,x) &= \frac{e^{-\sigx^2k^2 (2r^2-2 rx+1)}}{36 \left(r^2-2 r x+1\right)^2} \Bigg\{49\left[1+3 x^2+4\left(-rx+r^2x^2-rx^3\right)\right]\\
&\qquad\qquad\qquad\qquad -14 \left[7(1+3 x^2)+4\left(-4rx+4r^2x^2-10rx^3+3r^2x^4\right)\right] e^{\sigx^2k^2 \left(r^2-r x\right)}\\
\notag &\qquad\qquad\qquad\qquad +\left[49(1+3 x^2)+36r^2+4\left(-7rx-11r^2x^2-91rx^3+51r^2x^4\right)\right] e^{ 2\sigx^2k^2 \left(r^2-r x\right)} \Bigg\}\\
\tilde Q_1 (k,r,x) &=\lim_{\sigx\rightarrow 0}\tilde{\bar{Q}}_1(k,r,x)=\frac{r^2 \left(x^2-1\right)^2}{\left(r^2-2 r x+1\right)^2}\\
\tilde{\bar{Q}}_2(k,r,x) &= \frac{r (r x-1) e^{-\sigx^2 k^2  \left(2r^2-2r x+1\right)}}{6 \left(r^2-2 r x+1\right)^2} \left[ \left(7r-21 rx^2+34 x^3-20 x\right)e^{\sigx^2 k^2 (r^2-rx)}- \left(7r-21 r x^2+28 x^3-14 x\right) e^{\sigx^2 k^2 (r^2-1)}\right]\\
\tilde Q_2 (k,r,x) &=\lim_{\sigx\rightarrow 0}\tilde{\bar{Q}}_2(k,r,x)=\frac{r x \left(x^2-1\right) (r x-1)}{\left(r^2-2 r x+1\right)^2}\\
\tilde{\bar{Q}}_5(k,r,x) &= \frac{r e^{\sigx^2k^2 (-4 r^2 + 2rx-3)}}{6 \left(r^2-2 r x+1\right)^2} \left[\left(13r+14 x+34 r^2x^3-20 r^2x-41rx^2\right) e^{\sigx^2k^2 r^2 }- \left(7r+14 x+28 r^2x^3-14 r^2 x-35rx^2\right) e^{\sigx^2k^2 rx}\right]\\
\tilde Q_5 (k,r,x) &=\lim_{\sigx\rightarrow 0}\tilde{\bar{Q}}_5(k,r,x)= \frac{r^2 \left(x^2-1\right) (r x-1)}{\left(r^2-2 r x+1\right)^2}\\
\tilde{\bar{Q}}_8(k,r,x) &= \frac{r e^{-\sigx^2k^2 (2 r^2-2rx+1)}}{3 \left(r^2-2 r x+1\right)} \left[\left(7rx^2-7x\right)-\left(10 rx^2-3r-7 x\right) e^{\sigx^2k^2 (r^2-rx) } \right]\\
\tilde Q_8 (k,r,x) &=\lim_{\sigx\rightarrow 0}\tilde{\bar{Q}}_8(k,r,x)=\frac{r^2 \left(1-x^2\right)}{r^2-2 r x+1}
\end{align}
\end{subequations}

Furthermore all quantities which contain the linear power spectrum, more precisely Eqs. (B20,25,41,44) in \cite{CRW13}, have to be computed with the smoothed linear power spectrum, such that $P_L(k)\rightarrow \bar P_L(k) = \exp\left(-\sigx^2k^2\right)P_L(k)$.

In addition to the usual $\tilde{\bar{R}}_{n=1,2}$, which are modified, we had to define another kernel $\tilde{\bar{R}}_0$ which accounts for the fact, that in our case the quantities $C_i^{(3)}$ and $C_{ij}^{(13)}$ cannot be expressed in terms of $\tilde{\bar{R}}_1$. This is due to the different smoothing structure of $C_i^{(3)}$ and $C_{ij}^{(13)}$, which both contain two quantities, compared to $C_i^{(2)}$ and $C^{(12)}_{ij}$, which both contain three parts. As explained in \cite{M08} any transverse part of $C_{i_1\cdots i_m}^{(n_1\cdots n_m)}$ is irrelevant such that we only obtain longitudinal parts for $C_{i_1\cdots i_m}^{(n_1\cdots n_m)}$ for which, however, the transverse kernels $\v{T}^{(n)}$ have to be taken into account.
\begin{align}
\bar C^{(3)}_i(\v{k}) = \frac{5}{21}\frac{k_i}{k^2} \bar R_0(k)\,, \qquad
\bar C^{(13)}_{ij}(\v{k}) = \bar C^{(31)}_{ij}(\v{k}) = -\frac{5}{21} \frac{k_i k_j}{k^4} \bar R_0(k) \,.
\end{align}
This manifests itself in the following kernels which are instead of Eqs. (B26,43,46) in \cite{CRW13} then given by
\begin{align}
\bar U^{(3)}(q) &= \frac{1}{2\pi^2} \int_0^\infty dk\ k \left(-\frac{5}{21}\right) \bar R_0(k) j_1(kq) \\
\bar X^{(13)}(q)&=\frac{1}{2\pi^2} \int_0^\infty dk\ \frac{5}{21}\bar R_0(k) \left[\frac{2}{3}-2\frac{j_1(kq)}{kq} \right]\qquad
\bar Y^{(13)}(q)=\frac{1}{2\pi^2} \int_0^\infty dk\ \frac{5}{21}\bar R_0(k) \left[-2j_0(kq)+6\frac{j_1(kq)}{kq} \right]
\end{align}

\begin{subequations}
\label{tildeR}
\begin{align}
\notag \tilde{\bar{R}}_0(k,r) &= \frac{e^{-\sigx^2k^2 (r^2+1)}}{480 r^3 } \Bigg\{ 3 \left(r^2-1\right)^3 \Bigg[ 51 \left(r^2+1\right) e^{\frac{1}{2} \sigx^2k^2 (r^2+1) } \left(\text{Ei}\left[-\tfrac{1}{2} \sigx^2k^2 (r-1)^2 \right]-\text{Ei}\left[-\tfrac{1}{2} \sigx^2k^2 (r+1)^2 \right]\right)\\
\notag &\qquad\qquad\qquad\qquad\qquad -4 \log \left|\frac{r-1}{r+1}\right| \left(\left(7 r^2+2\right) e^{\sigx^2k^2 r^2}+21 \left(r^2+1\right)\right)\Bigg]\\
\notag &\qquad\qquad\qquad\qquad -168 r \left(3 r^6-5 r^4+9 r^2-3\right)  -8r \left(21 r^6-50 r^4+79 r^2-6\right) e^{\sigx^2k^2 r^2 }\\
\notag &\qquad\qquad\qquad\qquad -18e^{- \sigx^2k^2 r} \left(\frac{1}{(\sigx k)^2} (r+1) (-17 + 51 r + 44 r^2 + 68 r^3 - 51 r^4 + 17 r^5) \right.\\
&\qquad\qquad\qquad\qquad\qquad\qquad + \frac{2}{(\sigx k)^4} (17 + 44 r + 88 r^2 + 68 r^3 - 17 r^4) \\
\notag &\qquad\qquad\qquad\qquad\qquad\qquad\left.+\frac{8}{(\sigx k)^6} (17 r^2+24r+11) +\frac{192}{(\sigx k)^8}\right)\\
\notag &\qquad\qquad\qquad\qquad +18 e^{ \sigx^2k^2 r } \left(\frac{1}{(\sigx k)^2} (r-1) (17 + 51 r - 44 r^2 + 68 r^3 + 51 r^4 + 17 r^5) \right.\\
\notag &\qquad\qquad\qquad\qquad\qquad\qquad - \frac{2}{(\sigx k)^4} (-17 + 44 r - 88 r^2 + 68 r^3 + 17 r^4) \\
\notag &\qquad\qquad\qquad\qquad\qquad\qquad\left. + \frac{8}{(\sigx k)^6} (17 r^2-24r+11) + \frac{192}{(\sigx k)^8}\right)\Bigg\}\\
\notag \tilde{\bar{R}}_1(k,r) & = \frac{e^{-\sigx^2k^2 \left(r^2+1\right)}}{288 r^3 } \Bigg\{ 3  \left(r^2-1\right)^4 \left[28\log \left|\frac{r-1}{r+1}\right| -17 e^{\frac{1}{2} \sigx^2k^2 (r^2+1)} \left(\text{Ei}\left[-\tfrac{1}{2} \sigx^2k^2 (r-1)^2 \right]-\text{Ei}\left[-\tfrac{1}{2} \sigx^2k^2 (r+1)^2 \right]\right)\right]\\
&\qquad\qquad\quad -56  r \left(-3 r^6+11 r^4+11 r^2-3\right) \\
\notag &\qquad\qquad\quad -12  \left[\frac{2r}{(\sigx k)^2 } \left(17 r^4+22 r^2+17\right) +\frac{32r}{(\sigx k)^4} \left(r^2+1\right)\right]\cosh \left(\sigx^2k^2 r \right)\\
\notag &\qquad\qquad\quad -12   \left[\frac{1}{(\sigx k)^2 } \left(r^2+1\right) \left(17 r^4-90 r^2+17\right) -\frac{2}{(\sigx k)^4 } \left(17 r^4+22 r^2+17\right) -\frac{32}{(\sigx k)^6} \left(r^2+1\right)\right] \sinh \left(\sigx^2k^2 r \right) \Bigg\}\\
\tilde R_1 (k,r) &=\lim_{\sigx\rightarrow 0}\tilde{\bar{R}}_1(k,r)=\lim_{\sigx\rightarrow 0}\tilde{\bar{R}}_0(k,r)= \frac{-6 r^7+22 r^5+22 r^3-6 r-3 \left(r^2-1\right)^4 \log \left|\frac{r-1}{r+1}\right|}{48 r^3}
\end{align}
\begin{align}
\notag \tilde{\bar{R}}_2(k,r) &= \frac{\left(r^2-1\right) e^{-\sigx^2k^2 \left(r^2+1\right)}}{288 r^3} \Bigg\{3 \left(r^2-1\right)^2 \left(r^2+1\right) \left(28 \log \left|\frac{r-1}{r+1}\right| -17 e^{\frac{1}{2} P^2 (r^2+1) R^2} \left(\text{Ei}\left[-\tfrac{1}{2} \sigx^2k^2 (r-1)^2 \right]-\text{Ei}\left[-\tfrac{1}{2} \sigx^2k^2 (r+1)^2 \right]\right)\right)\\
&\qquad\qquad\qquad\qquad\quad +56 r \left(3 r^4-2 r^2+3\right) \\
\notag &\qquad\qquad\qquad\qquad\quad -12  \left[\frac{34r}{(\sigx k)^2}\left(r^2+1\right) +\frac{32r}{(\sigx k)^4}\right] \cosh \left(\sigx^2k^2 r\right)\\
\notag &\qquad\qquad\qquad\qquad\quad -12 \left[\frac{1}{(\sigx k)^2} \left(17 r^4-22 r^2+17\right) - \frac{34}{(\sigx k)^4} \left(r^2+1\right) -\frac{32}{(\sigx k)^6}\right] \sinh \left(\sigx^2k^2 r\right)\Bigg\} \\
\tilde R_2 (k,r) &=\lim_{\sigx\rightarrow 0}\tilde{\bar{R}}_2(k,r) = \frac{\left(1-r^2\right) \left[6 r^5-4 r^3+6 r+3 \left(r^2-1\right)^2 \left(r^2+1\right) \log \left|\frac{r-1}{r+1}\right|\right]}{48 r^3}
\end{align}
\end{subequations}
Note that, in the limit $\sigx\rightarrow 0$ we correctly recover the result of \cite{M08}.

\section{Contribution $K_2^\sigx$ to $K_2$ \eqref{K2} in cgCLPT}
\label{AppK2}
In the following we explicitly state the explicit result for the $\sigx$-correction term to $K_2$ which is given by the second tracer cumulant
\begin{align}
K^{\sigma_x}_{2,ij} &:= \exp\left[\sum_{N=1}^\infty \frac{i^N}{N!} \langle \tilde X^N_{\tilde J=0}\rangle_c\right]  \times \sum_{N=0}^\infty \frac{i^N}{N!} \Bigg\langle \tilde X^N_{\v{J}=0}\left[\left( \frac{\overline{(1+\delta)v_iv_j}}{1+\bar \delta}-\bar v_i\bar v_j \right)(\vx_1(\vq_1)) +\left(\frac{\overline{(1+\delta)v_iv_j}}{1+\bar \delta}-\bar v_i\bar v_j\right)(\vx_2(\vq_2))\right] \Bigg\rangle_c \ \Bigg|_{\mathcal O(P_L^2)} \,. \label{K2sigxapp}
\end{align}

Note that we used the notation $v_i$ for the dust velocity which yield the ordinary kernels for CLPT as given in \cite{CRW13}, \cite{WRW14}. In contrast, we use $\bar v_i := \overline{(1+\delta)v_i}/(1+\bar\delta)$ for the mass-weighted velocity which corresponds to our cgCLPT kernels computed for $\bar{\v{ v}}(\vx(\vq))=a\dot{\bar{\v{\varPsi}}}(\vq)$. We have performed a similar calculation to the one presented as pedagogical example B4 in \cite{CRW13} and rely on results given in \cite{M08}. For convenience, the derivation was carried out by computing mixed correlators between coarse-grained density and displacements, $\bar \delta_L$ and $\bar{\v{\varPsi}}$, and dust quantities, $\delta_L$ and $\v{\varPsi}$. Since the extra terms encoded in Eqs.\,\eqref{K2sigxapp} are given in Eulerian space we first had to perform a mapping to Lagrangian coordinates. This has been done according to $\vx(\tau)=\vq + \v{\varPsi}(\tau,\vq)$ by using the Jacobian $F_{ij}=\del x_i/\del q_j = \delta_{ij}+\varPsi_{i,j}$ with determinant $ J_F=\det F_{ij}$. More details concerning the mapping from Eulerian to Lagrangian space for the case of the coarse-grained dust model and explicit relations for the Lagrangian kernels up to third order can be found in \cite{UK14}. The result up to $\mathcal O (P_L^2)$ is 

 \begin{align}
 K_{2,ij}^\sigx  &=e^{-\tfrac{1}{2}K_{ij}k_ik_j} e^{-\tfrac{1}{2} (\lambda_1^2+\lambda_2^2)\sigma_R^2}  \times \Big\{ K_{2,ij}^{\sigma_x (\text{L})}+ K_{2,ij}^{\sigma_x (b0,2)} +i(\lambda_1+\lambda_2)K_{2,ij}^{\sigma_x (b1,2)}- \lambda_1\lambda_2\xi_L K_{2,ij}^{\sigma_x  (\text{L})}-(\lambda_1+\lambda_2)U_i^{10}k_i K_{2,ij}^{\sigma_x  (\text{L})}\Big\} 
 \end{align}
where
 \begin{align}
 K_{2,ij}^{\sigma_x (\text{L})} &:=\int \frac{\text{d} \,k}{2\pi^2} \cdot \frac{2}{3}\Big\{ [P_L(k)-\bar P_L(k)] \Big\}\delta_{ij}\\
\notag K_{2,ij}^{\sigma_x (b1,2)} &:= \int  \frac{\text{d} \,k}{2\pi^2} \frac{6}{7}\Big\{ \left[ \exp\left(-\sigma_x^2k^2\right)R_1(k)-\bar R_1(k)\right]\Big(1+j_0(k q)\Big)\delta_{ij}\\
&\qquad\qquad \  \  - \left[\exp\left(-\sigma_x^2k^2\right)[R_1(k)+2R_2(k)]-[\bar R_1(k)+2\bar R_2(k)] \right]\left(\frac{1}{3}+\frac{j_1(kq)}{kq}\right)\delta_{ij}\\
\notag &\qquad\qquad\  \ - \left[\exp\left(-\sigma_x^2k^2\right)[R_1(k)+2R_2(k)]-[\bar R_1(k)+2\bar R_2(k)] \right]\left(j_0(kq)-3\frac{j_1(kq)}{kq}\right)\hat{q}_i\hat{q}_j\Big\}\\
 K_{2,ij}^{\sigma_x (b0,2)} &:=\int  \frac{\text{d} \,k}{2\pi^2}\ k^2 \cdot \Bigg\{\left[\left[\mathcal Q_3(k)+\mathfrak{Q}_3(k)\right]+\frac{1}{3}\left[\mathcal Q_4(k)+\mathfrak{Q}_4(k)\right]+\left[\mathcal Q_1(k)+\mathfrak{Q}_1(k)\right]j_0(kq)+\left[\mathcal Q_2(k)+\mathfrak{Q}_2(k)\right]\left(\frac{j_1(kq)}{kq} \right)\right]\delta_{ij} \label{K2calQ}\\
\notag &\qquad\qquad\qquad +\frac{1}{k^2}\left[\frac{20}{21} [R_1(k)-\bar R_0(k)]+\frac{12}{49}[Q_1(k)-\bar Q_1(k)]\right]\delta_{ij}  +\left[\mathcal Q_2(k)+\mathfrak{Q}_2(k)\right]\left(j_0(kq)-3\frac{j_1(kq)}{kq} \right)\hat{q}_i\hat{q}_j \Bigg\} 
 \end{align}
with the spherical Bessel functions $j_0$ and $j_1$. The $R$ and $Q$ terms are the usual CLPT kernels from \cite{M08} whereas $\bar R$ and $\bar Q$ are our corresponding cgCLPT kernels given in \eqref{tildeQ} and \eqref{tildeR}. The additional kernels, $\mathfrak{Q}_{1-4}$ and $\mathcal{Q}_{1-4}$, appearing in \eqref{K2calQ} define two other classes of functions besides $R$ and $Q$,  according to
\begin{subequations}
\begin{align}
\mathcal{Q}_n(k) &= \frac{k^3}{4\pi^2} \int_0^\infty dr\, P_L(kr) \int_{-1}^1 dx \ P_L(k\sqrt{1-2rx+r^2})\ \tilde{\mathcal{Q}}_n(k,r,x) \,,\\
\notag \tilde{\mathcal{Q}}_n (k,r,x) &\stackrel{\sigx\rightarrow 0}{\longrightarrow} 0\\
\mathfrak{Q} _n(k) &= \frac{k^3}{4\pi^2} P_L(k)\int_0^\infty dr \int_{-1}^1 dx \ P_L(k\sqrt{1-2rx+r^2})\ \tilde{\mathfrak{Q}}_n(k,r,x) \,,\\
\notag \tilde{\mathfrak{Q}}_n(k,r,x) &\stackrel{\sigx\rightarrow 0}{\longrightarrow} 0\,, 
\end{align}
\end{subequations}
with
\begin{subequations}
\begin{align}
\tilde{\mathcal{Q}}_1(k,r,x) &= \frac{r \left(1-x^2\right) e^{-\sigx^2k^2 \left(2 r^2+1\right) }}{7 k^2 \left(r^2-2 r x+1\right)^2} \left(e^{\sigx^2k^2 r x}-e^{\sigx^2k^2 r^2}\right) \left[\left(10 rx^2-3r-7 x\right) e^{\sigx^2k^2 r^2}- 7x\left(rx-1\right) e^{\sigx^2k^2 r x}\right]\,, \\
\tilde{\mathcal{Q}}_2(k,r,x) &=\frac{\left(3 rx^2-r-2 x\right) e^{-\sigx^2k^2 \left(2 r^2+1\right)}}{7 k^2 \left(r^2-2 r x+1\right)^2} \left(e^{\sigx^2k^2 rx}-e^{\sigx^2k^2 r^2}\right) \left[\left(10rx^2-3r-7 x\right) e^{\sigx^2k^2 r^2 }- 7x\left(rx-1\right)e^{\sigx^2k^2 r x}\right]\,, \\
\tilde{\mathcal{Q}}_3(k,r,x) &= \frac{r \left(1-x^2\right) e^{-\sigx^2k^2 \left(2 r^2+1\right)}}{7 k^2 \left(r^2-2 r x+1\right)^2} \left[3 r \left(1-x^2\right) e^{\sigx^2k^2 \left(2 r^2+1\right)}+\left(10 rx^2-3r-7 x\right) e^{\sigx^2k^2 r(r+x)}-7x\left(rx-1\right) e^{2\sigx^2k^2 rx}\right]\,,\\
\tilde{\mathcal{Q}}_4(k,r,x) &= \frac{\left(3 rx^2-r-2 x\right) e^{-\sigx^2k^2 \left(2 r^2+1\right)}}{7 k^2 \left(r^2-2 r x+1\right)^2} \left[3 r \left(1-x^2\right) e^{\sigx^2k^2 \left(2 r^2+1\right)}+\left(10 rx^2-3r-7 x\right) e^{\sigx^2k^2 r (r+x)}-7x\left(rx-1\right) e^{2 \sigx^2k^2 rx}\right]\,,
\end{align}
\begin{align}
\tilde{\mathfrak{Q}}_1(k,r,x) &=-\frac{2 r^4 \left(1-x^2\right) e^{-\sigx^2k^2\left(r^2+2\right) }}{7 k^2 \left(r^2-2 r x+1\right)^2} \left[6 \left(x^2-1\right) e^{\sigx^2k^2 \left(r^2+1\right)}+\left(21 r x-34 x^2+6\right) e^{\sigx^2k^2 (r x+1)}-7\left(3rx-4x^2 \right) e^{2 \sigx^2k^2 r x}\right]\,,\\
\notag\tilde{\mathfrak{Q}}_2(k,r,x) &= \frac{2 r^3 e^{-\sigx^2k^2 \left(r^2+2\right)}}{7 k^2 \left(r^2-2 r x+1\right)^2} \left[6 \left(1-x^2\right) \left(3rx^2-r-2 x\right) e^{\sigx^2k^2 \left(r^2+1\right) }\right.\\
 &\qquad\qquad\qquad\qquad\quad +\left(7 r^2 \left(5-9 x^2\right) x+2 r \left(51 x^4-19 x^2-4\right)-68 x^3+40 x\right) e^{\sigx^2k^2 (r x+1)}\\
\notag &\qquad\qquad\qquad\qquad\quad\left.+7 \left(r^2 \left(9 x^2-5\right) x+2 r \left(-6 x^4+x^2+1\right)+8 x^3-4 x\right) e^{2\sigx^2k^2 rx}\right]\,,\\
\tilde{\mathfrak{Q}}_3(k,r,x) &= -\frac{2 r^4 \left(1-x^2\right) e^{-\sigx^2k^2 \left(r^2+2\right) }}{7 k^2 \left(r^2-2 r x+1\right)^2} \left[6 \left(x^2-1\right) e^{\sigx^2k^2 \left(r^2+2\right)}+\left(21 r x-34 x^2+6\right) e^{\sigx^2k^2 (r x+1)}-7 (3 rx-4 x^2) e^{2 \sigx^2k^2 rx}\right]\,,\\
\notag \tilde{\mathfrak{Q}}_4(k,r,x) &= \frac{2 r^3 e^{-\sigx^2k^2\left(r^2+2\right)}}{7 k^2 \left(r^2-2 r x+1\right)^2} \left[6 \left(1-x^2\right) \left(3 rx^2-r-2 x\right) e^{\sigx^2k^2 \left(r^2+2\right)}\right.\\
&\qquad\qquad\qquad\qquad\quad +\left(7 r^2 \left(5-9 x^2\right) x+2 r \left(51 x^4-19 x^2-4\right)-68 x^3+40 x\right) e^{\sigx^2k^2 (r x+1)}\\
\notag &\qquad\qquad\qquad\qquad\quad\left.+7 \left(r^2 \left(9 x^2-5\right) x+2 r \left(-6 x^4+x^2+1\right)+8 x^3-4 x\right) e^{2 \sigx^2k^2 r x}\right] \,.
\end{align}
\end{subequations}

\end{widetext}

\end{document}